\documentclass[11pt,urlcolor=blue, linkcolor=blue]{article} 
\usepackage{cite}

\usepackage{amsmath, amsthm, amssymb,slashed}
\usepackage{mathtools}

\usepackage{ifpdf}
\ifpdf
  \usepackage[pdftex]{graphicx}
  \usepackage{epstopdf}
\else
  \usepackage[dvips]{graphicx}
\fi
\usepackage[letterpaper]{geometry}
\parskip=\baselineskip

\usepackage{tablefootnote}

\usepackage[usenames, dvipsnames]{xcolor}

\allowdisplaybreaks[1]
\numberwithin{equation}{section}

\newcommand{\cblue}[1]{\textcolor{blue}{#1}}
\newcommand{\cred}[1]{\textcolor{black}{#1}}

\usepackage{upgreek}

\newcommand{\GSD}{\text{GSD}}
\usepackage{tabularx}

\DeclareMathAlphabet{\mathpzc}{OT1}{pzc}{m}{it}

\newcommand{\bea}{\begin{eqnarray}}
\newcommand{\eea}{\end{eqnarray}}
\def\be{\begin{equation}}
\def\ee{\end{equation}}

\newcommand{\re}{\hspace{1pt}\mathrm{e}}

\usepackage{color}
\usepackage[colorlinks,citecolor=blue]{hyperref}
\definecolor{red}{rgb}{1,0,0}
\definecolor{blue}{rgb}{0,0,1}
\definecolor{dblue}{rgb}{0,0,0.4}
\definecolor{green}{rgb}{0,1,0}
\definecolor{black}{rgb}{0,0,0}
\definecolor{white}{rgb}{1,1,1}

\definecolor{brn}{rgb}{.8,.4,.0}
\definecolor{redo}{rgb}{1,.5,.0}
\definecolor{ddgrn}{rgb}{0,0.4,0}
\definecolor{dgrn}{rgb}{0,0.55,0}
\definecolor{dbl}{rgb}{0,0,0.5}

\usepackage[bbgreekl]{mathbbol}
\usepackage{amscd}

\newcommand{\ii}{\hspace{1pt}\mathrm{i}\hspace{1pt}}
\newcommand{\dd}{\hspace{1pt}\mathrm{d}}

\newcommand{\Ref}[1]{Ref.~\cite{#1}}
\newcommand{\Eq}[1]{Eq.~(\ref{#1})} 
\newcommand{\eq}[1]{(\ref{#1})} 
\newcommand{\eqn}[1]{Eq.~(\ref{#1})} 
\newcommand{\Eqn}[1]{Eq.~(\ref{#1})} 

\newcommand{\Tr}{{\rm Tr}}

\newcommand{\prt}{\partial}

\newcommand{\bpm}{\begin{pmatrix}}
\newcommand{\epm}{\end{pmatrix}}
\newcommand{\bmm}{\begin{matrix}}
\newcommand{\emm}{\end{matrix}}

\newcommand{\cD}{ {\cal D} }

\newcommand{\cH}{ {\cal H} } 
 
\newcommand{\cL}{ {\cal L} }

\newcommand{\al}{\alpha} 
\newcommand{\bt}{\beta}

\newcommand{\ga}{\gamma}

\newcommand{\Z}{\mathbb{Z}}

\def\Z{{\mathbb{Z}}}

\def\R{{\mathbb{R}}}

\def\Tr{{\mathrm{Tr}}}

\def\bZ{{\mathbf{Z}}}

\def\diag{{\mathrm{diag}}}

\DeclareRobustCommand\sWang
{\cblue{\includegraphics[height=4.55ex]{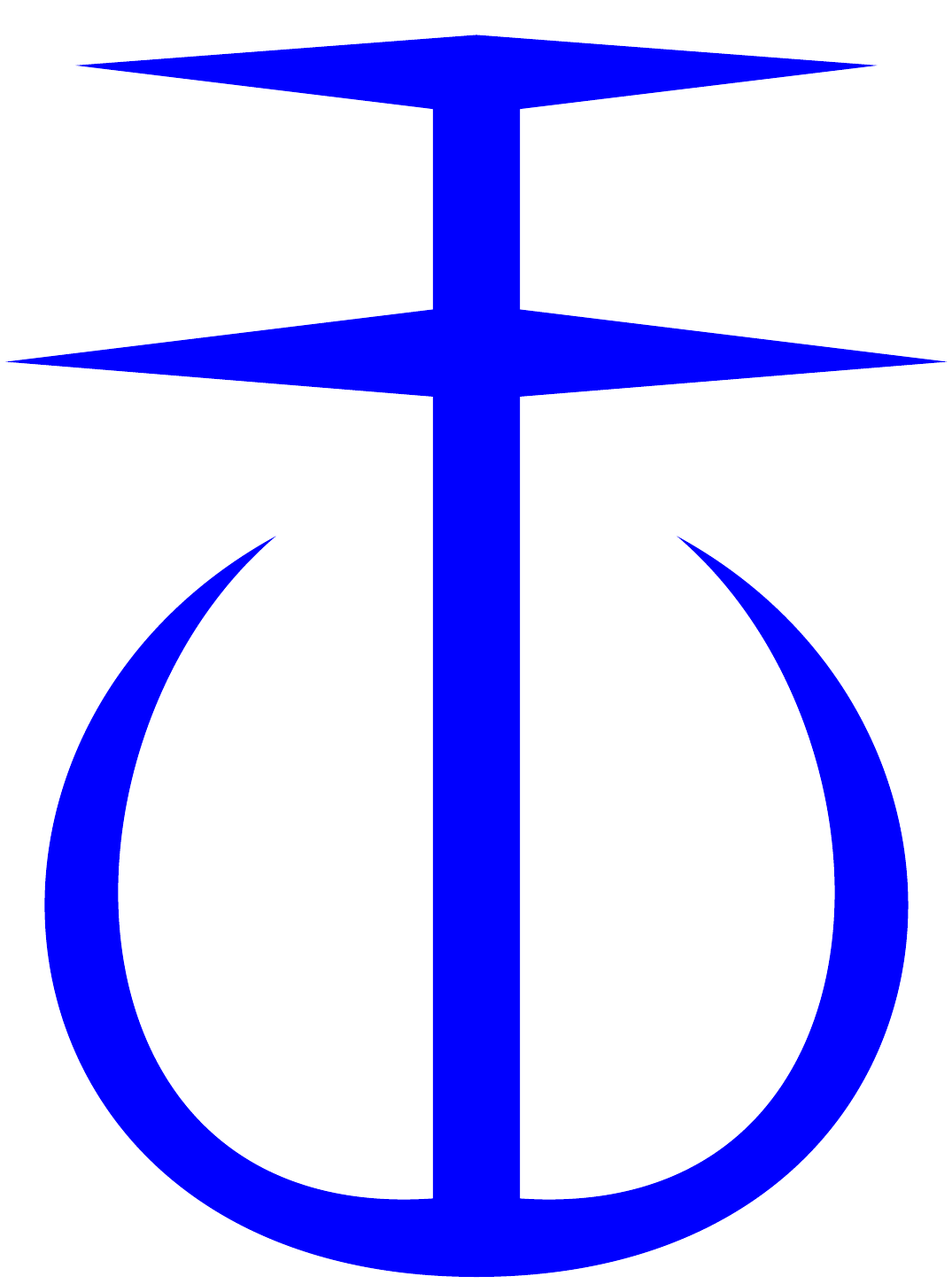}}}

\newcommand{\Wangfootnote}[1]{%
\let\oldthefootnote=\thefootnote%
\stepcounter{mpfootnote}%
\addtocounter{footnote}{-1}%
\renewcommand{\thefootnote}{\sWang}
\footnote{#1}%
\let\thefootnote=\oldthefootnote%
}

\DeclareRobustCommand\sXu%
{\cblue{\includegraphics[height=5.55ex]{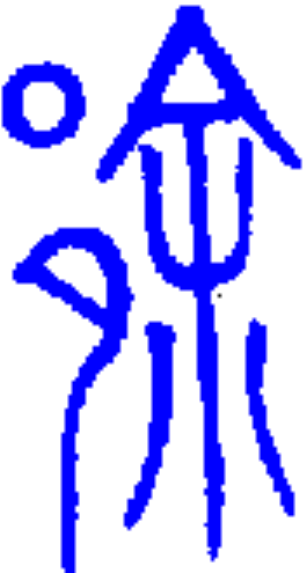}}}

\newcommand{\Xufootnote}[1]{%
\let\oldthefootnote=\thefootnote%
\stepcounter{mpfootnote}%
\addtocounter{footnote}{-1}%
\renewcommand{\thefootnote}{\sXu}
\footnote{#1}%
\let\thefootnote=\oldthefootnote%
}

\newcommand{\Wpfootnote}[1]{%
\let\oldthefootnote=\thefootnote%
\stepcounter{mpfootnote}%
\addtocounter{footnote}{-1}%
\renewcommand{\thefootnote}{{{W$^+$\,}}}
\footnote{#1}%
\let\thefootnote=\oldthefootnote%
}

\newcommand{\Xmfootnote}[1]{%
\let\oldthefootnote=\thefootnote%
\stepcounter{mpfootnote}%
\addtocounter{footnote}{-1}%
\renewcommand{\thefootnote}{{{X$^-$\,}}}
\footnote{#1}%
\let\thefootnote=\oldthefootnote%
}

\usepackage{enumitem,cleveref}
\newcommand{\nn}{\nonumber}

\usepackage{comment}

\def \- {\!\smallsetminus\!}

\def\B{\mathrm{B}}

\newcommand{\SO}{{\rm SO}}

\newcommand{\U}{{\rm U}}

\newcommand{\rN}{{\rm N}}

\def \rH{\operatorname{H}}

\def \Z{\mathbb{Z}}

\newcommand{\Sec}[1]{Sec.~\ref{#1}} 

\newcommand{\Fig}[1]{Fig.~\ref{#1}}

\usepackage{tikz}
\usetikzlibrary{calc}

\usepackage[all,cmtip]{xy}
\usepackage{tikz-cd}

\usetikzlibrary{matrix}
\usetikzlibrary{knots}

\usetikzlibrary{calc}

\usepackage[all,cmtip]{xy}
\usepackage{tikz-cd}

\usepackage{datetime}

\newenvironment{myfont}[2][]{\csname#2\endcsname[#1]}{}

\usepackage{makecell} 

\usepackage{slashed}
\usepackage[makeroom]{cancel}
\usepackage[normalem]{ulem} 
\usepackage{soul}

\usepackage[X2,T1]{fontenc}
\DeclareSymbolFont{cyrillic}{X2}{cmr}{m}{n}
\SetSymbolFont{cyrillic}{bold}{X2}{cmr}{bx}{n}
\DeclareMathSymbol{\khk}{\mathord}{cyrillic}{139}
\DeclareMathSymbol{\zh}{\mathord}{cyrillic}{117}
\DeclareMathSymbol{\zhen}{\mathord}{cyrillic}{182}
\DeclareMathSymbol{\zhe}{\mathord}{cyrillic}{134}

\def\ceq{\coloneqq}

\def\re{{\mathrm{e}}}

\def \U{\mathrm{U}}
\def \rO{\mathrm{O}}

\usepackage{esint}

\begin{document}
\begin{titlepage}
\begin{flushright}
\end{flushright}
\vskip 12.mm

\begin{center}

{\bf\LARGE{
Higher-Rank 
Tensor  
Field Theory  \\[5.5mm]  
of Non-Abelian Fracton and Embeddon 
\\[6.8mm] 
}}

\vskip.5cm
\quad\quad\quad
\Large{
Juven Wang$^{1,2}${
\Wpfootnote{e-mail: {\tt $^{1} $ jw@cmsa.fas.harvard.edu, $^{2}$ juven@ias.edu
(Corresponding Author) \href{http://sns.ias.edu/~juven/}{http://sns.ias.edu/$\sim$juven/} 
}}} 
\quad and  \quad Kai Xu$^{3}${
\Xmfootnote{e-mail: {\tt $^3$ kaixu@math.harvard.edu}
\hfill  August 2019 \\[2mm]
{
\begin{center}
  \emph{Dedicated to \\
90 years of Gauge Principle since Hermann Weyl} [Elektron und Gravitation, Zeit. f\"ur Physik 56, 330-352 (1929)]
 \\ 
\emph{and 65 years of Yang-Mills theory} [Phys. Rev. 96, 191 (1954)] \emph{in 2019}.\\ 
{\includegraphics[height=3.5ex]{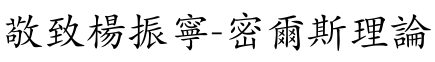}}
\end{center}
\flushleft .
}
} 
}}
 
\vskip.5cm
{\small{\textit{$^1${Center of Mathematical Sciences and Applications, Harvard University,  Cambridge, MA 02138, USA} \\}}
}
\vskip.2cm
{\small{\textit{$^2$School of Natural Sciences, Institute for Advanced Study,  Einstein Drive, Princeton, NJ 08540, USA}\\}}
\vskip.2cm
{\small{\textit{$^3$ Department of Mathematics, Harvard University, Cambridge, MA 02138, USA}\\}}

\end{center}
\vskip.35cm
\baselineskip 12pt
\begin{abstract}

We formulate a new class of tensor gauge field theories in any dimension that is a hybrid class between 
symmetric higher-rank tensor gauge theory (i.e., higher-spin gauge theory) and anti-symmetric tensor topological field theory. 
Our theory describes a mixed unitary phase interplaying between 
gapless and gapped topological order  phases (which can live with or without Euclidean, Poincar\'e 
or anisotropic symmetry, at least in ultraviolet high or intermediate energy field theory, but not yet to a lattice cutoff scale).
The ``gauge structure'' can be compact, continuous, abelian or non-abelian.
Our theory sits outside the paradigm of 
Maxwell electromagnetic theory in 1865 
and Yang-Mills isospin/color theory in 1954. 
We discuss its local gauge transformation in terms of the ungauged vector-like or tensor-like higher-moment global symmetry. 
The non-abelian gauge structure is caused by gauging the non-commutative symmetries: a higher-moment symmetry and a charge conjugation (particle-hole) symmetry.
Vector global symmetries along time direction may exhibit time crystals. 
We explore the relation of these long-range entangled matters 
to a 
non-abelian generalization of Fracton order in condensed matter,  a field theory formulation of foliation, 
 the spacetime embedding and Embeddon that we newly introduce, and possible fundamental physics applications to dark matter or dark energy. 


\end{abstract}
\end{titlepage}


\renewcommand{\eqref}{\eqn}

  \pagenumbering{arabic}
    \setcounter{page}{2}
    
\tableofcontents   


\section{Introduction
}
\label{sec:intro}

Gauge theory is a powerful tool in quantum field theory. 
One of the earliest gauge theories is the renown \emph{James Clerk Maxwell's dynamical} U(1) \emph{gauge theory of electromagnetism} in 1865 \cite{Maxwell1865zz}.
 \emph{Gauge principle} is the underlying principle of gauge theory, proposed by Weyl \cite{Weyl1929ZPhy} and pondered by Pauli and many others \cite{RevModPhys13.203Pauli}.
Chern introduced Chern characteristic classes laying the foundation of non-abelian gauge structure in terms of fiber bundles and connections in 1940s \cite{chern1946characteristic}.
Earlier pioneer works do make an impact, leading 
to another famous gauge theory attempting for a physics realization: \emph{Yang-Mills theory} in 1954 \cite{PhysRev96191YM1954} with  a non-abelian Lie gauge group 
(e.g. SU(N)). 
Yang-Mills theory incorporates the gauge principle over the isospin --- by promoting the internal global symmetry of isospin (such as flavor symmetry)
to a local symmetry. While its local symmetry transformation, say of matter fields, is compensated by that of the gauge field.
Thus gauge theory is an ideal framework to mathematically formulate the interactions between \emph{matters}, where the gauge field stands for the \emph{force mediator}.

In this work, we develop a new type of gauge theory,
obtained from a hybrid through the marriage between anti-symmetric tensor topological field theory (TQFT) and symmetric higher-rank tensor gauge theory.
We find that the gauge structure \footnote{Here the gauge structure is similar to the usual concept of the gauge group. 
But the gauge structure for our theories is not precisely the same as the familiar Lie group framework  of the well-known gauge theory.
There is still a notion of commutative-ness or non-commutative-ness generator of the  
gauge structure, which we shall still call
the commutative case as abelian,
and the
non-commutative case as non-abelian. 
} can be non-abelian.
Our work is inspired by the fracton order in condensed matter system (see review articles \cite{RahulNandkishore2018sel1803.11196, PretkoReview2020cko2001.01722}).
 To the best of our understanding, our theory is the first new example of \emph{higher-rank tensor compact non-abelian gauge theory} previously unknown to the literature.
 It is also a first example (in fact, we construct a web of families of many examples) 
 to have a continuous non-abelian gauge structure in any dimension in the fracton order literature. 

\noindent
\underline{\emph{Anti-symmetric tensor field theory}} 
has a long history as early as Kalb-Ramond work's on higher differential form gauge fields \cite{KalbRamond1974yc}.
The later development of higher form continuum field theory can be found in 
\cite{Banks2010zn1011.5120, Gaiotto2014kfa1412.5148} and References therein,
including higher-form gauge theory and higher-form global symmetries.
The anomalies of theories with ordinary global symmetries and 
higher-form global symmetries can also be systematically obtained by 
a cobordism theory \cite{Freed2016rqq1604.06527} and a generalized version of cobordism theory with higher classifying spaces \cite{Wan2018bns1812.11967}.
We will focus on a particular \emph{continuum} anti-symmetric tensor TQFT developed in \Ref{Wang1404.7854, Wang1405.7689, Gu2015lfa1503.01768,
YeGu2015eba1508.05689, Wang2016jdt1602.05569, 1602.05951WWY, Tiwari2016zru1603.08429, HeYQZheng2016xpi1608.05393, Putrov2016qdo1612.09298, Wang2018edf1801.05416, Wang1901.11537WWY} and references therein: 
A  \emph{continuum} gauge theory formulation \cite{Putrov2016qdo1612.09298, Wang2018edf1801.05416} of \emph{group cohomology} or \emph{higher group cohomology} type of TQFT
such as Dijkgraaf-Witten theory\cite{DijkgraafWitten1989pz1990}, with a discrete finite gauge group. We term
this finite gauge group for TQFT as $G_{\text{TQFT}}$.

\noindent
\underline{\emph{Symmetric higher-rank tensor field theory}} also has a long history
dating back to the study of arbitrary-integer-spin bosonic fields, including massive and massless boson fields, see
\cite{Dirac1936tgProcRoySocLond, FierzPauli1939ix, SinghHagen1974qz, Fronsdal1978rbPRD} and References therein.
Singh-Hagen \cite{SinghHagen1974qz} studies the Fierz-Pauli theory \cite{FierzPauli1939ix} for massive arbitrary-integer-spin $s$ boson fields.
Fronsdal \cite{Fronsdal1978rbPRD} studies their theory \cite{FierzPauli1939ix, SinghHagen1974qz} 
in the massless limit, then finds that the massless arbitrary-integer-spin $s$ boson fields
can be understood in terms of \emph{symmetric tensor bosonic field} of rank $s$, with certain constraints on the trace, double trace, and divergence, etc.
In this work, we will only focus on a certain \emph{higher-spin theory} 
in terms of a class of specific \emph{symmetric higher-rank tensor gauge theories}.

Our inspiration for  \emph{symmetric higher-rank tensor gauge theory} comes from the condensed matter systems studying the
quantum spin liquids described by emergent higher-rank symmetric tensor gauge fields, say a rank-$n$ tensor $A_{ijk\dots}$, where $n$-indices are symmetrized.
%
If the symmetric tensor gauge field is compact and abelian, 
then it is commonly referred this compact abelian theory as the ``higher-rank U(1) symmetric tensor gauge theory,''
or ``higher-rank U(1) spin liquids'' in the condensed matter literature.
However, we will refrain from using this terminology: 
 the U(1) may be an accidental misnomer.
We will see that there is no precise occurrence of an ordinary U(1) gauge group in the theory,
and the ungauged global symmetry is not simply a U(1) global symmetry group. 
(In facts, see the later \Eq{eq:Sym.Tensor.G}, there is only a \emph{group-analogous} structure that is abelian, instead of the familiar group structure.)
Thus, we instead name this type theory as  
\bea
\text{\emph{a higher-rank symmetric tensor compact abelian gauge theory}.}
\eea
\Ref{2016arXiv160108235RRasmussenYouXu} finds that
the
compact abelian \emph{symmetric} tensor gauge theory is unstable in 2+1D, 
but it becomes stable to be gapless-ness with deconfinement in 3+1D.\footnote{We denote $n+1$d for the $n+1$ spacetime dimensions, with $n$ spatial and 1 time dimensions.
We denote $m$d for the $m$ spacetime dimensions.
We denote $n+1$D for the $n$ spatial and 1 time dimensions.
}
In contrast, the \emph{anti-symmetric} tensor gauge theory
with a continuous gauge group is unstable and flows to
a gapped phase with confinement in 2+1d and 3+1d \cite{Orland1981kuInstantonsDisorderAntisymmetric1982, Pearson1981pkPRDPhaseStructureAntisymmetricTensorGaugeFields}.
We will follow closely a version of compact abelian higher-rank symmetric tensor gauge theory
developed and pioneered by Pretko and others, see for instance 
\Ref{Pretko2016kxt1604.05329, Pretko2016lgv1606.08857, Pretko2017xar1707.03838, Slagle2018kqf1807.00827, Pretko2018jbi1807.11479, Gromov2018nbv1812.05104}, where time and space indices
are treated in an unequal and non-interchangeable footing.
We name this type of theory as:
\bea \label{eq:Anisotropic-FT}
\text{\emph{Anisotropic-type higher-rank symmetric tensor gauge theory}.}
\eea
We also develop another generalization of two kinds of
higher-rank symmetric tensor gauge theory:
\bea \label{eq:E}
\text{\emph{Euclidean-type higher-rank symmetric tensor gauge theory}, or}\\
 \label{eq:L}
\text{\emph{Lorentzian-type higher-rank symmetric tensor gauge theory}.}
\eea

Let us give a quick overview and outlines where we are heading in this work.
Throughout this article, we only focus on the field theory living on a flat Euclidean or Minkowski spacetime $\R^{d+1}$ or $\R^{1,d}$ with Cartesian coordinates.
It will become clear soon that we can formulate non-abelian gauge structures (for example {$\left[\Z_2^C \ltimes \Big( \U(1)_{x_{(d+1)}} \Big) \right]$} introduced in \Sec{sec:ClassofNewNon-AbelianHigher-RankTensorGaugeTheories})  to a higher-rank tensor gauge theory (see later \Eq{eq:gauge-group-analogous}). 

\noindent
$\bullet$
The \emph{first ingredient} to obtain our non-abelian tensor gauge theory is by gauging the higher-moment global symmetry\footnote{For example, a vector global symmetry $ \U(1)_{x_{(d+1)}}$ in spacetime or $ \U(1)_{x_{(d)}}$ in space, see \Sec{sec:matter-vector-higher-moment-vector-global-symmetry}  for more details.} 
and a discrete $\Z_2^C$ charge conjugation symmetry together, relatively with a non-commutative semi-direct product ($\ltimes$) structure. 
This is done in \Sec{sec:gauge-Z2C}.

\noindent
$\bullet$ The \emph{second ingredient} to obtain  our non-abelian gauge structure is by coupling several copies of $(\Z_2^C)^N$ to a twisted gauge theory of TQFT.
The twisted-$(\Z_2^C)^N$ TQFT can be equivalent (strong and exactly dual) to a TQFT with a non-abelian gauge group with non-abelian anyonic excitations.
For example, certain twisted-$(\Z_2^C)^3$-TQFT in 3d (2+1D) can be dual to a non-abelian gauge theory of an order 8 dihedral D$_8$ or quaternion Q$_8$ gauge group \cite{deWildPropitius1995cf9511195, Wang1404.7854}. Similar non-abelian natures of field theories
hold in 4d (3+1D) and higher dimensions \cite{Putrov2016qdo1612.09298, Wang2018edf1801.05416}.

In the remaining of \Sec{sec:intro}, we will review the familiar gauge theories (U(1) Maxwell  and non-abelian Yang-Mills theories), 
in contrast of the new non-abelian higher-rank tensor gauge theories we develop in  \Sec{sec:ClassofNewNon-AbelianHigher-RankTensorGaugeTheories}.
Our gauge field theories in \Sec{sec:ClassofNewNon-AbelianHigher-RankTensorGaugeTheories} include the following properties: 
\begin{enumerate}[label=(\arabic*).]
\item Unitary (thus directly applicable to quantum matter and condensed matter systems),
\item Abelian (e.g. $\prod_{J=1}^N \U(1)_{x_{(d+1)}}$) or non-abelian (e.g. $\prod_{J=1}^N \left[\Z_2^C \ltimes \Big(  \U(1)_{x_{(d+1)}} \Big) \right]$),
\item Compact gauge structure (e.g. $\prod_{J=1}^N \left[\Z_2^C \ltimes \Big(  \U(1)_{x_{(d+1)}}\Big) \right]$ is compact),\footnote{
\cred{Here let us clarify the meaning of compactness of gauge group or gauge structure.\\ 
First, for the ordinary Maxwell's U(1) gauge theory, see \Sec{sec:U1maxwell-gauge},
the gauge field $A_\mu$ is a gauge connection of some principle-$G$ bundle, therefore  $A_\mu$ lives in the Lie algebra value, not the gauge group itself.
In this case, there is no difference between $u(1)$-valued Lie algebra or $\mathbb{R}$-valued Lie algebra for the local gauge connection $A$.
\\
However, there are differences for the structure group ($i.e.$, the gauge group) $G$ of the principle-$G$ bundle to be $G=\U(1)$ or $G=\R$.
\[
\left\{\begin{array}{l} 
\text{If the $G=\U(1) \simeq {\R}/{\Z}$, then the holonomy operator $\exp(\ii q \oint A)$ has its $\oint A$ being compact in U(1),} \\
\text{\quad\quad $i.e.$, $\oint A \sim \oint A + 2\pi$ is identified, while the charge $q \in \Z$ is quantized.}\\
\text{If the $G=\R$, then the holonomy operator $\exp(\ii q \oint A)$ has its $\oint A$ being non-compact in $\R$,}\\ 
\text{\quad\quad while the charge $q \in \R$ is also continuous and un-quantized.}
\end{array}\right.
\]
The large gauge transformations for $A$ exist in $G=\U(1) $, but do not exist in $G=\R$.
For the U(1) or $\R$ gauge theory, the  holonomy $\oint A = \iint \dd A$ can be arbitrary value in U(1) or $\R$ by Stokes theorem.
(In contrast, for a discrete gauge theory, such as the $\Z_n$-gauge theory, the holonomy $\oint A$ is zero for contractible cycles;
the holonomy $\oint A \in \frac{2 \pi}{n} \Z_n$ can be nonzero only winding around a non-contractible cycle [an element in a nontrivial homology group $\rH_1(M,\Z)$].)
\\[2mm]
Second, for the higher-rank tensor gauge theory, the compactness is more subtle. One way to justify the compactness is through the effect of the compact $\theta$-topological term and analogous Witten's effect \cite{Pretko2017xar1707.03838}. Another way to justify the compactness is to modify the previous condition 
$\oint A \sim \oint A + 2\pi$ for the ordinary U(1) gauge theory to
the symmetric tensor case: such that $A \sim A + 2\pi$ under a certain \emph{modified contour integration}, presumably compatible with a lattice cut-off.
The reason that we need to modify the contour integration, say in the rank-2 tensor case, is that the $\oiint$, which integrates over the anti-symmetric area or volume form,
cannot be paired up with the symmetric tensor $A_{ij}$ in a conventional way. See also Remark \ref{remark:Mathematical-ingredients}  in \Sec{sec:conclude}
for the required mathematical components of our field theory formulation.
} 
} 
\item Continuous gauge structure (e.g. thanks to $\U(1)_{x_{(d+1)}}$ or $ \U(1)_{x_{(d)}}$).
\end{enumerate}
In \Sec{sec:InterplayedBetweenGappedAndGaplessPhases},
we examine the {non-abelian tensor gauge theory interplayed between gapped and gapless phases},
including the topological degeneracy from the zero modes, the gapless gapless degrees of freedom and dispersion relations.
In \Sec{sec:Discussions:FractonEmbeddonFoliation}, 
we discuss how our theories can be related to
to a non-abelian generalization of Fracton order in condensed matter,  the spacetime embedding and the Embeddon that we newly introduce,
and a field theory formulation of foliation. 
 In \Sec{sec:conclude},
{we conclude with discussions and future directions on {the field theory quantization (either canonical quantization, or path integral), 
Feynman diagrams, and quantum Hamiltonian lattice models, and applications of our theories to time crystal and dark matter}.

\subsection{Overview of the familiar gauge theories} 
Gauge field allows a local transformation, depending on the spacetime coordinate.
This local transformation is known as the gauge transformation.
The whole partition function $\bZ$ or the path integral, defined through integrating out the gauge in-equivalent configuration phase space of gauge fields,
 is set to be invariant under the gauge transformation. We first review the familiar gauge theories and meanwhile set up our notations.
 For simplicity, we focus on the 4 dimensions.

\subsubsection{Abelian U(1) Maxwell gauge theory}
\label{sec:U1maxwell-gauge}

In the modern formulation of Maxwell's U(1) gauge theory \cite{Maxwell1865zz}, we have a gauge field $A(x^\mu)=A_\mu(x^\mu)\dd x^\mu $ (locally as a 1-form as a differential form,
but globally should be viewed as a U(1) gauge 1-connection), which under an infinitesimal gauge transformation becomes
\bea \label{eq:U1maxwell-gauge}
A_\mu(r) \to A_\mu(r) + \delta A_\mu(r ) = A_\mu(r) + \frac{1}{g} \partial_\mu \eta(r),
\eea
with the spacetime coordinate $r=(r_\mu) (\dd x^\mu)$ where $x^\mu :=(x^0, x^j) :=(x^0, \vec{x}) := (t, \vec{x})$.
Locally $\dd x^\mu$ is a differential 1-form, the $\mu$ runs through the indices of coordinate of spacetime $M$.
The coupling $g$ can be related to electromagnetic coupling $e$ as $g=-e$.
The $\eta(r)$ is locally 0-form with a spacetime dependence. 
Just like the relation of a local curvature to the 1-connection, 
we have the field strength 2-form to the 1-form U(1) gauge field via
\bea
F= \dd A =\frac{1}{2} F_{\mu\nu} (\dd x^\mu \wedge \dd x^\nu) =\frac{1}{2} (\partial_\mu A^\al_\nu -\partial_\nu A^\al_\mu ) (\dd x^\mu \wedge \dd x^\nu),
\eea
with the exterior derivative $\dd$ and the wedge product  $\wedge$.
The  field strength is gauge invariant:
\bea
F_{\mu\nu}(r) &=&F_{\mu\nu}(r) +  \delta F_{\mu\nu}(r ) \to F_{\mu\nu}(r).
\eea
The whole gauge invariant partition function of U(1) gauge theory is meant to be written systematically as:
\bea
\bZ_{\rm U(1),\text{EM}}
=\int [\mathcal{D}A] \re^{\ii\int \frac{1}{2}
F \wedge \star F} =
\int [\mathcal{D}A^\mu] \re^{\ii\int \dd^4x\frac{1}{4}
F^{\mu\nu}F_{\mu\nu}}.
\eea
The $[{\cal D} {A}]$ is the path integral measure, for a configuration of the gauge field $A$.
Here $\star F$ is $F$'s Hodge dual.
We integrated over all allowed gauge inequivalent configurations $\int [{\cal D} {A}]$, while gauge redundancy is 
mod out.
The integration is under a weight factor $\exp\big( \ii S \big)$.
%
In 4d (3+1D), we take the spacetime metric ${\rm g}_{\mu\nu}={\rm g}^{\mu\nu}=\diag(+,-,-,-)$, and $\tilde \epsilon^{\mu_1\mu_2\mu_3\mu_4}= -\tilde\epsilon_{\mu_1\mu_2\mu_3\mu_4}$,
while $\varepsilon_{ijk}=\varepsilon^{ijk}=\varepsilon^{i}{}_{jk}=\varepsilon^{ij}{}_{k}$ where raising or lowering indices does not affect.
Generally we take ${\mu,\nu},\mu_1,\dots$ are spacetime coordinates, while $i,j,k$ are space coordinates only.
This path integral may not be precisely mathematically well-defined, 
however it can be physically sensibly well-defined, for example being regularized via a higher energy cutoff such as a lattice cutoff scale.
Our present work only focus on the physics side of rigor.
%

 Indeed the above U(1) gauge theory follows the \emph{gauge principle} \cite{Weyl1929ZPhy} where the global U(1) symmetry
 of a point operator $\Phi$ 
 is promoted to a local symmetry variation that can be absorbed by
 local symmetry variation of the 1-form gauge field. Then the 1-form gauge field plays the role of 
making the charged matter (namely the matter carrying gauge charge) interacting with each other. Thus the 1-form gauge field behaves as a \emph{force mediator} between
 matter fields. The \emph{force mediator} is known as a spin-1 gauge boson in physics.

\subsubsection{Non-abelian SU(N) Yang-Mills gauge theories}
\label{sec:SUn-YM-gauge}

For the non-abelian \emph{Yang-Mills theory} (YM) \cite{PhysRev96191YM1954}, the gauge field
is locally a 1-form or a 1-connection
\bea
A&=&A_\mu \dd x^\mu= A^\al_\nu  T^\al  \dd x^\mu,
\eea
obtained from parallel transporting the SU(N) fiber of the principal principal-SU(N) {bundle} over the spacetime base manifold $M$.
Here $T^\alpha$ is the generator of Lie algebra {\bf{g}} for the gauge group (say SU(N), with N is called 
the number of \emph{color} for gauge theory, or the number of \emph{isospin} for global symmetry in physics), 
{with the commutator $[T^\alpha,T^\beta]=\ii f^{\alpha \beta \gamma} T^\gamma$, 
where $f^{\alpha \beta \gamma}$ is chosen to be a fully anti-symmetric structure constant.}
Then $A_\mu=A_\mu^\alpha T^\alpha$ is the Lie algebra valued gauge field, in the adjoint representation of the Lie algebra.
In physics, $A_\mu$ can represent the gluon vector field of quantum chromodynamics.
The non-abelian field strength is
$$
F=\dd A+A \wedge A
=\frac{1}{2} F^\al_{\mu\nu} T^\al(\dd x^\mu \wedge \dd x^\nu)
 = \frac{1}{2}(\partial_\mu A^\al_\nu -\partial_\nu A^\al_\mu + \ii f^{\bt\ga\al} A^\bt_{\mu} A^\ga_{\nu} ) T^\al(\dd x^\mu \wedge \dd x^\nu).
$$
For the convenience to bridge to the physics convention, 
we rescale and redefine $A = A^\al_\mu \frac{T^\al}{\ii}  g \dd x^\mu= A^\al_\mu  (-\ii g T^\al) \dd x^\mu= A_\mu  (-\ii g) \dd x^\mu$, thus
we also redefine 
the field strength
\bea
F=\frac{1}{2} F^\al_{\mu\nu}   (-\ii g T^\al)(\dd x^\mu \wedge \dd x^\nu)
 = \frac{1}{2}(\partial_\mu A^\al_\nu -\partial_\nu A^\al_\mu + g f^{\bt\ga\al} A^\bt_{\mu} A^\ga_{\nu}) T^\al(\dd x^\mu \wedge \dd x^\nu).
\eea
The $g$ is YM coupling constant. 
We also have the covariant derivative:
\bea
D_{\mu} = \partial_\mu - \ii g a^\al_\mu T^\al.
\eea
The field strength
$F^\al_{\mu\nu}$
 is not gauge invariant under the gauge transformation with $V=\exp( \ii \eta^\al T^\al)$:
 \bea \label{eq:amu-transf}
{A_{\mu}^\alpha T^\al  \to V(A_{\mu}^\alpha T^\al  + \frac{\ii}{g} \prt_\mu  )V^\dagger}
\simeq  (A_\mu^\alpha  + \frac{1}{g} \prt_\mu\eta^\alpha - f^{\bt \ga \alpha}  \eta^{\bt } A^{\ga}_{\mu}) T^\al + \dots, 
\eea
since the  field strength transforms to
\bea
F^{\al}_{\mu \nu} T^{\al} & \to& ( F^{\al}_{\mu \nu} T^{\al}) - (f^{\bt\ga\al} \eta^\bt F^{\ga}_{\mu \nu} T^\al)  + \dots.
\eea
There are color-electric field and  color-magnetic field related to the field strength
\bea
E^\al_{j} \ceq E^{\al j}  &\ceq&F^\al_{0j}
= \partial_0 A^\al_j -\partial_j A^\al_0 + g f^{\bt\ga\al} A^\bt_{0} A^\ga_{j}
=
- \partial_t A^{\al j} - \nabla_j A^{\al 0}- g f^{\bt\ga\al} A^{\bt 0}_{} A^{\ga j}_{}.\\
B^\al_{j} \ceq B^{\al j} &\ceq&
-\frac{1}{2}\varepsilon^{jkl}F^\al_{kl}
 =
-\frac{1}{2}\varepsilon^{jkl}F^{\al kl}
=
+\frac{1}{2}\varepsilon_{jkl}
(\partial_k A^{\al l} -\partial_l A^{\al k} - g f^{\bt\ga\al} A^{\bt {k}} A^{\ga {l}})  \nn\\
&=&
{
(\vec{\nabla} \times \vec{A})^{\al j}
-\frac{1}{2}
g f^{\bt\ga\al} (\vec{A}^{\bt } \times \vec{A}^{\ga })^j}.
\eea
Both color-electric field and color-magnetic field are \emph{not} gauge-invariant
\bea
E^{\al}_{j} T^{\al} & \to& ( E^{\al}_{j} T^{\al}) - (f^{\bt\ga\al} \eta^\bt E^{\ga}_{j}  T^\al) + \dots, \quad 
B^{\al}_{j} T^{\al} \to ( B^{\al}_{j} T^{\al}) - (f^{\bt\ga\al} \eta^\bt B^{\ga}_{j}  T^\al)  + \dots.
\eea
But the Yang-Mills action and its path integral $\bZ_{\text{YM}}$ are gauge invariant (we set a normalization factor ${\kappa}_{\text{YM}} =\frac{1}{2} c(G)^{-1}$ for the convenience
to match the standard convention):
\bea
\bZ_{\text{YM}} 
&=&\int [\mathcal{D}A] \re^{\ii  
 \int  -\Tr \big( \kappa_{\text{YM}} F \wedge \star F\big)} =
\int [\mathcal{D}A] \re^{\ii 
\int \dd^4x   \Tr \big( (-\frac{1}{2})\kappa_{\text{YM}} F_{\mu\nu}F^{\mu\nu}  \big) }\nn \quad\quad\\
&=&
\int [\mathcal{D}A ] \re^{\ii {c(G) } 
\int \dd^4x  \big(-  \frac{1}{2} \kappa_{\text{YM}}    F^{\al}_{\mu\nu}F^{\al, \mu\nu}   \big) }
=
\int [\mathcal{D}A ] \re^{\ii 
\int \dd^4x \;\frac{1}{4}  \big(  F^{\al,\mu\nu} F^\al_{\mu\nu} \big)  } \nn\\
&=&\int [\mathcal{D}A_\mu^\alpha] \re^{\ii 
 \int \dd^4x \;\frac{1}{4}   \big((F^{\al,0 j}F^\al_{0 j}) +  (F^{\al,jk} F_{\al,jk}) \big)}
=
\int [\mathcal{D}A_\mu^\alpha] \re^{\ii 
\int \dd^4x(-\frac{1}{2})  \big(  (E^\al_{j} )^2 -  (B^\al_{j} )^2  \big) }.
\eea
In the next section, we move on to develop 
our new type of exotic higher-rank tensor gauge theories.


\section{Class of New Non-Abelian Higher-Rank Tensor Gauge Theories}
\label{sec:ClassofNewNon-AbelianHigher-RankTensorGaugeTheories}

\subsection{Euclidean or Lorentz Invariant Non-Abelian Higher-Rank Tensor Gauge Theory}
\label{sec:Euclidean-Lorentz}

Below we can formulate theories in both the Euclidean spacetime $\R^{d+1}$ or in the Minkowski (or Lorentzian) spacetime $\R^{1,d}$.
The two versions of theories have the same form of path integral and the action, thus we will present them together.
We denote the space and time coordinates as $(t,\vec{x}_j)  = (x_\mu)$. Our formulation below can be regarded as
modifying Pretko's spacetime anisotropic theory \cite{Pretko2018jbi1807.11479} to a spacetime isotropic theory (up to the signature of Minkowski metric).

\subsubsection{Matter field theory and higher-moment vector global symmetry}
\label{sec:matter-vector-higher-moment-vector-global-symmetry}
We can start from a matter field theory in $(d+1)$d with the global symmetry including the spacetime symmetry and the internal symmetry.
We focus on first the scalar charge matter theory with a vector global symmetry.
Let the complex charge matter field called 
\bea\Phi (t,\vec{x})  =\Phi (x) \in \mathbb{C}, 
\eea

\begin{figure}[!h]
	\centering
	\includegraphics[width=14.cm]{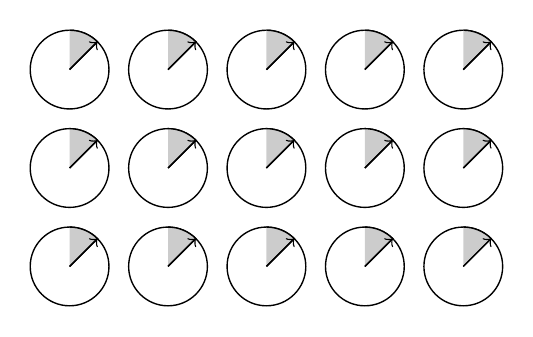}
	\caption[caption]{The ordinary (0-form) U(1) global symmetry transformation acts on the complex charged matter 
	$\Phi(x) \in \mathbb{C}$, e.g. the rotor field, distributed on the spacetime coordinate $x$. 
	For demonstration, here we show various $\Phi(x)$ fields sitting on discretized lattice points on $x$.
	Thus we get \Eq{eq:oglobal-1}: $\Phi \to e^{\ii \eta} \Phi$.
	}
	\label{clock-ordinary-global-1}
\end{figure}

\begin{figure}[!h]
	\centering
	\includegraphics[width=14.cm]{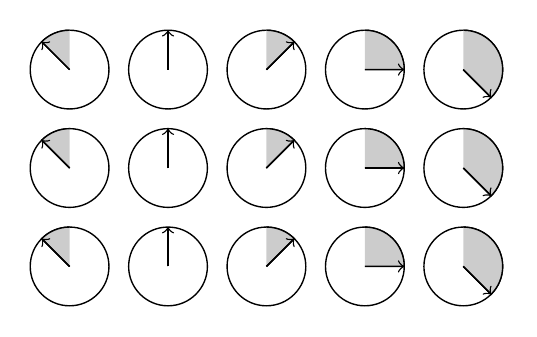}
	\caption[caption]{The {vector global symmetry} belongs to a generalized class of higher-moment symmetry. 
	The {vector} U(1) global symmetry transformation acts on the complex charged matter $\Phi(x)$, e.g. the rotor field, following the set-up in \Fig{clock-ordinary-global-1}.
	Thus we get \Eq{eq:vglobal-1}:
$\Phi \to e^{\ii \eta_v(x)} \Phi := e^{\ii  \Lambda \cdot x} \Phi$.
The angle $\Lambda \cdot x$ depends on a reference point (say $x=0$) and the distance $x$ away from the reference point.
	}
	\label{clock-vector-global-3}
\end{figure}

\begin{enumerate}
\item Internal symmetry:\\
The usual ordinary (0-form) U(1) global symmetry transformation acts on the charged matter as,
\bea \label{eq:oglobal-1}
\Phi \to e^{\ii \eta} \Phi,
\eea
where $\eta$ is a global parameter associated to the U(1) scalar charge, independent of spacetime $(t,\vec{x})$. See Fig. \ref{clock-ordinary-global-1}.

In addition, we like to impose an additional vector global symmetry
over the scalar charge matter
\bea \label{eq:vglobal-1}
\Phi \to e^{\ii \eta_v(x)} \Phi := e^{\ii  \Lambda \cdot x} \Phi .
\eea
Here $\Lambda$ is a $(d+1)$-vector on the spacetime, so the $\Lambda \cdot x$ takes the inner product with the spacetime coordinate.
We term this symmetry as 
\bea \label{eq:U1xd+1}
\U(1)_{x_{(d+1)}} \text{vector global symmetry} 
\eea 
due to the involvement of the ${d+1}$d spacetime vectors, $\Lambda$ and  $x$.
The {vector global symmetry} belongs to a generalized class of higher-moment global symmetry.
See Fig. \ref{clock-vector-global-3} for illustration of a symmetry transformation.

In fact, the $\U(1)_{x_{(d+1)}}$ involves several symmetry transformations under 
any $(d+1)$ linear-independent choice of 
$\Lambda=(\Lambda_0, \Lambda_1, \dots, \Lambda_d)$.
So
\Eqn{eq:vglobal-1} can be chosen to be  $e^{\ii  \Lambda_0 \cdot x_0}$, $e^{\ii  \Lambda_1 \cdot x_1}$, $\dots,
e^{\ii  \Lambda_{d} \cdot x_d}$. Thus \Eqn{eq:U1xd+1} contains
\bea \label{eq:U1xd+1-express}
\text{\Eqn{eq:U1xd+1}} \coloneqq \text{a set of $(d+1)$ independent }
\U(1)_{x_0}, 
\U(1)_{x_1}, \dots, \U(1)_{x_{d}}, 
\text{vector global symmetries}. 
\eea 
Note that however $\U(1)_{x_0}$ is not quite a standard global symmetry in terms of the Hamiltonian theory. 
Nonetheless we still abuse the name
global symmetry for $\U(1)_{x_0}$. The phenomenon here of $\U(1)_{x_0}$ can be potentially related to a generalization of \emph{time crystal} 
\cite{Shapere2012nq1202.2537, Wilczek2012jt1202.2539}, since
\bea \label{eq:time-crystal}
\Phi \to \Phi e^{\ii  \Lambda_{0} \cdot x_0}=\Phi e^{\ii  \Lambda_{0} \cdot t}, \quad 
\eea
while $x_0=t$ is time coordinate.
So the field configuration at a certain periodic time $t \simeq t + \frac{2\pi}{\Lambda_{0}} \Z$ is constrained. We will comment more in the Conclusion
in \Sec{sec:conclude}.

Now let us write down the matter field theory Lagrangian.
To recall, 
in order to have a Lagrangian kinetic term invariant under \emph{only the ordinary global symmetry},
we have the familiar kinetic term $\prt^\mu \Phi^\dagger \prt_\mu \Phi$.

In order to have a Lagrangian kinetic term invariant under the vector global symmetry (alternatively, under both the ordinary and vector global symmetry),
we should abandon the familiar kinetic term $\prt^\mu \Phi^\dagger \prt_\mu \Phi$ but
design a new Lagrangian term:\footnote{In the fracton literature, this is firstly propose by Pretko \cite{Pretko2018jbi1807.11479} for the vector global symmetry.
See a generalized formalism for field theory of any higher-moment global symmetry in 
\cite{Wang2019cbjJWKaiXu1911.01804, WXY3Wang2019mtt1912.13485}.}
\bea \label{eq:kinetic-Phi-1}
(\Phi^\dagger \prt^\mu \prt^\nu \Phi^\dagger -
 \prt^\mu \Phi^\dagger \prt^\nu \Phi^\dagger) 
 (\Phi \prt_\mu \prt_\nu \Phi -
 \prt_\mu \Phi \prt_\nu \Phi).
\eea
It is easy to check that
under \Eq{eq:vglobal-1} transformation, 
\bea \label{eq:kinetic-deri-Phi-1}
(\Phi \prt_\mu \prt_\nu \Phi -
 \prt_\mu \Phi \prt_\nu \Phi) \to
e^{\ii 2 \eta_v(x)}  (\Phi \prt_\mu \prt_\nu \Phi -
 \prt_\mu \Phi \prt_\nu \Phi + (\ii  \prt_\mu \prt_\nu \eta_v) \Phi^2),
 \eea
 thus this term is covariant up to a factor; if $\eta_v$ is in terms of a linear polynomial of $x$, namely $\eta_v=\Lambda \cdot x$,
 rather than higher-order polynomial, then $\prt_\mu \prt_\nu \eta_v=0$. 
 Hence \Eqn{eq:kinetic-Phi-1} is invariant  under both the ordinary and vector global symmetry \Eqn{eq:oglobal-1}
 and \Eqn{eq:vglobal-1}.
 
 There is also a discrete $\Z_2^C$ charge conjugation symmetry:
 \bea
 \Phi \to \Phi^\dagger.
\eea
It is easy to see that the ordinary U(1) and $\Z_2^C$ forms a non-abelian ${\rm U}(1) \rtimes \Z_2^{C}={\SO}(2) \rtimes \Z_2^{C}={\rm O}(2)$-symmetry.
The full internal symmetry in both Euclidean/Minkowski signature is:
\begin{itemize}
\item  Euclidean/Minkowski internal (including ordinary and higher-moment) global symmetry: \\
$$\Z_2^C \ltimes \Big( \U(1) \times \U(1)_{x_{(d+1)}}\Big).$$
\end{itemize}
\item Spacetime symmetry:
\begin{itemize}
\item Euclidean version's Poincar\'e symmetry: 
$
\R^{d+1} \rtimes \rO(d+1),
$
which includes Euclidean spacetime translation, rotation, and reflection (say $R$-symmetry, for Euclidean spacetime, this includes the parity $P$ 
and time reversal $T$ as the same symmetry).
Note that the reflection $R$-symmetry is a $\Z_2^R$ discrete symmetry, related to the 0-th homotopy group $\pi_0\rO(d+1)=\Z_2$,
so the reflection $R$ flips between two disconnected components of $\rO(d+1)$.
\item Minkowski  version's Poincar\'e symmetry: 
$
\R^{1,d}  \rtimes \rO(1,d),
$
which includes Minkowski spacetime translation, boost, rotation, and the parity $P$ and time reversal $T$.
Note that the $P$ and $T$ symmetry are both $\Z_2$ discrete symmetries, 
with $\Z_2^P$: $x_j \to -x_j$ for some $j$, and with $\Z_2^T$: $t \to -t$ over the spacetime coordinates.
They are related to the two $\Z_2$ generators of  $\pi_0\rO(1,d)=(\Z_2)^2$,
so the $P$ and $T$ flip between four disconnected components of $\rO(1,d)$.
\end{itemize}
\end{enumerate}
The $\U(1)_{x_{(d+1)}}$ has a nontrivial action on the Poincar\'e symmetry, thus we can define a semi-direct product structure:
$\Big( \R^{d+1} \rtimes \rO(d+1)\Big) \ltimes  \U(1)_{x_{(d+1)}}  $.
Combine the internal and spacetime symmetry together, we obtain the full global symmetry.
\begin{itemize}
\item Euclidean version's symmetry: 
\bea \label{eq:Euclidean-sym}
\Big( \R^{d+1} \rtimes \rO(d+1)\Big) \ltimes \Bigg( \Z_2^C \ltimes \Big( \U(1) \times \U(1)_{x_{(d+1)}}  \Big) \Bigg).
\eea
\item Minkowski version's symmetry is similar: 
\bea   \label{eq:Minkowski-sym}  
\Big(\R^{1,d}  \rtimes \rO(1,d)\Big)  \ltimes  \Bigg( \Z_2^C \ltimes \Big( \U(1) \times \U(1)_{x_{(d+1)}}\Big) \Bigg).
\eea
\end{itemize}
Note that the 0-form symmetry $\Z_2^C \ltimes \U(1)$ in fact commute with the Poincar\'e symmetry, 
thus we have the structure
$(\R^{d+1} \rtimes \rO(d+1) )\times ( \Z_2^C \ltimes  \U(1))$ if we disregard the vector global symmetry $ \U(1)_{x_{(d+1)}}$.


\subsubsection{Symmetric higher-rank tensor gauge theory and the gauging procedure}

Now we follow the gauge principle  to gauge the higher-moment symmetry $\U(1)_{x_{(d+1)}}$ in $( \U(1) \times \U(1)_{x_{(d+1)}}  )$.
To recall the standard gauging procedure of the ordinary 0-form global symmetry, we promote the global symmetry transformation parameter
$\eta$ in \Eq{eq:oglobal-1} to a spacetime-dependent local transformation $\eta(x)$. Thus under $\Phi \to e^{\ii \eta(x)} \Phi$,
the derivative term $ \prt_\mu \Phi $ is not covariant unless we revise it to a covariant 
derivative term $$ (\prt_\mu -  \ii g a_{\mu}) \Phi.$$
It is covariant under the familiar gauge transformation:
\bea \label{eq:0-gauge}
 \Phi \to e^{\ii \eta(x)} \Phi, \quad  a_{\mu }  \to a_{\mu }  +\frac{1}{g} \prt_\mu  ( \eta(x)).
 \eea
See \Fig{clock-ordinary-global-2-gauge} for a demonstration of the gauge fluctuation away from the 
0-form global symmetry.
 
Follow \cite{Pretko2018jbi1807.11479}, the way we do is to promote the global parameters $\eta$ and
 $\eta_v(x)= \Lambda \cdot x$ to depend on the spacetime:
 $\eta \to \eta(x)$ and  $\eta_v \to\eta_v(x)=  \Lambda(x) \cdot x$.
 In this case, in order to make \Eqn{eq:kinetic-deri-Phi-1} covariant,
 we revise it to a new term:
 \bea   \label{eq:vector-gauge-term}
(\Phi \prt_\mu \prt_\nu \Phi -
 \prt_\mu \Phi \prt_\nu \Phi - \ii g A_{\mu \nu} \Phi^2)  
 \eea
\begin{figure}[!t]
	\centering
	\includegraphics[width=14.cm]{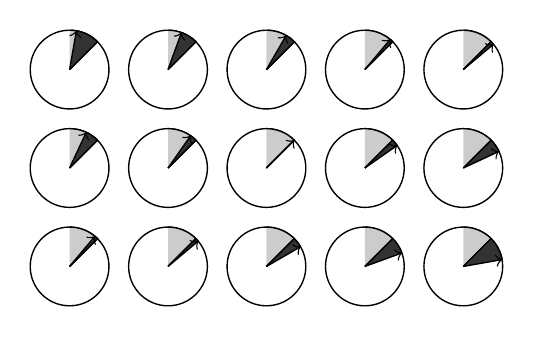}
	\caption[caption]{A demonstration of the gauge fluctuation (the dark gray color) deviates away from the 
0-form global symmetry transformation of \Fig{clock-ordinary-global-1} (the light gray color). See \Eq{eq:0-gauge}.
	}
	\label{clock-ordinary-global-2-gauge}
\end{figure}
\begin{figure}[!t]
	\centering
	\includegraphics[width=14.cm]{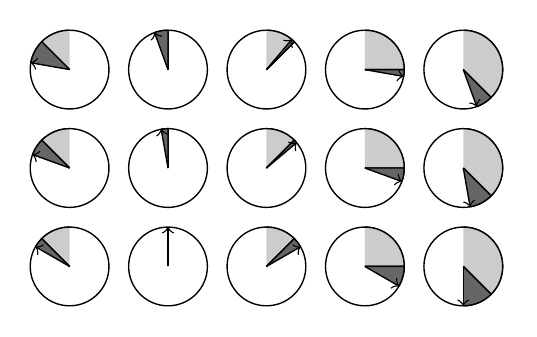}
	\caption[caption]{A demonstration of the gauge fluctuation (the dark gray color) deviates away from the 
 {vector global symmetry} of \Fig{clock-vector-global-3} (the light gray color, which belongs to a generalized class of higher-moment symmetry).
 See \Eq{eq:vector-gauge}.
	}
	\label{clock-vector-global-4-gauge}
\end{figure}
 such that it is covariant under the new gauge transformation 
 \bea \label{eq:vector-gauge}
 \Phi \to e^{\ii \eta_v(x)} \Phi, \quad  A_{\mu \nu}  \to A_{\mu \nu}  +\frac{1}{g} \prt_\mu \prt_\nu ( \eta_v(x)).
 \eea
The $A_{\mu \nu}$ is a symmetric rank-2 tensor of $(d+1) \times (d+1)$ components in a $(d+1)$d spacetime.
 
A major difference of ours apart from previous work is that we can also treat spacetime indices isotropically and equally (up to the metric signature)
instead of an-isotropically as in \cite{Pretko2018jbi1807.11479}.
After gauging $ \U(1)_{x_{(d+1)}}$ higher-moment symmetry, 
we have a ``gauge group-analogous structure'' from \Eq{eq:U1xd+1}:
\bea \label{eq:Sym.Tensor.G}
\left[  \U(1)_{x_{(d+1)}} \right],
\eea
where the big bracket $ \left[ \dots\right]$ stands for being dynamically 
gauged.\footnote{Again, it is only a ``group-analogous structure''  but not quite a group, because the 0-form symmetry ($\Z_2^C$) and higher moment symmetry ($ \U(1)_{x_{(d+1)}} $) are fundamentally different. Note that we only gauge $ \U(1)_{x_{(d+1)}} $ but we do not gauge U(1) in  $( \U(1) \times \U(1)_{x_{(d+1)}}  )$, 
but the U(1) global symmetry would be lost altogether after gauging $ \U(1)_{x_{(d+1)}} $. Importantly, we do not introduce the 
usual 1-form U(1) gauge field, because we do not gauge 0-form U(1) symmetry.
We introduce the symmetric rank-2 tensor $A_{\nu \xi}$ to only gauge the $ \U(1)_{x_{(d+1)}} $ symmetry.}
We can define the rank-3 field strength $F_{\mu \nu \xi}$ of the symmetric rank-2 tensor $A_{\nu \xi}$ as:
\bea \label{eq:U1x-F}
F_{\mu \nu \xi} =\prt_\mu A_{\nu \xi} -\prt_{\nu} A_{\mu \xi}. 
\eea
$F_{\mu \nu \xi}$ is anti-symmetric with respect to the first two indices $\mu \leftrightarrow \nu$.
And we can define the kinetic Lagrangian term for gauge fields as:
\bea
\cL_{A,\text{kinetic}}: =\frac{1}{g^2} |F_{\mu \nu \xi}|^2
: =\frac{1}{g^2} F_{\mu \nu \xi}F^{\mu \nu \xi}
=\frac{1}{g^2}(\prt_\mu A_{\nu \xi} -\prt_{\nu} A_{\mu \xi}) (\prt^\mu A^{\nu \xi} -\prt^{\nu} A^{\mu \xi}).
\eea
The equation of motion (EOM) for a pure gauge theory is
\bea
\prt^\mu F_{\mu \nu \xi}
=\prt^\mu(\prt_\mu A_{\nu \xi} -\prt_{\nu} A_{\mu \xi})
=0.
\eea
{We certainly can write down the classical field theory by giving the action, with or without matter field (which we can set $\Phi=0$):
\bea  \label{eq:U(1)vector-gauge-matter-action}
\int_{M^{d+1}} \dd^{d+1}x \Big( 
|F_{\mu \nu \xi}|^2  +
|(\Phi \prt_\mu \prt_\nu \Phi - \prt_\mu \Phi \prt_\nu \Phi - \ii g A_{\mu \nu} \Phi^2)|^2 +V(|\Phi|^2)) \Big).
\eea
Here $V(|\Phi|^2)$ is a potential term.
The ${M^{d+1}}$ is a ${d+1}$d spacetime manifold.
Ideally, we hope to discuss a quantum theory. Formally, 
we propose a schematic path integral:
\be \hspace{-10mm}
\label{eq:U(1)vector-gauge-matter}
\boxed{\bZ_{\text{rk-2-sym-$\Phi$}} =\int[\cD A_{\mu\nu}][\cD \Phi][\cD \Phi^\dagger]\exp(\ii \int_{M^{d+1}} \dd^{d+1}x \Big( 
 |F_{\mu \nu \xi}|^2 +
|(\Phi \prt_\mu \prt_\nu \Phi - \prt_\mu \Phi \prt_\nu \Phi - \ii g A_{\mu \nu} \Phi^2)|^2 +V(|\Phi|^2)\Big))}.
\ee
We shorthand ``{rk-2-sym-$\Phi$}'' for the rank-2 symmetric tensor gauge theory coupled to matters. 
Unfortunately, this field theory has no free nor quadratic term that we can start with to do a perturbative theory to extract the higher-order term effects.
In fact, due to the leading order term is already highly nontrivial interacting quartic interactions,
a fully fledged quantum field theory by a field quantization is still an open question. 
We comment more the quantum aspects of the theory in Conclusion \Sec{sec:conclude}.
}

\subsubsection{Field strength, electric and magnetic tensors: Independent components and representations}

{The rank-3 field strength $F_{\mu \nu \xi}$ can also be interpreted as the higher-rank generalized electric tensor field $\overline{\rm E}$ and magnetic tensor field $\overline{\rm B}$. 
To do so, we take those $F_{\mu \nu \xi}$ have an \emph{odd number of time indices} ($0$ or $t$)
as the electric tensor field $\overline{\rm E}$, while those 
$F_{\mu \nu \xi}$ have an \emph{even number of time indices} ($0$ or $t$)
as the magnetic tensor field $\overline{\rm B}$.
Above in this section, we have a full general discussion in any dimension.
Below, we focus on the 4d (3+1D) spacetime,
where $i,j,k,\ell \dots  \in \{1,2,3\}$ are space coordinates only without time.
Let us count independent ($\equiv$ indpt) components.
We define:
\bea \label{eq:EE}
\overline{\rm E} &:=&
\left\{\begin{array}{l} 
\overline{\rm E}^{}_{ij}:=
F_{0ij}=-F_{i0j}=
\prt_0 A_{i j} -\prt_{i} A_{0 j}, \text{with $3^2=9$-indpt components}\\
\overline{\rm E}^{}_{\ell } :=\frac{1}{2}\epsilon^{\ell ij} F_{ij0}, \text{ with $i,j$ summed.}\\
\quad\quad\text{where}\; \;F_{ij0}=-F_{ji0}=
\prt_i A_{j 0} -\prt_{j} A_{i 0}, \text{thus $i\neq j$ for $3$-indpt components}
 \end{array}\right.
\\
\overline{\rm B}&:=& \label{eq:BB}
\left\{\begin{array}{l} 
\overline{\rm B}^{}_{\ell k}:=
\frac{1}{2}\epsilon^{\ell ij} F_{ijk}=\frac{1}{2}\epsilon^{\ell ij} ( \prt_i A_{jk}-\prt_j A_{ik}), \text{ with $i,j$ summed.}\\ 
\quad\quad\text{where}\; \;F_{ijk}=-F_{jik}= \prt_i A_{jk}-\prt_j A_{ik}, \text{thus $i\neq j$ with $\frac{3\cdot2}{2}3=9$-indpt components}.\quad\quad\quad \\
\overline{\rm B}^{}_{j}
:= F_{0j0}=-F_{j00}=
\prt_0 A_{j 0} -\prt_{j} A_{0 0}, \text{with $3$-indpt components}.
 \end{array}\right.
\eea
Thus both
$\overline{\rm E}$ and $\overline{\rm B}$ tensor contains 12 components.
Here the 9 means the spatial rank-2 tensor (say $i,j  \in \{1,2,3\}$), and the 3 means spatial vector (say $j  \in \{1,2,3\}$).
We can understand the above in terms of representation of spatial rotation symmetry group SO(3).
The SO(3) has a $(2n+1)$-dimensional representation, we need that $n=0$ gives the trivial representation ${\bf 1}$, $n=1$ which gives the vector representation ${\bf 3}$, 
and the $n=2$ gives the 5-dimensional representation ${\bf 5}$.
The $\overline{\rm E}^{}_{\ell }$ and $\overline{\rm B}^{}_{j}$ give the  the vector representation ${\bf 3}$ with 3 components.
The $\overline{\rm E}^{}_{ij}$ and $\overline{\rm B}^{}_{\ell k}$ give the decomposition of representations $( {\bf 1} +  {\bf 3} + {\bf 5})$ with 9 components.
 Thus, we have both tensor fields decomposed in terms of spatial rotational SO(3) representations:
 \bea  \label{eq:EB-ML-SO3}
\begin{array}{lll}
\overline{\rm E} &=&{\bf 3}  +( {\bf 1} +  {\bf 3} + {\bf 5}),\\
\overline{\rm B} &=&{\bf 3}  +( {\bf 1} +  {\bf 3} + {\bf 5}).
 \end{array}
\eea}
\subsubsection{Gauge the $\Z_2^C$-charge conjugation symmetry:\\ 
Non-abelian higher-moment continuous gauge structure}
\label{sec:gauge-Z2C}
For both the ungauged theory (e.g. \Eq{eq:kinetic-Phi-1}) and the gauged theory (e.g. \Eq{eq:U(1)vector-gauge-matter}), there is a discrete charge conjugation $\Z_2^C$ global symmetry, or the so-called particle-hole symmetry in condensed matter system.\footnote{The $\Z_2^C$-symmetry is an ordinary unitary 0-form global symmetry in the context of \cite{Gaiotto2014kfa1412.5148}.}
 For both the ungauged theory and the gauged theory,  
 the $\Z_2^C$ symmetry flips the complex scalar $\Phi$ to its complex conjugation, 
   \bea 
   \Phi\to   \Phi^\dagger,
 \eea 
It is easy to see that the ordinary U(1) global symmetry and $\Z_2^C$ symmetry  of
\Eq{eq:Euclidean-sym} and \Eq{eq:Minkowski-sym} do \emph{not} commute.
Take these two symmetry transformations, as $U_{{\rm{U}}(1)}$
and  $U_{\Z_2^C}$ operators respectively, we see that the transformations act on $\Phi$ as:
\bea
U_{\Z_2^C} U_{{\rm{U}}(1)}  \Phi &=&U_{\Z_2^C} ( e^{\ii \eta} \Phi)
=e^{\ii \eta} \Phi^\dagger.\nn\\
 U_{{\rm{U}}(1)} U_{\Z_2^C} \Phi &=&U_{\Z_2^C} (\Phi^\dagger)
=e^{-\ii \eta} \Phi^\dagger.
\eea
For the ordinary U(1) global symmetry, we fix $\eta$ to be a spacetime-independent parameter as a constant.
Thus $U_{\Z_2^C} U_{{\rm{U}}(1)} \neq U_{{\rm{U}}(1)}  U_{\Z_2^C} $, we have a non-abelian symmetry group structure
the $\Z_2^C \ltimes \U(1)=\Z_2^C \ltimes \SO(2)={\rm {O}}(2)$.

For the vector global symmetry in \Eq{eq:U1xd+1},
$e^{\ii \eta_v(x)} \Phi = e^{\ii  \Lambda \cdot x} \Phi$, so
\bea
U_{\Z_2^C} U_{{\rm{U}}(1)_{x_{(d+1)}}}  \Phi &=&U_{\Z_2^C} ( e^{\ii \eta_v} \Phi)
=e^{\ii \eta_v} \Phi^\dagger.\nn\\
 U_{{\rm{U}}(1)_{x_{(d+1)}}} U_{\Z_2^C} \Phi &=&U_{\Z_2^C} (\Phi^\dagger)
=e^{-\ii \eta_v} \Phi^\dagger.
\eea
We still have the non-commutative structure $U_{\Z_2^C} U_{{\rm{U}}(1)_{x_{(d+1)}}} \neq U_{{\rm{U}}(1)_{x_{(d+1)}}}  U_{\Z_2^C} $, resulting in a non-abelian symmetry structure
$\Z_2^C \ltimes \U(1)_{x_{(d+1)}}$.

 For the gauged theory,  
 the $\Z_2^C$ symmetry also flips the sign of the real-valued gauge field $A_{\mu \nu }$,
 and the vector ``gauge symmetry transformation'' parameter $\eta_v(x)$
 as follows
   \bea \label{eq:Amunu-eta-Z2C-transf}
   \quad A_{\mu \nu} 
\to -A_{\mu \nu},  \quad \eta_v(x)\to - \eta_v(x).
 \eea
 Thus, the \Eq{eq:vector-gauge-term} under $\Z_2^C$-symmetry transformation becomes its complex conjugation:
  \bea   
 (\Phi \prt_\mu \prt_\nu \Phi -
 \prt_\mu \Phi \prt_\nu \Phi - \ii g A_{\mu \nu} \Phi^2)
 \to
(\Phi^\dagger \prt_\mu \prt_\nu \Phi^\dagger -
 \prt_\mu \Phi^\dagger \prt_\nu \Phi^\dagger + \ii g A_{\mu \nu} (\Phi^\dagger)^2)  ,
 \eea
 such that the action \Eq{eq:U(1)vector-gauge-matter-action} is complex conjugation $\Z_2^C$-symemtric invariant.
 
Below we work out the non-commutative-ness of gauge transformations from $\Z_2^C$ and $\U(1)_{x_{(d+1)}}$:
\bea
U_{\Z_2^C} U_{{\rm{U}}(1)_{x_{(d+1)}}} A_{\mu \nu} &=&U_{\Z_2^C} ( A_{\mu \nu}  +\frac{1}{g} \prt_\mu \prt_\nu  \eta_v)
=- A_{\mu \nu} +\frac{1}{g} \prt_\mu \prt_\nu  \eta_v.\nn\\
 U_{{\rm{U}}(1)_{x_{(d+1)}}} U_{\Z_2^C} A_{\mu \nu} &=&U_{\Z_2^C} ( -A_{\mu \nu})
=- A_{\mu \nu} -\frac{1}{g} \prt_\mu \prt_\nu  \eta_v.
\eea
Namely, if we gauge $ \U(1)_{x_{(d+1)}}$ higher-moment symmetry in \Eq{eq:vector-gauge}, and we  further gauge $\Z_2^C$ ordinary 0-form symmetry,
then we can obtain a gauge theory with the new ``gauge structure'' as:
\bea \label{eq:gauge-group-analogous}
 \left[\Z_2^C \ltimes \Big( 
 \U(1)_{x_{(d+1)}} \Big) \right].
\eea
Here the big bracket $ \left[ \dots\right]$ indicates the global symmetry inside is being dynamically gauged. So among the global symmetry of
\Eq{eq:Euclidean-sym} and \Eq{eq:Minkowski-sym},
the ordinary $\U(1)$-global symmetry of \Eq{eq:Euclidean-sym} and \Eq{eq:Minkowski-sym} is 
\emph{not a
global symmetry anymore} after gauging $ \U(1)_{x_{(d+1)}}$ as the way we did in \Eq{eq:vector-gauge}.
Again $\left[\Z_2^C \ltimes \Big(  \U(1)_{x_{(d+1)}} \Big) \right]$ is not quite a group, which is definitely not just $\Z_2^C \ltimes \U(1)={\rm {O}}(2)$, but only
as what we call a ``gauge structure'' or ``gauge group-analogous structure.''

Let us temporarily \emph{turn off the gauged matter} $\Phi$  in \Eq{eq:U(1)vector-gauge-matter}, 
then focus on promoting the global $\Z_2^C$ symmetry to a local symmetry that can be gauged in a pure tensor gauge theory.
This means that we promote the $\Z_2^C$ symmetry \Eq{eq:Amunu-eta-Z2C-transf} to a dynamical local transformation involving a gauge parameter. 
\cred{It is useful to first embed $\Z_2^C \subset \U(1)^C$, such that we consider an enlarged $ \U(1)^C$ gauge symmetry with a gauge parameter $\gamma_c(x)$:}
\bea \label{eq:Z2C-vector-gauge}
A_{\mu \nu}  \to  e^{\ii \gamma_c(x)} A_{\mu \nu}, \quad 
C_\nu \to C_\nu +\frac{1}{g_c} \prt_\nu \gamma_c(x).
\eea
This \Eq{eq:Z2C-vector-gauge} deserves explanations.
We introduce a new $g_c$ coupling and the new 1-form $\Z_2^C$-gauge field $C$ coupling to the
0-form symmetry $\Z_2^C$-charged $A_{\mu \nu}$. We know that $A_{\mu \nu}$ has an odd $\Z_2^C$ charge because \Eq{eq:Z2C-vector-gauge} says
$A_{\mu \nu} \to - A_{\mu \nu}$ under $\Z_2^C$.
Note $A_{\mu \nu} \in \R$ is real-valued, but a generic $e^{\ii \gamma_c(x)}$ complexifies the $A_{\mu \nu} \in \mathbb{C}$. However, what we really mean is to restrict 
 gauge transformation so it is only $\Z_2^C$-gauged (not $\U(1)^C$-gauged),
 \bea \label{eq:Z2-phase}
 e^{\ii \gamma_c(x)} := (-1)^{\gamma_c'(x)} \in \{\pm 1\}, \quad\gamma_c(x) \in \pi \Z, \quad {\gamma_c'(x)} \in \Z,
 \eea
 so ${\gamma_c'(x)} \in \Z$ is an integer, and $A_{\mu \nu} \in \R$ stays in real. Thus ${\gamma_c'(x)}$ can jumps between even or odd integers, while the 
 $\Z_2^C$-gauge transformation is better formulated on a lattice. We can directly rewrite the above \Eq{eq:Z2C-vector-gauge} on a simplicial complex and a triangulable manifold, e.g. 
following the Dijkgraaf-Witten's discretized topological gauge theory on a lattice \cite{DijkgraafWitten1989pz1990}.
  
We also define a new covariant derivative with respect to $\Z_2^C$:
\bea \label{eq}
D_\mu^c :=(\prt_\mu - \ii g_c C_\mu).
\eea
We need to modify $ \U(1)_{x_{(d+1)}}$-gauge transformation \Eq{eq:vector-gauge} 
and combine with 
the $\Z_2^C$-gauge transformation \Eq{eq:Z2C-vector-gauge} to:\footnote{
One may also naively consider 
\bea
   A_{\mu \nu}  &\to& e^{\ii \gamma_c(x)} A_{\mu \nu}  +\frac{1}{g} D_\mu^c D_\nu^c ( \eta_v(x))
   \nn\\
  &=&{
  e^{\ii \gamma_c(x)}A_{\mu \nu}   +\frac{1}{g} \Big(\prt_\mu \prt_\nu ( \eta_v) 
 - \ii g_c  (\prt_\mu C_\nu) \eta_v 
 - \ii g_c ( C_\nu (\prt_\mu \eta_v)+ C_\mu   (\prt_\nu\eta_v)) -(g_c)^2 C_\mu C_\nu \eta_v\Big)}, 
 \label{eq:vector-gauge-charge-asym}
\eea
but in this case, the $A_{\mu \nu}$ is not symmetric under this specific gauge transformation --- 
even for this case however does not affect the gauge covariance of
\bea
\hat F^c_{\mu \nu \xi}  &\to&
 e^{\ii \gamma_c(x)} \hat F^c_{\mu \nu \xi} 
- \ii \frac{g_c}{g} (C_\mu \prt_\nu   \prt_\xi \eta_v-  C_\nu  \prt_\mu \prt_\xi  \eta_v)
+\frac{(-\ii g_c)^2}{g} (
C_\mu \prt_\nu (C_\xi \eta_v) 
-C_\nu \prt_\mu (C_\xi \eta_v) 
)
\nn\\
&&  - \ii \frac{g_c}{g} (\prt_\mu (C_\nu   (\prt_\xi\eta_v))- \prt_\nu (C_\mu   (\prt_\xi\eta_v)))  
 -  \frac{g_c^2}{g} (\prt_\mu(C_\nu C_\xi \eta_v) -\prt_\nu(C_\mu C_\xi \eta_v)  )
\nn \\
    &=&
   e^{\ii \gamma_c(x)} \hat F^c_{\mu \nu \xi}
   {- \ii \frac{g_c}{g}(\prt_\mu C_\nu - \prt_\nu C_\mu) D_\xi^c\eta_v} \nn\\
    &=& e^{\ii \gamma_c(x)} \hat F^c_{\mu \nu \xi}.   \label{eq:F-change-asym}
\eea
Here $(\prt_\mu C_\nu) \eta_v$ term in the variation \Eq{eq:vector-gauge-charge-asym}
can be shown to be equivalent to $\frac{(\prt_\mu C_\nu+\prt_\nu C_\mu)}{2} \eta_v$ when it is \emph{on-shell}, i.e., the on-shell theory satisfies EOM:
 $\dd C=0 \Rightarrow (\prt_\mu C_\nu - \prt_\nu C_\mu)=0$ so $C$ is locally flat --- a local constraint for $\Z_2$-gauge field $C$.
 So \Eq{eq:vector-gauge-charge} and \Eq{eq:vector-gauge-charge-asym} are equivalent on-shell.
Moreover, in \Eq{eq:F-change-asym}, we can eliminate ${(\prt_\mu C_\nu - \prt_\nu C_\mu) D_\xi^c\eta_v}$ via 
again $(\prt_\mu C_\nu - \prt_\nu C_\mu)=0$ because $\dd C=0$.

 On the other hand, if we wish to make sense of the \emph{off-shell} symmetric tensor gauge field $A_{\mu \nu}$, we will use the definition \Eq{eq:vector-gauge-charge} instead.
}
 \bea \label{eq:vector-gauge-charge}
   A_{\mu \nu}  &\to& e^{\ii \gamma_c(x)} A_{\mu \nu}  + \frac{1}{2g}(D_\mu^c D_\nu^c+ D_\nu^c D_\mu^c) ( \eta_v(x))
   \nn\\
   &=&e^{\ii \gamma_c(x)}A_{\mu \nu}   +\frac{1}{g} \Big(\prt_\mu \prt_\nu ( \eta_v) 
 - \ii g_c  \frac{\prt_\mu (C_\nu\eta_v) +\prt_\nu (C_\mu\eta_v)  }{2}
 - \ii g_c \frac{C_\mu (\prt_\nu\eta_v)+C_\nu (\prt_\mu\eta_v)}{2} -(g_c)^2 C_\mu C_\nu \eta_v\Big) \nn\\
  &=&{
  e^{\ii \gamma_c(x)}A_{\mu \nu}   +\frac{1}{g} \Big(\prt_\mu \prt_\nu ( \eta_v) 
 - \ii g_c  \frac{(\prt_\mu C_\nu+\prt_\nu C_\mu)}{2} \eta_v 
 - \ii g_c (  C_\mu   (\prt_\nu\eta_v) + C_\nu (\prt_\mu \eta_v)) -(g_c)^2 C_\mu C_\nu \eta_v\Big).} \nn\\
 C_\nu &\to& C_\nu +\frac{1}{g_c} \prt_\nu \gamma_c(x).
 \eea
 We re-define \Eq{eq:U1x-F}'s $ F_{\mu \nu \xi}$ into the new gauge covariant field strength $\hat F^c_{\mu \nu \xi}$: 
\bea
\boxed{\hat F^c_{\mu \nu \xi} :=D_\mu^c A_{\nu \xi} -D_{\nu }^c A_{\mu \xi }
:={(\prt_\mu - \ii g_c C_\mu) A_{\nu \xi} -(\prt_{\nu} - \ii g_c C_\nu ) A_{\mu \xi}}}.
\eea
We can show that the field strength $\hat F^c_{\mu \nu \xi}$  is covariant 
 under both 
the modified $ \U(1)_{x_{(d+1)}}$ gauge transformation
and  $\Z_2^C$ gauge transformation  \Eq{eq:vector-gauge-charge}:  
\bea\hspace{-10mm}
\hat F^c_{\mu \nu \xi}  &\to&
 e^{\ii \gamma_c(x)} \hat F^c_{\mu \nu \xi}
 - \ii \frac{g_c}{g} \big(  \prt_\mu( \frac{(\prt_\nu C_\xi+\prt_\xi C_\nu)}{2} \eta_v ) - \prt_\nu( \frac{(\prt_\mu C_\xi+\prt_\xi C_\mu)}{2} \eta_v )  \big) \nn\\
  &&- \ii \frac{g_c}{g}  \prt_\mu  (  C_\nu   (\prt_\xi\eta_v) + C_\xi (\prt_\nu \eta_v))
  - \ii \frac{g_c}{g}  \prt_\nu  (  C_\mu   (\prt_\xi\eta_v) + C_\xi (\prt_\mu \eta_v))
  - \frac{g_c^2}{g}   (\prt_\mu (C_\nu C_\xi \eta_v) -\prt_\nu (C_\mu C_\xi \eta_v))
  \nn\\
  &&   - \ii  \frac{g_c}{g}  ( C_\mu \prt_\nu \prt_\xi ( \eta_v)  - C_\nu  \prt_\mu \prt_\xi ( \eta_v) ) 
  +\frac{(-\ii g_c)^2}{g} \big( (C_\mu \frac{(\prt_\nu C_\xi+\prt_\xi C_\nu)}{2}  - C_\nu \frac{(\prt_\mu C_\xi+\prt_\xi C_\mu)}{2} )\eta_v  \big) \nn\\
  && +\frac{(-\ii g_c)^2}{g} \big(C_\mu (  C_\nu   (\prt_\xi \eta_v) + C_\xi  (\prt_\nu \eta_v)) 
  -C_\nu (  C_\mu   (\prt_\xi \eta_v) + C_\xi  (\prt_\mu \eta_v)) \big)\nn\\
  &=&
   e^{\ii \gamma_c(x)} \hat F^c_{\mu \nu \xi}
 + \ii \frac{g_c}{g} \big(  \frac{{(\prt_\nu C_\xi-\prt_\xi C_\nu )( D_\mu^c \eta_v)}+ {( \prt_\xi C_\mu - \prt_\mu C_\xi   )(D_\nu^c \eta_v )} }{2} \big)  
       \nn\\
    &&   +
   \frac{ \prt_\xi (\prt_\mu C_\nu- \prt_\nu C_\mu )}{2} \eta_v 
   - \ii \frac{g_c}{g}   (  \prt_\mu  C_\nu -\prt_\nu    C_\mu)  (D_\xi^c\eta_v)       \nn\\
    &=& e^{\ii \gamma_c(x)} \hat F^c_{\mu \nu \xi},
    \label{eq:F-change}
\eea
which is covariant (invariant up to a $\pm$ phase \Eq{eq:Z2-phase} $e^{\ii \gamma_c(x)} := (-1)^{\gamma_c'(x)} \in \{\pm 1\}$).
Here we can eliminate 
$(\prt_\mu C_\nu - \prt_\nu C_\mu)=(\prt_\nu C_\xi-\prt_\xi C_\nu )=( \prt_\xi C_\mu - \prt_\mu C_\xi   )=0$,
because $\dd C=0$ is locally flat via the local constraint for $\Z_2$-gauge field $C$.
Let us put things altogether in the next paragraph.

Based on the continuum formulation of discrete gauge theories (e.g. \cite{Putrov2016qdo1612.09298}), since $C$ is a 1-form $\Z_2$-gauge field, we can either (1) treat $C$ as a $\Z_2$-valued 1-cochain, or
(2) treat it as a U(1) gauge field but impose a locally flat condition $\dd C = 0$ and $\oint C = m \pi \mod 2 \pi,$ with $m \in \Z_2$.
For the later purpose for (2), this can be done by introducing a Lagrange multiplier $(d-1)$-form $B$ field (in 3+1d, we have $d=3$).
We introduce the famous level-2 anti-symmetric ($\equiv$ asym) tensor  BF theory,
\bea \label{eq:Z2C-gauge}
\bZ_{\text{level-2-BF}} = \int [\cD B][\cD C]\exp(\ii \int_{M^{d+1}}  \ii\frac{ 2}{2 \pi} B \dd C).
\eea
So far we have a $(d-1)$-form $\Z_2$-gauge $B$ field and 1-form $\Z_2$-gauge field $C$, obviously we abbreviate $B \dd C : =B \wedge \dd C$.
Note it is also a standard convention to omit the wedge product $\wedge$
when we are taking the wedge product between the differential forms.
Furthermore, ideally and for simplicity, by leaving out the matter $\Phi$, we can aim for a full quantum theory for a 
gauge theory of a gauge-analogous structure \Eq{eq:gauge-group-analogous}.
Formally, we propose a schematic path integral:
\bea \label{eq:U(1)vector-Z2C-gauge}
\bZ_{\underset{\text{asym-BF}}{\text{rk-2-sym}}}: =\int[\cD A_{\mu\nu}][\cD B][\cD C]\exp(\ii \int_{M^{d+1}}   \Big( 
|\hat F^c_{\mu \nu \xi}|^2 (\dd^{d+1} x) + \frac{ 2}{2 \pi} B \dd C\Big)), 
\eea
\emph{which is non-trivially fully gauge-invariant} under \Eq{eq:vector-gauge-charge}, 
the gauge transformation of $B$ field via a local $(d-2)$-form $ \eta_B$:
\bea \label{eq:B-transf-change}
B \to B + \dd \eta_B
\eea
and \Eq{eq:F-change}. 
The term
\begin{multline}
\label{eq:|F|^2}
|\hat F^c_{\mu \nu \xi}|^2:= \hat F^c_{\mu \nu \xi} \hat F^{c,{\mu \nu \xi \dagger}}\\
={\big((\prt_\mu - \ii g_c C_\mu) A_{\nu \xi} -(\prt_{\nu} - \ii g_c C_\nu ) A_{\mu \xi}\big)}
{\big((\prt^\mu + \ii g_c C^\mu) A^{\nu \xi} -(\prt^{\nu} + \ii g_c C^\nu ) A^{\mu \xi}\big)}
\end{multline}
pairs $\hat F^c_{\mu \nu \xi}$ with its complex conjugation $ \hat F^{c,{\mu \nu \xi \dagger}}$ contracting all indices.
We shorthand ``{rk-2-sym}'' for the rank-2 symmetric tensor gauge theory. 
We shorthand ``{asym-BF}'' for the anti-symmetric tensor BF theory. 
Our theory is unitary. 
At this stage, we only deal with the kinematics of the theory, we discuss its possible quantum dynamics in Conclusion in \Sec{sec:conclude}.

\subsubsection{Coupling to anti-symmetric tensor topological field theories}
\label{sec:coupled-to-DW}

Another new ingredient of our approach is that we can introduce $\Z_2^C$-gauge theory \Eq{eq:U(1)vector-Z2C-gauge} which is
an anti-symmetric tensor topological field theory. A formal way to introduce this $\Z_2^C$-gauge theory in $(d+1)$d is that
it is a group cocycle element $\omega_{d+1}$ of the topological gauge theory specified by the group cohomology $\rH^{d+1}$ of the gauge group $G_g$ \cite{DijkgraafWitten1989pz1990}:
\bea
\omega_{d+1} \in \rH^{d+1}(G_g, \R/\Z) := \cH^{d+1}(\B G_g,\R/\Z  ),
\eea 
or the topological cohomology $\cH^{d+1}$ of the classifying space of $\B G_g$ in the second expression \cite{DijkgraafWitten1989pz1990}.
The cohomology group $\rH^{d+1}(G_g, \R/\Z)$ always forms an abelian group.

If we only have a single copy of an abelian symmetric higher-rank tensor gauge theory
with a single gauge structure $\left[\Z_2^C \ltimes \Big( \U(1)_{x_{(d+1)}}\Big) \right]$,
then we can determine possible bosonic $\Z_2^C$-gauge theory via
$
\rH^{d+1}(\Z_2^C, \R/\Z). 
$
More generally, we can consider 
$N$ copies of such abelian symmetric higher-rank tensor gauge theory
with $N$ copies of gauge structure 
\bea
\prod_{J=1}^N \left[\Z_2^{C} \ltimes \Big(  \U(1)_{x_{(d+1)}}^{}\Big) \right].
\eea
So we have the ordinary gauge group 
\bea\label{eq:GgZ2C}
G_g=[(\Z_2^C)^N]
\eea
thus a $[(\Z_2^C)^N]$-gauge theory in addition to some higher-moment gauge structure (formally not a group),
then we can specify a group cocycle \cite{DijkgraafWitten1989pz1990}, which is computed systematically in \Ref{Wang1404.7854, Wang1405.7689}
 \bea
\omega_{d+1} \in  \rH^{d+1}(G_g, \R/\Z)=  \rH^{d+1}((\Z_2^C)^N, \R/\Z). 
\eea
See Table \ref{table:Hgroup} for explicit results for a generic cohomology group.
\begin{itemize}
\item
If the $\omega_{d+1}$ is a trivial cocycle, namely it satisfies the coboundary relation
$$
\omega_{d+1} =\delta \beta_{d},
$$
then which is a coboundary term under the coboundary operator 
$\delta$. The $\beta_{d} \in C^{d}(G_g, \R/\Z)$ is a lower dimensional $d$-cochain.
This means that $\omega_{d+1} \simeq 1$ is identically to the identity in the cohomology group.
\item If the $\omega_{d+1}$ is a nontrivial cocycle, namely
$\omega_{d+1} \neq \delta \beta_{d}$, which is not exact for any cochain $\beta_{d}$, 
the gauge theory is commonly called as the \emph{twisted group cohomology gauge theory},
or \emph{Dijkgraaf-Witten gauge theory} ($\equiv$ DW) of gauge group $G_g$ \cite{DijkgraafWitten1989pz1990}.
 \Ref{Wang1404.7854, Wang1405.7689,  1602.05951WWY, Putrov2016qdo1612.09298, Wang2018edf1801.05416, Wang1901.11537WWY}
 had systematically studied these topological gauge theories as 
 (i) discrete cocycle partition functions, and,
 (ii) continuum TQFTs.
 See Table \ref{table:TQFT} and \Sec{sec:asymDWTQFT} for the overview of the continuum TQFT formulations
  \Ref{Wang1405.7689,  1602.05951WWY, Putrov2016qdo1612.09298, Wang1901.11537WWY}.
\end{itemize}

\begin{table}[!h]
\centering
\hspace*{-7mm}
\begin{tabular}{|c||c|c|c|c|c|c|c|c|c|}
\hline
      &Type I& Type II & Type III &Type  IV &Type  V  &
      $\dots$ & \\[0mm]  \hline
    Cohomology group  &$\Z_{N_i}$& $\Z_{N_{ij}}$& $\Z_{N_{ijl}}$ &  $\Z_{N_{ijlm}}$ &  $\Z_{{\gcd} \otimes^5_{i}(N^{(i)})}$
      & $\Z_{{\gcd} \otimes^m_{i}(N_i)}$ & $\Z_{{\gcd} \otimes^d_{i}N^{(i)}}$\\[0mm]  \hline
$\rH^1(G,\R/\Z)$ &$1$& $$& $$ & $$ & $$ & $$ 
& $$ \\[0mm]  \hline
$\rH^2(G,\R/\Z)$&$0$  &$1$ & $$ & $$ & $$ & $$ 
 & $$\\[0mm]  \hline
$\rH^3(G,\R/\Z)$ &$1$ & $1$ & $1$ & $$ & $$ & $$ 
  & $$\\ \hline
$\rH^4(G,\R/\Z)$ &$0$ & $2$ & $2$ & $1$ & $$ & $$ 
 & $$\\ \hline
$\rH^5(G,\R/\Z)$ & $1$ & $2$ & $4$ & $3$ & $1$ 
& $$ & $$\\ \hline
$\rH^6(G,\R/\Z)$ & $0$ & $3$ & $6$ & $7$ & $4$ 
& $$  & $$\\ \hline
$\rH^d(G,\R/\Z)$ & $\frac{(1-(-1)^d)}{2}$ & $\frac{d}{2}-\frac{(1-(-1)^d)}{4}$ & $\dots$ & $\dots$ & $\dots$ 
& $\dots$  & $1$\\ \hline
\end{tabular}
\caption{Cohomology group:
Table from \Ref{Wang1405.7689} exhibits the exponent of the $\Z_{{\gcd} \otimes^m_{i}(N_i)}$ class in the cohomology group
$\rH^d(G_g,\R/\Z)$ for a finite Abelian group $G_g=\underset{u=1}{\prod^k} \Z_{N_u}$.
We abbreviate the greatest common divisor $\equiv$ gcd.
We define a shorthand of $\Z_{{\gcd} (N_{i_1},N_{i_2},\dots, N_{i_m})}\equiv
\Z_{N_{i_1,\dots,i_m}} \equiv  \Z_{{\gcd} \otimes^m_{i}(N_i)}$, etc also for other higher
 gcd.  
 The Type $m$ in the top row of the Table is given by the number ($m$) of cyclic gauge
groups in the gcd class $\Z_{{\gcd} \otimes^m_{i}(N_i)}$.  The number of
exponents can be systematically obtained by K\"unneth formula and Universal Coefficient Theorem.
This table can also be independently derived by a field theory approach \cite{Wang1405.7689}.
For example, we obtain
$\mathrm{H}^5(G_g,\R/\Z) =  \underset{1 \leq i < j < l<m <n \leq k}{\prod} \Z_{N_i} \times (\Z_{N_{ij}})^2\times (\Z_{N_{ijl}})^4 \times (\Z_{N_{ijlm}})^3  \times  \Z_{N_{ijlmn}} $, etc.
We can derive the continuum field theory from the group cohomology result, or the other way around 
\cite{Wang1405.7689,  1602.05951WWY, Putrov2016qdo1612.09298}.
For the purpose of present work, we simply take $G_g=[(\Z_2^C)^N]$ as in \Eq{eq:GgZ2C}.
}
\label{table:Hgroup}
\end{table}
Now that $\omega_{d+1}(\{C_I\}) \in  \rH^{d+1}(G_g, \R/\Z)$ maps to the $\R/\Z = \U(1)$ coefficient, a complex U(1) phase, physically $\omega_{d+1}(\{C_I\})$ 
is the weight of 
the quantum amplitude in the path integral.
Thus, summing over such a $\omega_{d+1}(\{C_I\})$ is also related to the orbifold construction.
\be \label{eq:ZABC}
\hspace*{-4mm}
\boxed{
\bZ_{\underset{\text{asym-tw}}{\text{rk-2-sym}}}:=\int (\prod_{I=1}^{N}[\cD A_{I,\mu\nu}] [\cD B_I]  [\cD C_I])\exp(\ii \int_{M^{d+1}} ( 
\sum_{I=1}^N |\hat F^{c,I}_{\mu \nu \xi}|^2 (\dd^{d+1} x) + \frac{ 2}{2 \pi} \sum_{I=1}^{N} B_I \dd C_I)) \cdot \omega_{d+1}(\{C_I\})},
\ee
where
$\omega_{d+1} \in \rH^{d+1}((\Z_2^C)^N, \R/\Z)$.
We shorthand ``asym-DW'' for the anti-symmetric tensor twisted ($\equiv$ tw) Dijkgraaf-Witten (DW) gauge theory.
Adding the DW topological term, \Eq{eq:B-transf-change} needs to be modified to $B \to B + \dd \eta_B + \dots$,
some examples of the additional $\dots$ terms are shown in Table \ref{table:TQFT}.
We can apply the variation principle on the $A,B$ and $C$ fields to obtain their \text{EOM}s (i.e., Euler-Lagrange equation): 
\bea
\label{eq:eom-A}
\frac{\delta{}}{\delta A^{\nu \xi}}(\dots)  \Rightarrow D^{c \mu}  \hat F^c_{\mu \nu \xi} 
&=&(\prt^\mu - \ii g_c C^\mu) \Big((\prt_\mu - \ii g_c C_\mu) A_{\nu \xi} -(\prt_{\nu} - \ii g_c C_\nu ) A_{\mu \xi} \Big)=0.\\
\label{eq:eom-B}
\frac{\delta{}}{\delta B}(\dots)  \Rightarrow \;\quad\quad  \frac{ 2}{2 \pi}\dd C&=&0,\\
\label{eq:eom-C}
\frac{\delta{}}{\delta C}(\dots)  \Rightarrow \;\quad\quad   \frac{ 2}{2 \pi} \dd B&=&\# C_I \wedge C_J  \wedge \dots \wedge \dd C_K  - \ii 4 g_c 
 \#  \star
 (A^{\nu \xi} \hat F^{c}_{\mu \nu \xi} \dd x^\mu). 
\eea
Here $ \#$ are some factors depending on the data $\omega_{d+1}$ in terms of the continuum TQFTs given in Table \ref{table:TQFT}. 
In \Eq{eq:eom-A}, \Eq{eq:eom-B} and \Eq{eq:eom-C},
we only list down the schematic result since we do not yet precisely provide the data of the group cohomology cocycle $\omega_{d+1}$,
which will be given later explicitly in \Sec{sec:asymDWTQFT}. The allowed \emph{topological term} (or the twisted term)
$C_I \wedge C_J  \wedge \dots \wedge \dd C_K$ depends on the spacetime dimensions  and the copies of $G_g$, see Table \ref{table:Hgroup}
and Table \ref{table:TQFT}.
Here are some comments about EOM:\\[-10mm]
\begin{itemize}
\item In \Eq{eq:eom-A}, the dependence of $A$ and $C$ reveals the first ingredient of the \emph{non-abelian} gauge structure (see \Sec{sec:intro}):
  A higher-moment symmetry $(\U(1)_{x_{(d+1)}}^{})$ and a charge conjugation (particle-hole) symmetry $\Z_2^C$ form the non-commutative symmetries.
{Notice that this EOM is \emph{linear} respect to the solutions of $A$, but \emph{nonlinear} respect to the solutions of $C$.}\footnote{This means that
if $A$ is a solution of EOM and $A'$ is a solution of EOM, then their linear combination $A+A'$ is also a solution of EOM.
However, if $C$ is a solution of EOM and $C'$ is a solution of EOM, then their linear combination $C+C'$ in general is not a solution of EOM.
\label{footnote:A-C-linear}}
\item  In \Eq{eq:eom-B}, in any case, $\dd C=0$, so $C$ is always a locally flat $\Z_2$ gauge field.
\item In \Eq{eq:eom-C}, the dependence of $B$ to $A$ and $C$ reveals a second ingredient for the possible
{\emph{non-abelian} gauge structure (see \Sec{sec:intro}): By coupling the symmetric tensor gauge theory to a non-abelian TQFT. See more details in \Sec{eq:Top-Type}}.
{Notice that this EOM is \emph{nonlinear} respect to both the solutions of $A$ and of $C$.}
\end{itemize}

Of course, the discussions in this whole subsection have two versions, simple by changing the spacetime metric signatures:
{\emph{Euclidean-type higher-rank symmetric tensor gauge theory} \Eq{eq:E}} and
{\emph{Lorentzian-type higher-rank symmetric tensor gauge theory}  \Eq{eq:L}.} 
Their physics interpretations would be different.

\subsection{Anisotropic Non-Abelian Higher-Rank Tensor Gauge Theory for Space and Time}

\label{sec:Anisotropic}

Let us consider the \emph{anisotropic-type higher-rank symmetric tensor gauge theory}  \Eq{eq:Anisotropic-FT}.
Let us list down the modification for this new case \Eq{eq:Anisotropic-FT} from the previous subsection.

\subsubsection{Electric and magnetic tensors: Independent components and SO(3) representations}
First, we need to introduce a symmetric rank-2 tensor  $A_{ i j}$  of $d \times d$ components in a $(d+1)$d spacetime.
We also need a scalar potential $ A_0$.
Instead of the $F_{\mu \nu \xi}$ in \Eq{eq:U1x-F}, or the electromagnetic tensor
$\overline{\rm E}$ in \Eq{eq:EE} and
$\overline{\rm B}$ in \Eq{eq:BB}, we redefine the
	electric field and magnetic field tensors similar to Pretko's \Ref{Pretko2016kxt1604.05329, Pretko2016lgv1606.08857, Pretko2017xar1707.03838, Slagle2018kqf1807.00827, Pretko2018jbi1807.11479}:
\bea \label{eq:EB-aniso}
\tilde {\rm E}_{ij} &:=& -\partial_0A_{ij}+\partial_i\partial_j A_0 = -\partial_tA_{ij}+\partial_i\partial_j A_0.\\
\tilde {\rm B}_{ij}&:=&\varepsilon_{i \ell m} \partial_\ell A_{mj}.
\eea
(Naively we can also write $\tilde {\rm B}_{ijk} \propto \partial_j A_{ik}-\partial_kA_{ij}$ as a rank-3 tensor, but it is not necessary.)
Now that $\tilde {\rm E}_{ij}$ and $\tilde {\rm B}_{ij}$ are rank-2 tensors. 
In $(3+1)$d spacetime, we have $3 \times 3=9$ components for each. 
Let us count the independent $\tilde {\rm E}_{ij}$ and $\tilde {\rm B}_{ij}$ in terms of the spatial rotational SO(3) representations as we did in \Eq{eq:EB-ML-SO3}.
Note that $\tilde {\rm E}_{ij}$ is symmetric, thus we have $\tilde {\rm E}_{ii}$ 3 components and
$\tilde {\rm E}_{i\neq j}$ another 3 components. 
Note that $\tilde {\rm B}_{ij}$ is neither symmetric nor anti-symmetric, thus 9 components.
We have both tensor fields decomposed in terms of SO(3) representations:
 \bea \label{eq:EB-Aniso-SO3}
 \begin{array}{lll}
\tilde {\rm E}_{}  &=&( {\bf 1} + {\bf 5}),\\
\tilde {\rm B}_{} &=& ( {\bf 1} +  {\bf 3} + {\bf 5}).
 \end{array}
\eea
The ${\bf 1}$ in $\tilde {\rm E}_{}$ is from the identity component $\frac{1}{3}{\sum_{i=1}^3 \tilde {\rm E}_{ii} }(\mathbb{I}_{3})$.
The ${\bf 5}$ in $\tilde {\rm E}_{}$ are from other components.

\subsubsection{Matter field theory and symmetric higher-rank tensor gauge theory}
We restrict our global symmetry transformation to the space in contrast to \Eq{eq:vglobal-1}:
\bea \label{eq:vglobal-1-space}
\Phi \to e^{\ii \eta_v(x)} \Phi := e^{\ii  {\vec{\Lambda} \cdot \vec{x}}} \Phi .
\eea
Here $\vec\Lambda$ is a $(d)$-vector on the space, so the ${\vec{\Lambda} \cdot \vec{x}}$ takes the inner product with the space coordinate.
In contrast to \Eq{eq:U1xd+1}, we term this
\bea \label{eq:U1xd}
\U(1)_{x_{(d)}} \text{vector global symmetry}. 
\eea 
Promoting the vector global symmetry transformation to a gauge transformation, follow \Sec{sec:matter-vector-higher-moment-vector-global-symmetry}, 
this yields Pretko's field theory \cite{Pretko2017xar1707.03838} with some couplings $g_0$ and $\lambda$:\\[-14mm]
\begin{center}
\hspace*{-6mm}
\fbox{\parbox{7.5in}{\parindent=0pt 
\begin{multline} \label{eq:U(1)vector-gauge-matter-space}
{\bZ'_{\text{rk-2-sym-$\Phi$}}: =\int [\cD A_{0}][\cD A_{ij}][\cD \Phi][\cD \Phi^{\dagger}]\exp(\ii \int_{M^{d+1}} \dd^{d+1}x }\\
{\Big( 
(|\tilde {\rm E}_{ij} |^2-|\tilde {\rm B}_{ij} |^2) +
|(\prt_t- \ii g_0 A_0 )\Phi|^2 - \lambda |(\Phi \prt_i \prt_j \Phi - \prt_i \Phi \prt_j \Phi - \ii g A_{i  j} \Phi^2)|^2 +V(|\Phi|^2) + \dots\Big))}.
\end{multline}
}}
\end{center}
The theory is gauge invariant under:
 \bea \label{eq:vector-gauge-space} 
 \Phi \to e^{\ii \eta_v(x)} \Phi ,
 \quad
 A_0  \to A_0 +\frac{1}{g} \prt_0  ( \eta_v(x)) = A_0 +\frac{1}{g} \prt_t  ( \eta_v(x)), \quad  A_{ij }  \to A_{ij }  +\frac{1}{g} \prt_i \prt_j ( \eta_v(x)).
 \eea
 There are other possible terms listed in $\dots$, see, for instance, \Ref{Pretko2017xar1707.03838}.
 
\subsubsection{Gauge $\Z_2^C$ symmetry for another non-abelian higher-moment-gauged tensor theory}

\label{sec:gauge-Z2C-space}

Similar to \Sec{sec:gauge-Z2C}, we aim to obtain a non-abelian tensor gauge theory by gauging the ordinary 0-form $\Z_2^C$ charge conjugation symmetry:
   \bea 
   \Phi\to   \Phi^\dagger,
  \quad A_0  \to - A_0,
   \quad A_{ij } 
\to -A_{ ij },  \quad \eta_v(x)\to - \eta_v(x).
 \eea
 Under the same $\Z_2^C$-covariant derivative  \Eq{eq}'s $D_\mu^c :=(\prt_\mu - \ii g_c C_\mu)$, 
similar to \Eq{eq:vector-gauge-charge} and \Eq{eq:Z2-phase}, 
we couple the gauged theory to DW topological terms as in \Sec{sec:coupled-to-DW}.
We have the combined gauge transformation\footnote{Naively, we may define:
\bea
   A_{i j}  &\to& e^{\ii \gamma_c(x)} A_{i j}  +\frac{1}{g} D_i^c D_j^c ( \eta_v(x))\nn\\
   &=&e^{\ii \gamma_c(x)}A_{i j}   +\frac{1}{g} \Big(\prt_i \prt_j ( \eta_v) 
 - \ii g_c  \prt_i (C_j\eta_v) 
 - \ii g_c C_i   (\prt_j\eta_v) -(g_c)^2 C_i C_j \eta_v\Big). \nn
 \eea
However, this results in $A_{i j}$ is not symmetric under this specific gauge transformation --- 
this case however does not affect the gauge covariance of 
$\tilde {\rm E}_{ij}^c $ and $\tilde {\rm B}_{ij}^c$ tensor
at least for the on-shell case.}
 \bea
    A_{0}  &\to& e^{\ii \gamma_c(x)}     A_{0}+\frac{1}{g} D_0^c  ( \eta_v(x)) = e^{\ii \gamma_c(x)}     A_{0}+\frac{1}{g} D_t^c  ( \eta_v(x)). \nn\\
   A_{i j}  &\to& e^{\ii \gamma_c(x)} A_{i j}  +\frac{1}{2g} (D_i^c D_j^c +  D_j^c D_i^c) ( \eta_v(x))\nn\\
  &=&e^{\ii \gamma_c(x)}A_{i j}   +\frac{1}{g} \Big(\prt_i \prt_j ( \eta_v) 
 - \ii g_c  \frac{(\prt_i C_j+\prt_j C_i)}{2}\eta_v 
 - \ii g_c C_i   (\prt_j\eta_v)- \ii g_c C_j   (\prt_i\eta_v) -(g_c)^2 C_i C_j \eta_v\Big). 
 \nn\\
 C_j &\to& C_j +\frac{1}{g_c} \prt_j \gamma_c(x). \nn\\
 e^{\ii \gamma_c(x)} &:=& (-1)^{\gamma_c'(x)} \in \{\pm 1\}.\nn\\
 B &\to& B + \dd \eta_B + \dots.
 \eea 
A $\Z_2^C$-gauge transformation can be much easier defined in a discretized spacetime \cite{DijkgraafWitten1989pz1990}.
Examples of the additional $\dots$ terms are shown in Table \ref{table:TQFT}.
We also refine the $\tilde {\rm E}_{ij}, \tilde {\rm B}_{ij}$ to the covariant derivative \Eq{eq} version's $\tilde {\rm E}_{ij}^c, \tilde {\rm B}_{ij}^c$:
\bea \label{eq:aniso-E}
\tilde {\rm E}_{ij}^c &:=&-D_0^c A_{ij}+\frac{D_i^c D_j^c+ D_j^c D_i^c }{2}A_0 = -D_t^cA_{ij}+\frac{D_i^c D_j^c+ D_j^c D_i^c }{2} A_0.\\
\label{eq:aniso-B}
\tilde {\rm B}_{ij}^c&:=&\varepsilon_{i \ell m} D_\ell^c A_{mj}.
\eea
Again, we can check the covariance of $\tilde {\rm E}_{ij}^c$ and $\tilde {\rm B}_{ij}^c$ tensor under the gauge transformation, after some long calculations and subtle cancellations:
\bea
\tilde {\rm E}_{ij}^c &\to &  e^{\ii \gamma_c(x)} \tilde {\rm E}_{ij}^c, \nn\\
\tilde {\rm B}_{ij}^c &\to & e^{\ii \gamma_c(x)} \tilde {\rm B}_{ij}^c, \nn
\eea
up to a $ e^{\ii \gamma_c(x)} := (-1)^{\gamma_c'(x)} \in \{\pm 1\}$ phase,
as long as we can use $\dd C \sim \varepsilon_{i \ell m} \prt_\ell C_m =0$ for the on-shell flat gauge $C$ condition.
Again we couple a symmetric tensor gauge theory to a DW term $\omega_{d+1}(\{C_I\}) \in  \rH^{d+1}(G_g, \R/\Z)$ shown in \Ref{Wang1405.7689,  1602.05951WWY,  Putrov2016qdo1612.09298, Wang2018edf1801.05416, Wang1901.11537WWY}  to gain
another unitary field theory with anisotropic space time symmetry:\footnote{Readers should beware that the definitions of
$B$  and  ${\rm B}$ are fundamentally different.
We use $B$ for the any-symmetric tensor field (here $(d-2)$-form gauge field),
here we use $\tilde {\rm B}$ to represent the magnetic tensor field in its anisotropic version.} 
\begin{center}
\hspace*{-6mm}
\fbox{\parbox{7in}{\parindent=0pt 
\begin{multline}
\label{eq:ZABC-aniso}
\bZ'_{\underset{\text{asym-tw}}{\text{rk-2-sym}}}:=\int (\prod_{I=1}^{N}[\cD A_{I,0}][\cD A_{I,ij}] [\cD B_I]  [\cD C_I])
\exp(\ii \int_{M^{d+1}} 
(\sum_{I=1}^{N} (|\tilde {\rm E}_{ij}^{c,I} |^2-|\tilde {\rm B}_{ij}^{c,I} |^2))
(\dd^{d+1} x) \\+ \frac{ 2}{2 \pi} \sum_{I=1}^{N} B_I \dd C_I) \cdot \omega_{d+1}(\{C_I\})
\end{multline}
}}\\[-8mm]
\end{center}
Again we apply the variation principle on the $A,B$ and $C$ fields to gain anisotropic \text{EOM} (for $I=1$ only): 
\bea
\label{eq:eom-A0}
\frac{\delta{}}{\delta A^{0}}(\dots)  &\Rightarrow&  D^{c i} D^{c j} \tilde {\rm E}_{ij}^{c}
= D^{c i} D^{c j}  \Big(  -D_t^cA_{ij}+\frac{D_i^c D_j^c+ D_j^c D_i^c }{2} A_0 \Big)=0,\\
\label{eq:eom-Aij}
\frac{\delta{}}{\delta A^{ij}}(\dots)  &\Rightarrow& D_t^c  \tilde {\rm E}_{ij}^{c}- \varepsilon_{i \ell m} D_\ell^c \tilde {\rm B}_{mj}^c
=D_t^c( -D_t^cA_{ij}+\frac{D_i^c D_j^c+ D_j^c D_i^c }{2} A_0) 
- \varepsilon_{i \ell m} D_\ell^c(\varepsilon_{m \ell' m'} D_{\ell'}^c A_{m' j})=0,\quad\quad\quad\quad\\
\label{eq:eom-B-aniso}
\frac{\delta{}}{\delta B}(\dots)  &\Rightarrow&  \frac{ 2}{2 \pi}\dd C=0,\\
\label{eq:eom-C0-aniso}
\frac{\delta{}}{\delta C^0}(\dots)  &\Rightarrow&    \frac{ 2}{2 \pi} \dd B=\# C_I \wedge C_J  \wedge \dots \wedge \dd C_K  - \ii g_c 
  \star
 (\# A^{i j } \tilde {\rm E}_{ij}^{c} \dd x^0), \\
 \label{eq:eom-Cij-aniso}
\frac{\delta{}}{\delta C^{i}}(\dots)  &\Rightarrow&    \frac{ 2}{2 \pi} \dd B=\# C_I \wedge C_J  \wedge \dots \wedge \dd C_K  - \ii g_c 
   (( \# (D^{cj} A^{0}) \tilde {\rm E}_{ij}^{c} +\#  
   (\varepsilon^{i' }{}_{i m} A^{mj}) \tilde {\rm B}^c_{i' j} )\dd x^i). \quad\quad
\eea
Several factors $\#$ can be determined fully via the given data and Table \ref{table:TQFT} later. Again
$\varepsilon_{ijk}=\varepsilon^{ijk}=\varepsilon^{i}{}_{jk}=\varepsilon^{ij}{}_{k}$, raising or lowering indices does not affect its sign for spatial indices. 
Comments on EOMs:\\[-10mm]
\begin{itemize}
\item In \Eq{eq:eom-A0} and \Eq{eq:eom-Aij}, 
{this EOM is \emph{linear} respect to solutions of $A$, but \emph{nonlinear} respect to solutions of $C$, see footnote \ref{footnote:A-C-linear}.}
\item  
In \Eq{eq:eom-B-aniso}, in any case, $\dd C=0$ is always a locally flat $\Z_2$ gauge field.
\item In \Eq{eq:eom-C0-aniso} and \Eq{eq:eom-Cij-aniso}, the dependence of $B$ to $A$ and $C$ reveals a second ingredient for the possible
{\emph{non-abelian} gauge structure (see \Sec{sec:intro}): By coupling the symmetric tensor gauge theory to a non-abelian TQFT.  See more details in \Sec{eq:Top-Type}}.
{Notice that this EOM is \emph{nonlinear} respect to both the solutions of $A$ and of $C$.}
\end{itemize}

\section{Non-Abelian Tensor Gauge Theory Interplayed Between Gapped And Gapless Phases}
\label{sec:InterplayedBetweenGappedAndGaplessPhases}

Now we provide evidence that our theories in \Eq{eq:ZABC} and \Eq{eq:ZABC-aniso}
are in the interplay between gapped phases (see \Sec{sec:asymDWTQFT}) and gapless phases (see \Sec{sec:symTgauge}). 
Let us organize and summarize our statements line by line.
\begin{enumerate}
\item 
	\underline{The \emph{phases}} mean that  \emph{the states of matter in the phase diagram} of the quantum matter or of the condensed matter.

\item \underline{The \emph{gapped} phases}  mean that the excitations in the phases are massive and highly energetic expansive.
In fact, in the TQFT limit, it costs infinite energy to break up the extend operators (such as line and surface operators) with open ends ---
only the open ends can host massive particles or heavy objects.

\item \underline{The \emph{gapless} phases}  mean that the dispersion of excitations is continuous and massless in the large system size limit.

\item  The anti-symmetric tensor TQFTs describe gapped phases shown in \Sec{sec:asymDWTQFT}.
The \emph{gapped excitations} are the anyonic particles (described by
the worldline of $C$ field), or the anyonic strings/$(d-3)$-branes 
(described by the worldsheet/volume of $B$ field).
 
\item The symmetric tensor gauge theories describe gapless phases that will be shown in the following \Sec{sec:symTgauge}.

\end{enumerate}

\subsection{Anti-Symmetric Tensor Topological Field Theory and Gapped Phase} 
\label{sec:asymDWTQFT}
 
Shown in Table \ref{table:TQFT}, here we quickly overview the continuum field theory formulation \cite{1602.05951WWY, Putrov2016qdo1612.09298, Wang2018edf1801.05416} of topological gauge theory from a cohomology group \cite{DijkgraafWitten1989pz1990}: $\rH^d(G_g,\R/\Z)$ for a finite Abelian group $G_g=\underset{u=1}{\prod^k} \Z_{N_u}$, or specifically 
a gauge group $G_g=(\Z_2^C)^m$ of $m$-layers of gauged charge conjugation symmetries.

The first column of Table \ref{table:TQFT} 
shows the  continuum anti-symmetric tensor TQFT actions and their \emph{gauge transformations to all order in analytic exact forms}.
Namely, Table \ref{table:TQFT} contains a finite gauge transformations, instead of only infinitesimal gauge transformations.
There is no further higher order term that we can add to the \emph{gauge transformations} beyond what we had found in Table \ref{table:TQFT}. 

In fact, the term 
in the path integral of
\Eq{eq:ZABC} and \Eq{eq:ZABC-aniso} 
$
\int ([\cD B_I]  [\cD C_I])\exp(\ii \int_{M^{d+1}}  \ii\frac{ 2}{2 \pi} \sum_{I=1}^{N} B_I \dd C_I)) \cdot \omega_{d+1}(\{C_I\})
$
can be replaced to this continuum theory \cite{1602.05951WWY, Putrov2016qdo1612.09298, Wang1901.11537WWY}, e.g.,
\bea \label{eq:cocylce-to-continuum-QFT}
\int ([\cD B_I]  [\cD C_I])\exp(\ii \int_{M^{d+1}}  \ii\frac{ 2}{2 \pi} (\sum_{I=1}^{N} B_I \wedge \dd C_I) + \# C_I \wedge C_J  \wedge \dots \wedge \dd C_K). 
\eea
with some coefficients $\#$, shown in Table \ref{table:TQFT}.

In Table \ref{table:TQFT}, we provide the link configurations from the extended operators 
linked in the spacetime\cite{1602.05951WWY, Putrov2016qdo1612.09298, Wang1901.11537WWY}.
The extended operators include, for example, line operators on circles $S^1$,
and the surface operators on $S^2$ or $T^2$.

Readers can find the explicit derivation of: 
$$\left\{\begin{array}{l} 
\text{\emph{link invariants and braiding statistics} 
with extended operators from a continuum TQFT in \cite{Putrov2016qdo1612.09298}.}\\
\text{dim($\cH_d$) = GSD associated to $d$-dimensional space 
(counting zero energy modes)  in \cite{Wang2018edf1801.05416}.}
 \end{array}\right.
 $$
 The EOMs follow \Eq{eq:eom-A},\Eq{eq:eom-B}, and \Eq{eq:eom-C}. The first two do not depend on the topological term $\omega_{d+1}$ while the third EOM does:
\bea
\frac{\delta}{\delta C}=0\implies \frac{2}{2\pi}\dd B=+{\xi_\omega}-4\ii g_c \star(A^{\nu\xi}F^c_{\mu\nu\xi}dx^\mu)=0.
\eea
The $\xi_\omega$ depends on $\omega_{d+1}$. 
For the five cases listed in Table \ref{table:TQFT},
the $\xi_\omega$ corresponds to, or, 
is proportional to:\\[-10mm]
\begin{enumerate}[label=(\arabic*).]
	\item $\frac{p_{IJ}}{2\pi}\dd C^J$.
	\item $c_{123}C^2\wedge C^3+\text{permutations ($1\leftrightarrow2\leftrightarrow3$)}$.
	\item $0$.
	\item $\frac{N_IN_Jp_{IJK}}{(2\pi)^2N_{IJ}}C^J\wedge \dd C^K$.
	\item $c_{1234}C^2\wedge C^3\wedge C^4 +\text{permutations ($1\leftrightarrow2\leftrightarrow3\leftrightarrow4$)}$.\\[-8mm]
\end{enumerate}

\newgeometry{left=1.2cm, right=1.4cm, top=0.4cm, bottom=2.cm}
%

%
%
\begin{table}[!h]
\noindent
\fontsize{9}{9}\selectfont
\hspace*{6mm}
{
\makebox[\textwidth][r]{
\begin{tabular}{ccc} 
\hline
$\begin{matrix}\text{(i). TQFT actions}\\  
\end{matrix}$ 
& 
$\begin{matrix}\text{(ii). Group-cohomology cocycles}\\ 
\text{associated to (i)} 
\end{matrix}$ & 
$\begin{matrix}
\text{(iii). Path-integral link 
invariants;}\\
\text{Quantum statistic spacetime-braiding} \\
\text{data characterized by (i)}
\end{matrix}$ 
\\
\hline\\[-2mm]
\multicolumn{3}{c}{3d (2+1D) spacetime} \\
\cline{1-3}\\[-2mm]
$\begin{matrix}
  \int 
\overset{}{\underset{{I}}{\sum}} \frac{ N_I}{2\pi}{B^I  \wedge \dd C^ I} + { \frac{ p_{IJ}}{2 \pi}} C^ I \wedge \dd C^ J\\[2mm]
C^ I \to C^ I+ \frac{ \dd \gamma_c^I}{g_c}, \\[2mm]
 B^I  \to  B^I + \dd \eta_B^ I.
 \end{matrix}
$
& $\begin{matrix}
{\omega}_3=\\
\exp \Big( \frac{2 \pi \ii {p_{IJ} }  }{N_{I} N_{J}} \; a_{I}(b_{J} +c_{J} -[b_{J} +c_{J}]_{N_J}) \Big) \\[2mm]
 \in \rH^{3}((\Z_2^C)^2, \R/\Z)
  \end{matrix}$  &  
$\begin{matrix}
\bZ \bpm \includegraphics[scale=0.3]{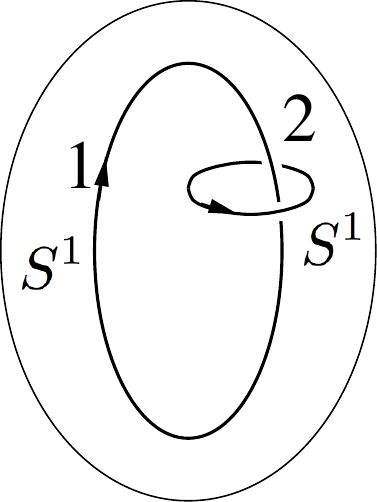} \includegraphics[scale=0.24]{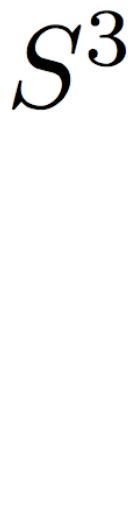}\epm  
 \end{matrix}$
  \\
\hline\\[-2mm]
$\begin{matrix}
  \int 
\overset{}{\underset{{I}}{\sum}} \frac{ N_I}{2\pi}{B^I  \wedge \dd C^ I}+{{} c_{123}} C^ 1 \wedge C^ 2 \wedge  C^ 3\\[2mm]
C^ I \to C^ I+  \frac{ \dd \gamma_c^I}{g_c}, \\[2mm]
 B^I  \to  B^I + \dd \eta_B^I+ 2\pi \frac{{\tilde{c}}_{IJK}}{N_I}  C^ J  \frac{  \gamma_c^K}{g_c} \\[2mm] 
- \pi \frac{{\tilde{c}}_{IJK}}{N_I}  \frac{  \gamma_c^J}{g_c}  \frac{ \dd \gamma_c^K}{g_c}.
\end{matrix}$ 
 &   
 $\begin{matrix} 
{\omega}_3^{\text{Top}}=\\[2mm]
 \exp \Big( \frac{2 \pi \ii p_{123}  }{N_{123}} \;  a_{1}b_{2}c_{3} \Big)\\[2mm]
  \in \rH^{3}((\Z_2^C)^3, \R/\Z)
  \end{matrix}$    & 
  $\begin{matrix}
\bZ \bpm \includegraphics[scale=0.35]{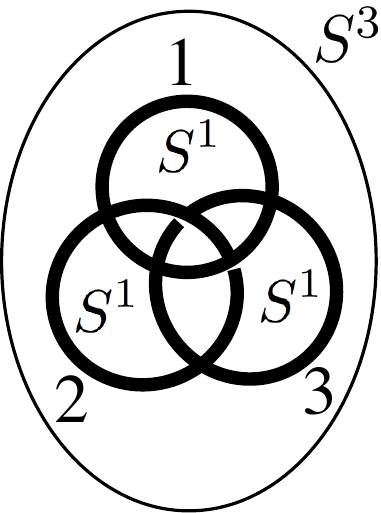} \epm
 \end{matrix}$  
     \\
\hline\\[-2mm]
\multicolumn{3}{c}{4d (3+1D) spacetime} \\
\cline{1-3}\\[-2mm]
$
 \begin{matrix}
  \int \frac{ N_I}{2\pi}{B^I  \wedge \dd C^ I}\\[2mm]
C^ I \to C^ I+  \frac{ \dd \gamma_c^I}{g_c}, \\[2mm]
 B^I  \to  B^I + \dd \eta_B^I.
 \end{matrix}
$
& ${\omega}_4=1$ & 
$\begin{matrix}
\bZ \bpm \includegraphics[scale=0.35]{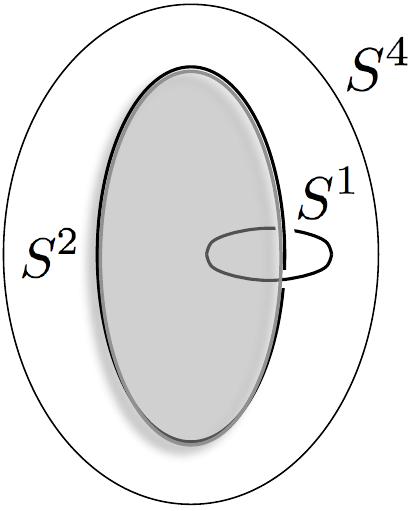} \epm 
\end{matrix}$
 \\
\hline\\[-2mm]
$\begin{matrix}
  \int 
\overset{}{\underset{{I}}{\sum}} \frac{ N_I}{2\pi}{B^I  \wedge \dd C^ I} {{+}}  
\overset{}{\underset{{I,J}}{\sum}}
\frac{ N_I N_J \; p_{IJK}}{{(2 \pi)^2 } N_{IJ}}   
C^ I \wedge C^ J \wedge \dd C^K \\[2mm]
C^ I  \to C^ I +  \frac{ \dd \gamma_c^I}{g_c}, \\[2mm]
  B^I  \to  B^I  + \dd \eta_B^I +    \varepsilon_{IJ}\frac{  N_J \; p_{IJK}}{{2 \pi } N_{IJ}}  \frac{ \dd \gamma_c^J }{g_c}\wedge C^ K, \\[2mm] 
\text{here $K$ is fixed.} 
\end{matrix}$  
&   
$\begin{matrix}
{\omega}_4=\\[2mm]
{\exp \big( \frac{2 \pi \ii p_{{IJK}}^{} }{ (N_{IJ} \cdot N_K  )   }    (a_I b_J )( c_K +d_K - [c_K+d_K  ]_{N_K}) \big)}\\[2mm]
  \in \rH^{4}((\Z_2^C)^3, \R/\Z)
\end{matrix}$   & 
$\begin{matrix}
\bZ \bpm \includegraphics[scale=0.35]{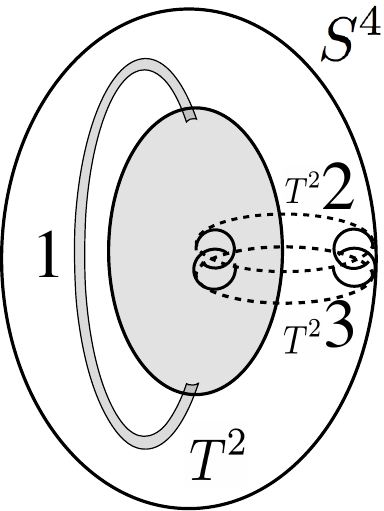} \epm 
\end{matrix}$     
\\ 
\hline\\[-2mm]
$\begin{matrix}
 \int  \overset{}{\underset{{I}}{\sum}} \frac{ N_I}{2\pi}{B^I  \wedge \dd C^ I} + {{} c_{1234}} C^ 1 \wedge C^ 2 \wedge C^ 3 \wedge C^ 4 \\[2mm]
 C^ I  \to C^ I +  \frac{ \dd \gamma_c^I}{g_c}, \\[2mm]
  B^I  \to  B^I + d\eta_B^I -\pi \frac{{\tilde{c}}_{IJKL}}{N_I}  C^J  C^ K \frac{  \gamma_c^L}{g_c} \\[2mm]
 + \pi \frac{{\tilde{c}}_{IJKL}}{N_I}  C^ J  \frac{  \gamma_c^K}{g_c}  \frac{ \dd \gamma_c^L }{g_c}
- \frac{\pi}{3} \frac{{\tilde{c}}_{IJKL}}{N_I}  \frac{  \gamma_c^J}{g_c}  \frac{ \dd \gamma_c^K}{g_c}  \frac{ \dd \gamma_c^L}{g_c}. \\[2mm]
\end{matrix}
$
 & 
$\begin{matrix}{\omega}_4^{\text{Top}}=\\[2mm]\exp \big( \frac{2 \pi \ii p_{1234}}{ N_{1234} }  a_1 b_2 c_3 d_4 \big)\\[2mm]
  \in \rH^{4}((\Z_2^C)^4, \R/\Z)
  \end{matrix}$ 
  & 
  $\begin{matrix}
\bZ \bpm \includegraphics[scale=0.35]{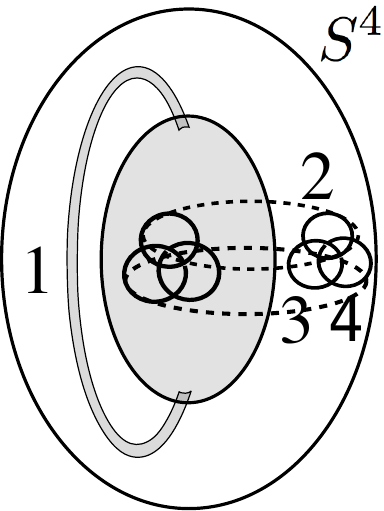} \epm
\end{matrix}$
 \\
\hline
\end{tabular}
}
\caption{Continuum field theory formulation \cite{1602.05951WWY, Putrov2016qdo1612.09298, Wang2018edf1801.05416} of topological gauge theory from cohomology group \cite{DijkgraafWitten1989pz1990} of Table 
\ref{table:Hgroup}.
Previously we list down a gauge group $G_g=(\Z_2^C)^m$ because of the limitation of our model's charge conjugation symmetry, but we can replace it to a more generic gauge group
$(\Z_{N_I})^m$, see
\Ref{Wang1405.7689,  1602.05951WWY,  Putrov2016qdo1612.09298, Wang2018edf1801.05416, Wang1901.11537WWY}. 
The first column shows the proposed continuum 
anti-symmetric tensor 
TQFT actions for these theories and their \emph{gauge transformations to all order in analytic exact forms}.
The second column shows the group-cohomology cocycle data $\omega$ as a certain partition-function solution of Dijkgraaf-Witten theory, where $\omega$ belongs to
the group-cohomology group, $\omega \in \rH^{d+1}(G,\R/\Z)=\rH^{d+1}(G,\mathrm{U}(1))$.
The notations such as the mod $N_J$ summation for $[\dots]_{N_J}$ are introduced in \Ref{Wang1405.7689}.
The third column shows the path integral form which encodes the braiding process of particles (on the open ends of 1-worldline) and strings (on the open ends of 2-worldsheet) in the spacetime.
%
In  2+1D (3d), $C$ and $B$ are 1-forms, while $\gamma_c$ and $\eta_B$ are 0-forms.
In 3+1D (4d) , $B$ is a 2-form, $C$ and $\eta_B$ are 1-forms, while $\gamma_c$ is a 0-form.
The $I,J,K \in \{1,2,3,\dots\}$ belongs to the gauge subgroup indices, 
$N_{12\dots u} \equiv  \gcd(N_1,N_2, \dots, N_u)$ is defined as the greatest common divisor (gcd) of $N_1,N_2, \dots, N_u$.
Here $p_{IJ} \in \Z_{N_{IJ}}, p_{123} \in \Z_{N_{123}}, p_{IJK} \in \Z_{N_{IJK}}, p_{1234} \in \Z_{N_{1234}}$ are integer coefficients. The $c_{IJ}, c_{123}, c_{IJK}, c_{1234}$ are quantized coefficients labeling distinct topological gauge theories, where
$c_{12}=\frac{1}{(2 \pi)} \frac{N_1 N_2\;
p_{12 }}{N_{12}}$, $c_{123}=\frac{1}{(2 \pi)^2 } \frac{N_1 N_2 N_3\;
p_{123}}{N_{123}}$,  $c_{1234}=\frac{1}{(2 \pi)^3}  \frac{N_1 N_2
N_3 N_4\; p_{1234 }}{N_{1234}}$.
The quantization labelings are described and derived in \cite{{Wang1404.7854},{Wang1405.7689}}.
Be aware that we define
both $p_{IJ \dots}$ and $c_{IJ \dots}$ as constants with \emph{fixed-indices} ${I,J, \dots}$ without summing over those indices. 
We additionally define
${\tilde{c}}_{IJ \dots} \equiv \epsilon_{IJ \dots} c_{12 \dots}$ with the $\epsilon_{IJ \dots}= \pm 1$ as an anti-symmetric Levi-Civita alternating tensor where
${I,J, \dots}$ are \emph{free indices} needed to be Einstein-summed over. The $c_{12 \dots}$ is fixed, not summed over.
%
%
}
\label{table:TQFT}
}
\end{table}
%
\subsubsection{Abelian TQFTs v.s. Top-Type Non-Abelian TQFTs} 
\label{sec:vsNon-Abelian-TQFTs}

\restoregeometry
In fact, when we have the number of $N$-copies of charge conjugation symmetry $(\Z_2^C)^N$ as the same as ${N={d+1}}$ of total spacetime dimensions,
then we are allowed to input an exotic cocycle $\omega_{d+1}^{\text{Top}}$ $\in$ $\rH^{d+1}((\Z_2^C)^{N={d+1}},\R/\Z)$ into
the path integral. In the form of continuum gauge theory, \Eq{eq:cocylce-to-continuum-QFT}
can be written as 
\bea \label{eq:combine-twist}
\int (\prod_{I=1}^{N}[\cD B_I]  [\cD C_I])\exp(\ii \int_{M^{d+1}}   (\sum_{I=1}^{N} \frac{ N_I}{2\pi}{B^I \wedge  \dd C^I}) + 
p_{}{ \frac{N_1 N_2 \dots N_{d+1}\;
}{{(2 \pi)^{d} } N_{1\dots d,d+1} }} C^1 \wedge \dots \wedge C^{d+1}) \cdot \omega_{d+1}^{\text{Lower}} 
\eea
with $p \in \Z_{N_{1\dots d,d+1}}=\Z_2$, with all of $N_I=2$ thanks to  $(\Z_2^C)$,
such that the cocycle $\omega_{d+1}^{\text{Top}}$ is known as the Top Type. (For this Top Type cocycle, 
see References therein \cite{deWildPropitius1995cf9511195} and \cite{Wang1404.7854, Wang1405.7689,  1602.05951WWY, Putrov2016qdo1612.09298, Wang2018edf1801.05416}.) These Top Type TQFTs are \emph{non-abelian} topological orders in nature,
while other Lower types $\omega_{d+1}^{\text{Lower}}$ of cocycle twists \emph{alone} produce \emph{abelian} topological orders.
Combine both the Lower and Top Types of cocycle twists as in \Eq{eq:combine-twist}
\emph{also} produce \emph{additional non-abelian} topological orders.

By non-Abelian topological orders, 
we mean that some of the following properties are matched:\\[-10mm]
\begin{itemize}  
\item  The $\dim \cH_{S^{d}; \sigma_1, \sigma_2, \sigma_3, \dots}= \bZ(S^{d} \times S^1; \sigma_1, \sigma_2, \sigma_3, \dots) =$ GSD
on a sphere $S^{d}$ with operator insertions (or the insertions of anyonic particle/string excitations on $S^{d}$)
show the following behavior:\\
(1) $\bZ(S^{d} \times S^1; \sigma_1, \sigma_2, \sigma_3, \dots)= \dim \cH_{S^{d}; \sigma_1, \sigma_2, \sigma_3, \dots} \propto k^n$
will grow exponentially as $k^n$ for a certain set of large $n$ number of insertions, for some number $k$.
These anyonic particles 
are called non-Abelian anyons or non-Abelian quasiparticles.
The anyonic string that causes this behavior can be called non-Abelian string \cite{Wang1404.7854, Wang2014wkaLevin1412.1781}.
(2)  $\bZ(S^{d} \times S^1; \sigma_1, \sigma_2, \sigma_3)= \dim \cH_{S^{d}; \sigma_1, \sigma_2, \sigma_3}>1$
for a certain set of three insertions as $(\sigma_1, \sigma_2, \sigma_3)$.
\item Prepare a certain configuration (with a degenerate Hilbert space $\dim \cH_{T^{d}}>1$, e.g. as the previous remark),
the adiabatic braiding process of gapped excitations will rotate the ground state-vector $|\Psi_{\text{initial}}\rangle$ in the Hilbert space into
a new final state-vector $|\Psi_{\text{final}}\rangle$ \emph{not} parallel \emph{nor} up to a phase respect to the initial state $|\Psi_{\text{initial}}\rangle$. Namely,
$|\Psi_{\text{final}}\rangle = U_{\text{braid}} |\Psi_{\text{initial}}\rangle \not\propto \re^{\ii \theta_{\text{abelian}}} |\Psi_{\text{initial}}\rangle$.
This means the $U_{\text{braid}}$ matrix is generically non-commutative thus non-abelian.

\item  The $ \dim \cH_{T^{d}}=\bZ(T^{d} \times S^1)=$ GSD,
for a discrete gauge theory of a gauge group $G$ on a spatial torus  $T^{d}$, behaves as GSD$<|G_g|^d$. The GSD is reduced to a smaller number than the Abelian GSD. 
This criterion however works only for a 1-form gauge theory (here for the 1-form $C$ field). See the detail explanation in the footnote 7 of \cite{Wang2018edf1801.05416}.
In Table \ref{Table:GSD}, we give these examples of abelian TQFTs v.s. non-abelian TQFTs in 2+1D and in 3+1D.

\end{itemize}  
\begin{table}[h!]
\centering
	\begin{tabular}{ |c|c| c|} 
	\hline
	$\bmm 
	 \dim \cH_{T^{d}} = \GSD_{T^d},\\
	G_g=(\Z_2^C)^{N={d+1}} \\
	\text{in }{d+1}\text{D}
	\emm$ & Abelian TQFT GSD$_{T^d}$$=|G_g|^d$ & Non-abelian TQFT GSD$_{T^d}$$<|G_g|^d$  \\
	\hline
	2+1D & $(2^3)^d=64$ & 22  \\
	\hline
	3+1D & $(2^4)^d=4096$  & 1576 \\
		\hline
	\end{tabular}
	\nonumber
	\caption{Data of zero energy modes counting below the TQFT energy gap, as $\dim \cH_{T^{d}} = \GSD_{T^d}$ 
	 from \cite{Wang1404.7854, Wang2018edf1801.05416}.}
	\label{Table:GSD}
\end{table}

\subsubsection{Non-abelian tensor gauge theory with gapless modes, massive particles and gapped non-abelian anyonic strings
in 3+1D (4d),
and gapped non-abelian anyonic branes
in 4+1D (5d)}

Below let us write down the  3+1D (4d) and 4+1D (5d) models explicitly that have non-abelian anyonic excitations. 
Both
are non-abelian tensor gauge theory with gapless modes. Both have massive particles (from the end of worldline of $C$ field).
The  3+1D (4d) model has gapped anyonic strings (the end of worldsheet of $B$ field).
The  4+1D (5d) model has gapped anyonic branes (the end of worldvolume of $B$ field).
%
%
Here are the 3+1D (4d) and 4+1D (5d) models with schematic path integrals written in generality:\footnote{Readers should beware that the definitions of
$B$  and  ${\rm B}$ are fundamentally different.
We use $B$ for the any-symmetric tensor field (here $(d-2)$-form gauge field),
while we use ${\rm B}$, ${\overline {\rm B}}$, $\tilde {\rm B}$, $\hat {\rm B}$, etc., to represent the magnetic tensor field from the abelian field strength $F_{\mu \nu \xi}$ or its anisotropic version.
Similarly, we use $\hat {\rm B}_{}^c$, etc., to represent the magnetic tensor field from the nonabelian field strength $\hat F^c_{\mu \nu \xi}$.
}
\hspace*{-6mm}
\begin{multline}
\label{eq:ZABC-aniso-4}
{\bZ}_{\underset{\text{asym-tw}}{\text{rk-2-sym}}}^{\text{3+1D}}:=\int (\prod_{I=1}^{N}[\cD A_{I,\mu\nu}] [\cD B_I]  [\cD C_I])
\exp(\ii \int_{M^{d+1}}  
( \sum_{I=1}^{N} (|\hat {\rm E}_{}^{c,I} |^2-|\hat {\rm B}_{}^{c,I} |^2))
(\dd^{d+1} x) \\+ 
 \sum_{I=1}^{N}  \frac{ N_I}{2\pi}{B^I  \wedge\dd C^I} + 
p_{}{ \frac{N_1 N_2 N_3 N_4\;
}{{(2 \pi)^3 } N_{1234}}} C^1 \wedge  C^2 \wedge  C^3 \wedge C^4) \cdot \omega_{4}^{\text{Lower}}(\{C_I\}).
\end{multline}\\[-6mm]
\hspace*{-26mm}
\begin{multline}
\label{eq:ZABC-aniso-5}
{\bZ}_{\underset{\text{asym-tw}}{\text{rk-2-sym}}}^{\text{4+1D}}:=\int (\prod_{I=1}^{N}[\cD A_{I,\mu\nu}] [\cD B_I]  [\cD C_I])
\exp(\ii \int_{M^{d+1}} 
( \sum_{I=1}^{N} (|\hat {\rm E}_{}^{c,I} |^2-|\hat {\rm B}_{}^{c,I} |^2))
(\dd^{d+1} x) \\+ 
 \sum_{I=1}^{N}  \frac{ N_I}{2\pi}{B^I  \wedge\dd C^I} + 
p_{}{ \frac{N_1 N_2 N_3 N_4N_5\;
}{{(2 \pi)^4 } N_{12345}}} C^1 \wedge C^2 \wedge  C^3 \wedge C^4 \wedge C^5) \cdot \omega_{5}^{\text{Lower}}(\{C_I\}).
\end{multline}
%
Here $p \in \Z_{N_{1\dots d,d+1}}=\Z_2$ and $N_1 = N_2 = N_3 = N_4 = N_5= \dots =2$.
For models of Euclidean/Lorentz version in \Sec{sec:Euclidean-Lorentz}, we have 
$$
\hspace*{-6mm}
(|\hat {\rm E}_{}^{c,I} |^2-|\hat {\rm B}_{}^{c,I} |^2) := |\hat F^{c,I}_{\mu \nu \xi}|^2
=\Big((\prt_\mu - \ii g_c C_{\mu,I}) A_{\nu \xi,I} -(\prt_{\nu} - \ii g_c C_{\nu,I}  ) A_{\mu \xi,I} \Big)
\Big((\prt^\mu + \ii g_c C^{\mu,I} ) A^{\nu \xi, I}  -(\prt^{\nu} + \ii g_c C^{\nu,I}  ) A^{\mu \xi,I} \Big),$$
where $\hat {\rm E}_{}^c, \hat {\rm B}_{}^c$ should be the covariant version of $\overline{\rm E}, \overline{\rm B}$ in \Eq{eq:EE} and \Eq{eq:BB}.
For  models of anisotropic version in \Sec{sec:Anisotropic}, we have $(|\hat {\rm E}_{}^c |^2-|\hat {\rm B}_{}^c |^2) := (|\tilde {\rm E}_{ij}^c |^2-|\tilde {\rm B}_{ij}^c |^2)$
given by \Eq{eq:aniso-E} and \Eq{eq:aniso-B}, and  $[\cD A_{I,\mu\nu}]$ should be replaced by $[\cD A_{I,0}][\cD A_{I,ij}]$.

\label{eq:Top-Type}


\subsection{Symmetric Higher-Rank Tensor Gauge Theory  and Gapless Phase}

\label{sec:symTgauge}

We discuss that the symmetric higher-rank tensor gauge theory with an action alone
$\int_{M^{d+1}} 
 |\hat F^c_{\mu \nu \xi}|^2 (\dd^{d+1} x)$
in $d+1$ dimensions (at least $d+1 \geq 3+1$) 
 can contain gapless phase with massless modes.

\subsubsection{Degrees of freedom for gapless modes}

\emph{Degrees of freedom ($\equiv$ dof) counting for gapless modes}:\\[-8mm]
%
%
\begin{enumerate}
\item U(1) gauge theory of \Sec{sec:U1maxwell-gauge}:\\
Let us recall the standard dof counting for the gapless (or massless) modes, e.g., the photons, 
in the spin-1 Maxwell theory in 3+1D (4d) in \Sec{sec:U1maxwell-gauge}. 
Let a single photon moves in the speed of light say with a momentum $\vec{p}_z$ along an arbitrary $z$ direction.

Let us first count the photon dof physically. Naively the photon can either oscillate along the longitudinal direction (1 dof along $\vec{p}_c$) 
and the transverse direction (2 dof perpendicular to $\vec{p}_z$). However, the longitudinal mode does not make sense because the photon
cannot oscillate forward faster than the speed of light. Thus there are only 2 transverse modes:
say along with the $p_x$ and $p_y$ directions. Then their linear combinations $p_x + \ii p_y$ and $p_x - \ii p_y$ 
are effectively the two eigenvectors for the spin angular momentum operator $\hat{S}_z$ of the photon. Namely, there is a map of 2 dof:
$$
p_x \pm \ii p_y \Longrightarrow  \langle \hat{S}_z \rangle = \pm 1 \Longrightarrow |  {S}, S_z \rangle = | 1, \pm 1 \rangle.
$$
These are in fact also the 2 dof from the 2 helicity from the 2 transverse modes
$$
\hat{S}_z \cdot \hat{P}_z = \pm 1.
$$

Let us also count the dof mathematically.
We see that a photon field is a 4-vector $A_\mu$ with $\mu=0,1,2,3$. We can choose the gauge on $\eta$, such that \Eq{eq:U1maxwell-gauge}
$A_0 \to A_0 +\frac{1}{g} \prt_0 \eta=0$, which gives 1 dof out of 4 dof. Furthermore, the gauge transformation $A_0 \to A_0 +\frac{1}{g} \prt_0 \eta$
can still hold if $\eta \to \eta + \eta'(x_j)$  where $\eta'(x_j)$ is a pure function of space indices $x_j$.
Therefore the real physical degrees of freedom (dof) for gapless modes of photon can be at most 
$4-1-1=2$ dof, which agrees with the physics story given earlier.

\item SU(N) YM gauge theory of \Sec{sec:SUn-YM-gauge}:\\
Similarly, the dof counting of gluon field $A_{\mu}^{\al}$ is similar to the previous counting.
We have 2 dof for each  $A_{\mu}^{\al}$ and there are $\rN^2-1$ independent of gluons with index $\al$ in the adjoint representation.  
However, as we know due to the color confinement \cite{Wilson1974skPhysRevD.10.2445ConfinementofQuarks},
the gapless dof of gluon field $A_{\mu}^{\al}$ \emph{do not} appear at the low energy, but only above the confinement energy gap scale $\geq \Lambda_{\text{YM}}$.
At low energy $< \Lambda_{\text{YM}}$, we only observe a confinement energy gap without massless modes.

\item {Euclidean or Lorentz invariant non-abelian higher-rank tensor gauge theory} of \Sec{sec:Euclidean-Lorentz}:

\Ref{2016arXiv160108235RRasmussenYouXu} finds that the compact abelian \emph{symmetric} tensor gauge theory is unstable to confinement in 2+1D, 
but it becomes stable to be gapless-ness with deconfinement in 3+1D. We thus focus on 3+1D (or above) for its richer physics in a gapless phase.

Let us count the dof in 3+1D.
We see that the rank-2 symmetric tensor field is $A_{\mu \nu}$ with $\mu,\nu=0,1,2,3$ naively with 10 dof. We can choose the gauge on $\eta^v$ such that \Eq{eq:vector-gauge}
$A_{00} \to A_{00} +\frac{1}{g} \prt_0\prt_0 \eta^v=0$, which kills 1 dof out of 10 dof. Furthermore, the gauge transformation $A_{00} \to A_{00} +\frac{1}{g} \prt_0\prt_0 \eta^v=0$
can still hold if $\eta^v \to \eta^v + \eta'(x_j) + t \cdot\eta''(x_j)$  where the two additional redundant dof $\eta'(x_j)$ and $ \eta''(x_j)$ are pure functions of space indices $x_j$.
Since the shifted function has a linear $t$ dependence, we have $ \prt_0\prt_0 \eta^v \to  \prt_0\prt_0 \eta^v$.
Therefore the real physical degrees of freedom (dof) for gapless modes of photon can be at most 
$10-1-2=7$ dof.

Let us also count the dof in $d+1$D. The rank-2 symmetric tensor field is $A_{\mu \nu}$ naively with $\frac{(d+2)(d+1)}{2}$ dof.
We can choose the gauge on $\eta^v$ such that \Eq{eq:vector-gauge}
$A_{00} \to A_{00} +\frac{1}{g} \prt_0\prt_0 \eta^v=0$, which reduces 1 dof. 
Furthermore, the gauge transformation $A_{00} \to A_{00} +\frac{1}{g} \prt_0\prt_0 \eta^v=0$
can still hold if $\eta^v \to \eta^v + \eta'(x_j) + t \cdot\eta''(x_j)$ 
since the shifted function (with 2 dof) has a linear $t$ dependence, we have $ \prt_0\prt_0 \eta^v \to  \prt_0\prt_0 \eta^v$.
Therefore the real physical degrees of freedom (dof) for gapless modes of photon can be at most 
$\frac{(d+2)(d+1)}{2}-1-2=\Big(\frac{(d+2)(d+1)}{2}-3\Big)$ dof.

\item {Anisotropic non-abelian higher-rank tensor gauge theory for space and time} of \Sec{sec:Anisotropic}.

Again due to 3+1D may becomes stable to be gapless-ness with deconfinement \cite{2016arXiv160108235RRasmussenYouXu},
let us count the dof in 3+1D or in $d+1$D.
The rank-2 symmetric tensor field is $A_{i j}$ naively with $\frac{(d+1)d}{2}$ dof and $A_0$ for 1 dof.
We can choose the gauge on $\eta^v$ such that \Eq{eq:vector-gauge-space}
$A_{0} \to A_{0} +\frac{1}{g} \prt_0 \eta^v=0$, which reduces 1 dof. 
Furthermore, the gauge transformation $A_{0} \to A_{0} +\frac{1}{g} \prt_0 \eta^v=0$
can still hold if $\eta^v \to \eta^v + \eta'(x_j)$ 
since the shifted function (with 1 dof) has a linear $t$ dependence, we have $ \prt_0 \eta^v \to \prt_0 \eta^v$.
Therefore the real physical degrees of freedom (dof) for gapless modes of photon can be at most 
$\frac{(d+1)d}{2}+1-1-1=\Big(\frac{(d+1)d}{2}-1\Big)$ dof.
In 3+1D, we have 5 dof.
\end{enumerate}
%
%
See a summary in Table \ref{Table:dof-dispersion}.

\subsubsection{Dispersions of gapless modes}

Now we determine the dispersions of gapless modes, as the scaling of energy $\omega_E(k) \propto k^{\rm z}$
and the momentum $k$ to the dispersion's dynamical exponent ${\rm z}$. For the dimensional analysis, below we denote the scaling of
the spatial dimensions as $[L]$.

\begin{table}[h!]
\centering
	\begin{tabular}{ |c|c| c|} 
	\hline
	$\bmm 
	\text{gapless dof}
	\emm$ & 
{Euclidean/Lorentz invariant} of \Sec{sec:Euclidean-Lorentz}	
	 &  
{Anisotropic for space and time} of \Sec{sec:Anisotropic}
	  \\
	\hline
	3+1D & 7  & 5 \\
        \hline
	$d$+1D & $\Big(\frac{(d+2)(d+1)}{2}-3\Big)$  & $\Big(\frac{(d+1)d}{2}-1\Big)$ \\
		\hline
		\hline
		$\bmm 
	\text{dispersion}\\
		\text{3+1D}
	\emm$& 
	$\omega_E(k) \propto c_v k$ &  
	  $\omega_E(k) \propto c_v k$ \\
	  \hline	 
	  $\bmm 
	\text{E and B tensor}\\
\text{SO(3) Rep}
	\emm$
	  & 
	  $\begin{array}{lll}
\overline{\rm E} &=&{\bf 3}  +( {\bf 1} +  {\bf 3} + {\bf 5}),\\
\overline{\rm B} &=&{\bf 3}  +( {\bf 1} +  {\bf 3} + {\bf 5}).
 \end{array}$   \text{in \Eq{eq:EB-ML-SO3}}
	   &
	    $ \begin{array}{lll}
\tilde {\rm E}_{}  &=&( {\bf 1} + {\bf 5}),\\
\tilde {\rm B}_{} &=& ( {\bf 1} +  {\bf 3} + {\bf 5}).
 \end{array}$
   \text{in \Eq{eq:EB-Aniso-SO3}}
 \\
	  \hline
	\end{tabular}
	\nonumber
	\caption{We collect the data of the degrees of freedom (dof) and dispersion relation for gapless modes, and the electric E and magnetic B tensors
	in terms of the spatial rotational SO(3) representation.}
	\label{Table:dof-dispersion}
\end{table}
\begin{enumerate}
\item Obviously both Maxwell and Yang-Mills theories have photons and gluons
with linear dispersion relation $\omega_E(k) \propto c k$, where $c$ is known as the speed of light constant (alternatively,
the effective speed of light in the quantum matter material). This is easy to derive based on the scaling between
E$^2 = c^2$ B$^2$, where the ratio gives the (square of) the speed of light.

\item {Euclidean or Lorentz invariant non-abelian higher-rank tensor gauge theory} of \Sec{sec:Euclidean-Lorentz}:\\
We see that from 
\Eq{eq:vector-gauge}, \Eq{eq:EE} and \Eq{eq:BB},
\bea
\left\{\begin{array}{lll} 
[ A_{\mu \nu}]   &\sim& \frac{1}{g} [\prt_\mu \prt_\nu ( \eta_v(x))] \sim [L]^{-2},\\[0mm] %
[\overline{\rm E}^{}_{ij}] &\sim&[\prt_0 A_{i j} -\prt_{i} A_{0 j}] \sim [L]^{-3},\\[0mm] %
[\overline{\rm B}^{}_{\ell k}] &\sim& [\frac{1}{2}\epsilon^{\ell ij} ( \prt_i A_{jk}-\prt_j A_{ik})] \sim [L]^{-3}. \\[0mm] 
[\overline{\rm E}^{}_{ij}]^2 &\sim&[\overline{\rm B}^{}_{\ell k}]^2, \text{ as } [\prt_0]^2  \sim (v_c)^2 [\prt_{i}]^2.
 \end{array}\right.
\eea
So $[t] \sim [x]\sim [L]$, thus it is a linear dispersion relation $\omega_E(k) \propto v_c k$ with ${\rm z}=1$.

\item {Anisotropic non-abelian higher-rank tensor gauge theory for space and time} of \Sec{sec:Anisotropic}:\\
From \Eq{eq:vector-gauge-space} and \Eq{eq:EB-aniso},
\bea
\left\{\begin{array}{lll} 
[ A_0]   &\sim&  \frac{1}{g} [\prt_0  ( \eta_v(x))] \sim \frac{1}{g} [\prt_t  ( \eta_v(x))] \sim  [L]^{-1},\\[0mm]
[A_{ij }]   &\sim& \frac{1}{g} [\prt_i \prt_j ( \eta_v(x))] \sim  [L]^{-2},\\[0mm]
[\tilde {\rm E}_{ij}] &\sim& [-\partial_0A_{ij}+\partial_i\partial_j A_0 ]\sim  [L]^{-3},\\[0mm]
[\tilde {\rm B}_{ij}]&\sim& [\varepsilon_{i \ell m} \partial_\ell A_{mj}] \sim  [L]^{-3}. \\[0mm]
[\tilde {\rm E}_{ij}]^2 &\sim& [\tilde {\rm B}_{ij}]^2,  \text{ as } [\prt_0]^2  \sim (v_c)^2 [\prt_{i}]^2.
 \end{array}\right.
\eea
So $[t] \sim [x]\sim [L]$, it is still a linear dispersion relation $\omega_E(k) \propto v_c k$ with ${\rm z}=1$.
\end{enumerate}
See a summary in Table \ref{Table:dof-dispersion}.
In fact, for other tensor gauge models, there are other possibilities of gapless modes with different dispersion 
of $\omega_E(k) \propto  k^{\rm z}$ with ${\rm z}=1,2,3$, etc., e.g., see \Ref{2016arXiv160108235RRasmussenYouXu, Pretko2016lgv1606.08857}.

\section{Discussions: Fracton, Embeddon and Foliation}
\label{sec:Discussions:FractonEmbeddonFoliation}

\subsection{Fracton}
\label{sec:Fracton}

The name Fracton is introduced in \cite{Vijay2015mka1505.02576VijayHaahFu, Vijay2016phm1603.04442}.
Famous examples include the Chamon quantum glass model \cite{2005PRL0404182Chamon, 1108.2051CastelnovoChamon}, the Haah's cubic code \cite{2011PhRvAHaah1101.1962} and others.
The characterizations and physics definitions of fracton orders are 
that the excitations (including gapless or gapped excitations) have restricted mobility when acted by local operators \cite{RahulNandkishore2018sel1803.11196}.
The fractonic excitations have either of the following:\\[-10mm]
\begin{enumerate}[label=\textcolor{blue}{(\arabic*)}., ref={(\arabic*)}]
\item \label{p:fractonic}
The excitations cannot move without creating additional excitations (other fractonic excitations).
\item  \label{p:sub}
The excitations can move only in  sub-dimensions or certain directions. These excitations are also known as sub-dimensional particles. 
\end{enumerate}
It is easy to see that the theory in \Eq{eq:U(1)vector-gauge-matter} and \Eq{eq:U(1)vector-gauge-matter-space}, indeed have the property \ref{p:fractonic}. 
By looking at one term, for instance, we obtain
\bea
&&(\Phi \prt_x \prt_y \Phi - \prt_x \Phi \prt_y \Phi - \ii g A_{xy} \Phi^2) \\
&&\simeq 
\Phi(x,y) \big((\Phi(x+\Delta x,y+\Delta y) - \Phi(x,y+\Delta y) -\Phi(x+\Delta x,y) + \Phi(x,y) \big)   \nn\\
&& -(\Phi(x+\Delta x,y)- \Phi(x,y))(\Phi(x,y+\Delta y)- \Phi(x,y)) - \ii g A_{xy}(x,y) \Phi^2(x,y)  \nn\\
&&=\Phi(x,y)\Phi(x+\Delta x,y+\Delta y)-\Phi(x+\Delta x,y)\Phi(x,y+\Delta y)- \ii g A_{xy}(x,y) \Phi^2(x,y)
\eea in a discretized manner. To minimize the action by the variational principle, say this $\Phi(x,y)\Phi(x+\Delta x,y+\Delta y)-\Phi(x+\Delta x,y)\Phi(x,y+\Delta y)$ term,
it means that the $\U(1)_{x_{(d+1)}}$ or  $\U(1)_{x_{(d)}}$ vector global symmetry charge should be arranged in a higher-moment quadrupole manner.
Indeed, the property that the \emph{dipole} should not be created but the
\emph{quadrupole} can be created, is noticed and analyzed in various Pretko's work \cite{Pretko2018jbi1807.11479}.

We hope to explore \ref{p:sub} in a new setting in terms of the spacetime embedding in the next subsection.

\subsection{Embedding to Embeddon}
	\label{sec:embedding}

\begin{figure}[!h]
	\centering
	\includegraphics[width=10.cm]{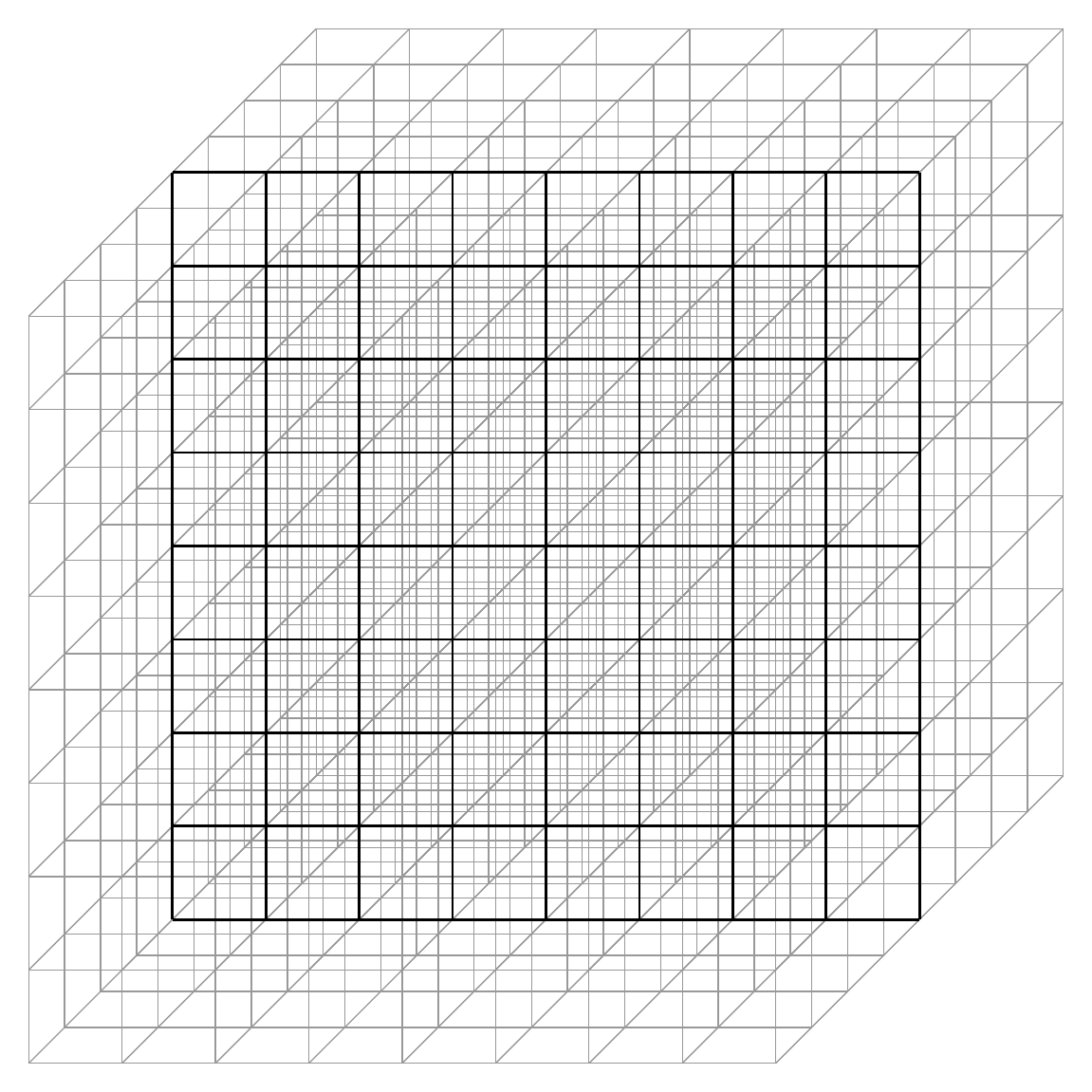}
	\caption[caption]{Interpretations of the embeddon:
	(1) Quantization of spacetime embedding \cite{JWang2018Embeddon}. 
(2) The anyonic objects (e.g. particles/strings/branes) live in the embedded manifold ${M^{n}_{\text{sub-$M$}}} \subset {M^{d+1}}$ in $n$d, thus the 
anyonic objects are embedded inside the sub-dimensions.  
	}
	\label{fig:embedding}
\end{figure}
Let us construct another infinite series of new theories similar to \Eq{eq:ZABC} and \Eq{eq:ZABC-aniso}, by the idea of {embedding}.
We can devise new kinds of path integral as:
\begin{multline} \label{eq:ZABC-embed}
\int(\prod_{I=1}^{N}[\cD A_{I,\mu\nu}] [\cD B_I]  [\cD C_I])
\exp(\ii \int_{M^{d+1}} 
(\sum_{I=1}^{N} |\hat F^{c,I}_{\mu \nu \xi}|^2) +\\
\int_{M^{n}_{\text{sub-$M$}}} \ii(\frac{ 2}{2 \pi} \sum_{I=1}^{N} B_I \dd C_I +\# C_I \wedge C_J  \wedge \dots \wedge \dd C_K ) \underbrace{\dots}_{\text{NO ambient space}} ). 
\end{multline}
Here we introduce a notation
$\underbrace{\dots}_{\text{NO ambient space}}$  means that there is no volume form of the ambient space.
We introduce the lower dimensional $n$d anti-symmetric TQFT on the submanifold ${M^{n}_{\text{sub-$M$}}} \subset {M^{d+1}}$.
For example, the submanifold ${M^{n}_{\text{sub-$M$}}} \subset {M^{d+1}}$ is a darken slice in \Fig{fig:embedding}.
We thus have a constraint $N \leq n \leq {d+1}$ for the twisted topological term
$\omega_{n}(\{C_I\})$ that we can add from Table \ref{table:Hgroup}
and Table \ref{table:TQFT}, here $\omega_{n} \in \mathrm{H}^{n}((\Z_2^C)^N, \R/\Z)$ for a lower-dim twisted cohomology group.

Physically, what we are doing is to embed an anti-symmetric TQFT in a lower dimensional space, into a higher dimensional space
where the symmetric tensor gauge theory resides. Mathematically, this can be done by gauging the sub-sector of the \Eq{eq:GgZ2C}'s $G_g=[(\Z_2^C)^N]$.
The anyonic objects (e.g. particles/strings/branes) described by $\omega_{n}$ live in $n$d, thus the 
anyonic objects are embedded inside the sub-dimensions. We shall name those embedded objects in the quantum theory as 
\emph{embeddon}.

\subsection{Foliation}
	\label{sec:Foliation}

\begin{figure}[!h]
	\centering
	\includegraphics[width=10.cm]{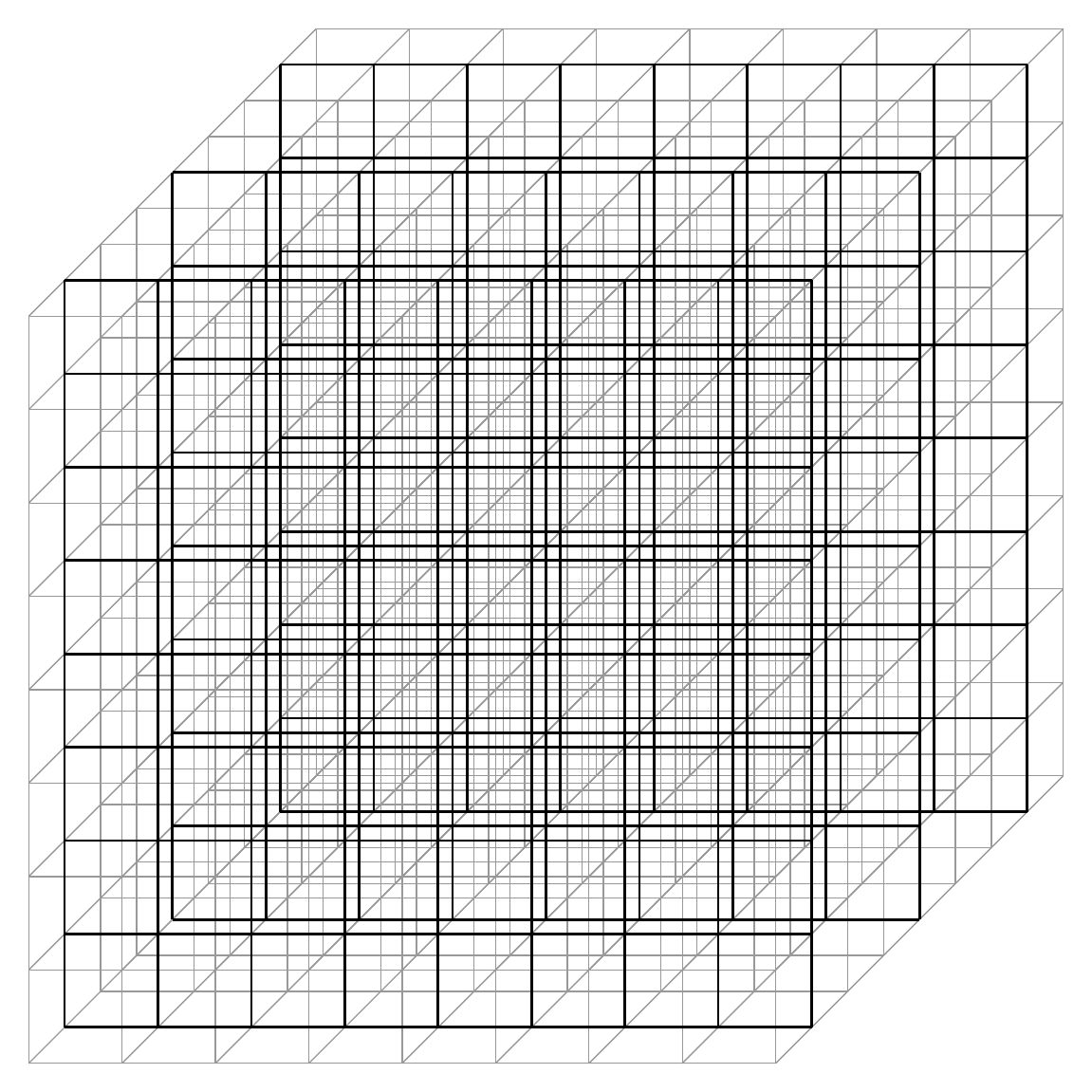}
	\caption[caption]{From the sequences of embedding to foliation in the spacetime.
}
	\label{fig:foliation-plane}
\end{figure}

We design new kinds of path integral as another infinite series of new theories similar to \Eq{eq:ZABC} and \Eq{eq:ZABC-aniso}, similar to \Eq{eq:ZABC-embed} 
 but different in spirit:
\begin{multline} \label{eq:ZABC-foliation}
\int(\prod_{I=1}^{N}[\cD A_{I,\mu\nu}] [\cD B_I]  [\cD C_I])
\exp(\ii \int_{M^{d+1}} 
(\sum_{I=1}^{N} |\hat F^{c,I}_{\mu \nu \xi}|^2) +\\
(\frac{ 2}{2 \pi} \sum_{I=1}^{N} B_I \dd C_I +\# C_I \wedge C_J  \wedge \dots \wedge \dd C_K ) \underbrace{(\wedge \dots \wedge \dd x')}_{\text{ambient space}} ). 
\end{multline}
Here $\underbrace{(\wedge \dots \wedge \dd x')}_{\text{ambient space}}$ is the volume of the ambient space.
We gain another theory by a sequence of embedding of submanifolds into a new picture of {foliation}.
By this stacking of embedded TQFT picture along any codimension direction (in the dual $d-n$-dimensions in space), 
we can also easily reproduce the exponential growth of  ground state degeneracy (GSD) associated to the $d$-dimensional space (counting zero energy modes) 
and thus exponential growth of the ground state subspace dim($\cH_d$).

 For the physics picture, our theory thus potentially corresponds to the exponential growth of GSD counting respect to the system size in \Ref{Slagle2017wrc1708.04619}
and the \emph{foliation picture} \cite{Shirley2017suz1712.05892} on a wider class of manifolds.
We may also construct the sub-dimensional sectors with BF theory or Chern-Simons theory in $n$d with $n \leq d+1$, either via 
the embedding in \Sec{sec:embedding} or the foliation in \Sec{sec:Foliation}, potentially bridge to another recent work \cite{You2019ciz1904.11530}.
Our theory can also combine the \emph{foliation} (in this subsection) and the \emph{anisotropic properties} (\Sec{sec:Anisotropic}), likely reproduce a field theory interpretation 
for a lattice construction similar to \cite{Fuji2019qea1908.02257}.

\section{Conclusions: Quantization, Feynman diagram, Lattice Model, and Dark Matter}
\label{sec:conclude}

\begin{enumerate}[label=\textcolor{blue}{\arabic*}., ref={\arabic*}]
\item We have constructed a new family of hybrid classes between symmetric higher-rank tensor gauge theories and anti-symmetric tensor topological field theories (TQFTs)
in \emph{any dimension}.
In condensed matter community, it is believed that the compact abelian symmetric rank-2 tensor field theory in \Eq{eq:U(1)vector-gauge-matter}
without matter field, in 3+1D (4d) or above dimensions, describes a gapless deconfined phase \cite{2016arXiv160108235RRasmussenYouXu}. 
Thus our theory in 3+1D (4d) could describe a unitary mixture phase including the
 gapless sectors (which can live with or without Euclidean, Poincaré or anisotropic symmetry, from
 compact abelian symmetric tensor gauge theories, thus possibly can be relativistic or non-relativistic), and the gapped topological order phases (with anti-symmetric tensor TQFTs at low energy).

However, the mixture of such gapless phase and gapped phase is only the effective field theory (EFT) description, at least in the ultraviolet high energy (UV) or intermediate energy,
but not yet to the lattice cutoff scale.

What are the infrared low energy (IR) fates and quantum dynamics of our theories?
First of all, the continuous non-abelian gauge structure of our theories is 
{$\left[\Z_2^C \ltimes \Big( 
 \U(1)_{x_{(d+1)}} \Big) \right]$} in \Eq{eq:gauge-group-analogous},
 a higher-moment continuous $ \U(1)_{x_{(d+1)}} $ twisted by the discrete gauged $\Z_2^C$. So one may expect that the 
\emph{gapless} and \emph{deconfinement} of compact abelian symmetric higher-rank tensor gauge theories should be maintained.
We may speculate the gapped anti-symmetric tensor TQFT sector would not severely interfere the gapless sector, but only introduce additional massive degrees of freedom
of anyonic particles (in 2+1D)/strings (in 3+1D)/branes (in 4+1D), etc. as massive high-energetic  excitations above certainly TQFT energy gap $\Delta_E$.

To confirm the speculation, we need to know the quantum dynamics of our theories. It may be challenging.
To make an analogy, writing down a path integral form of Yang-Mills theory  \cite{PhysRev96191YM1954}
is still quite far from to understanding its quantum dynamics and 
confinement mechanism \cite{Wilson1974skPhysRevD.10.2445ConfinementofQuarks}. Let us make a few remarks before coming back to 
the quantization issue in the Remark \ref{eq:Feynman-graph-quantization}.

 \item
\emph{Causes of non-abelian higher-moment global symmetry}:\\
 It is easy to see that the non-abelian global symmetry is due to the non-commutative nature of $( \Z_2^C \ltimes \Big( \U(1) \times \U(1)_{x_{(d+1)}}  \Big) )$
displayed in \Eq{eq:Euclidean-sym} and \Eq{eq:Minkowski-sym}.  Our non-abelian-ness  is however \emph{not} due to a non-abelian Lie group and 
its non-commutative Lie algebra as the familiar case like Yang-Mills isopsin theory \cite{PhysRev96191YM1954}.
 
 \item
\emph{Causes of non-abelian gauge structure}:\\
There are two ingredients in our construction to cause the non-abelian gauge structure.\\[-6mm]
\begin{itemize}[leftmargin=2.0mm]
\item
The first ingredient is due to the ``\emph{continuous} gauge group-analogous structure'' is \Eq{eq:gauge-group-analogous}
$\left[\Z_2^C \ltimes \Big(  \U(1)_{x_{(d+1)}} \Big) \right]$ once we gauge both $ \U(1)_{x_{(d+1)}}$ and $\Z_2^C$.
Even without any topological twisted term $\omega_{d+1}$, our theory is still a non-abelian tensor gauge theory (say in \Eq{eq:ZABC} and others).\\

In contrast, recent works in the fracton literature construct only non-abelian fracton order with \emph{discrete} gauge structure on a lattice, 
via distinct mechanisms (distinct from ours):
either through gauging the permutation symmetry of $N$-layer systems \cite{BulmashMaissamBarkeshli2019taq1905.05771, PremWilliamson2019etl1905.06309},
or coupling to non-abelian TQFTs/topological orders \cite{Vijay2017cti1706.07070, Song2018gbb1805.06899,  PremHuangSong2018jsn1806.04687}.\\

\item
The second ingredient is due to the topological twisted term $\omega_{d+1}$ involving the Top Type explained in \Sec{sec:vsNon-Abelian-TQFTs}.
\end{itemize}

\item \label{remark:Mathematical-ingredients}
\cred{\emph{Mathematical ingredients for the field theories}: \\
Let us discuss the mathematical interpretations of field contents in the abelian gauged fractonic matter theory \Eq{eq:U(1)vector-gauge-matter} and the non-abelian gauge theory
\Eq{eq:ZABC}.
\begin{itemize}[leftmargin=2.0mm]
\item
The complex scalar $\Phi$ is a section (or a cross section) of field bundle which is a complex line bundle over $M^{d+1}$.
\item
The $A$ is a rank-2 symmetric tensor gauge field, see the explicit math formulation and generalization in \cite{2007arXiv0711.0977LJohnLoftin, Wang2019cbjJWKaiXu1911.01804, WXY3Wang2019mtt1912.13485, LWXYfracton}. (This is rather different, in contrast, from the ordinary 1-form gauge field which is a gauge connection of the principal U(1)-bundle.
The ordinary 1-form gauge field can be also the gauge connection of the complex line bundle of $\Phi$.)
\item The $B$ is an anti-symmetric tensor $(d-1)$-form gauge field. The $B$ can also be mapped to the $\Z_2$-valued $(d-1)$-cochain, in that case
$B \in \rH^{d-1}(M^{d+1},\Z_2)$, mapping from $M^{d+1}$ to the higher classifying space of $\Z_2$: $M^{d+1} \to \B^{d-1} \Z_2$.
The $B$ can also be regarded as the Lagrange multiplier to set $C$ to be locally flat.\footnote{\cred{In the homotopy theory,
the classifying space of $\B G_g$ \cite{DijkgraafWitten1989pz1990}
of a group $G_g$ 
is the quotient of a weakly contractible space E$G_g$ 
($i.e.$, a topological space whose all homotopy groups are trivial) 
by a proper free action of $G_g$. The 
classifying space $\B G_g =K(G_g,1)$ is the Eilenberg-MacLane space.
For an abelian $G_g$, the higher
classifying space $\B^{d-1} G_g$ is the Eilenberg-MacLane space $K(G_g,d-1)$.}
}
\item The $C$ is a $1$-form gauge field. The $C$ can also be regarded as the $\Z_2$-valued 1-cochain, 
in that case
$C \in \rH^1(M^{d+1},\Z_2)$, mapping from $M^{d+1}$ to the classifying space of $\Z_2$: $M^{d+1} \to \B \Z_2$.
It becomes the $\Z_2$-valued 1-cocycle under the constraint of $B$'s EOM.
\end{itemize}
}

\item \emph{Fracton, Embeddon, and Foliation}: We explain how our theories can incorporate the phenomenon of Fracton in \Sec{sec:Fracton}, almost 
directly given from \Ref{Pretko2018jbi1807.11479}.
We introduce the spacetime embedding in terms of field theory setup and the Embeddon in \Sec{sec:embedding},
and how it bridges to build up the foliation  in \Sec{sec:Foliation}.

\item
\emph{Feynman diagrams (Dyson graphs), path integral and (canonical) quantization}: \label{eq:Feynman-graph-quantization}\\
A fully-fledged quantum field theory, even of \Eq{eq:U(1)vector-gauge-matter} or \eq{eq:U(1)vector-gauge-matter-space}, with matter and with or without gauge fields,
by a field-theoretic quantization is still an open question. The usual canonical quantization by identifying 
the field variable $\Phi$ and the conjugate momentum variable may not be transparent for this quartic higher-order intrinsic 
interacting field theory.\footnote{Note that there is a very recent work by Seiberg \cite{SeibergF2019} commenting about the quantization of the quartic matter field theory
and its spectrum, however without the gauge fields.}
On the other hand, the path integrals we formulated are so far schematic.
It is important to notice that the matter theory
(\Eq{eq:U(1)vector-gauge-matter} or \eq{eq:U(1)vector-gauge-matter-space}) has \emph{none} quadratic (free) kinetic term,
thus it is associated with the non-Gaussian integral, instead of the familiar Gaussian integral.
It also suggests that the two-point correlation function (or Green's function) as the free propagator may not exist,
$\langle \Phi^\dagger(x_1) \Phi(x_2)  \rangle \simeq 0$.
Likely we should think about this quantum field theory starting from the three-point correlation function
\bea
\langle \Phi^\dagger(x_1) \Phi^\dagger(x_2)  \Phi(x_3) \rangle
\eea
or the 
four-point correlation function
of matter fields:
\bea \label{eq:4-point-correlation-function}
\langle \Phi^\dagger(x_1) \Phi^\dagger(x_2)  \Phi(x_3)  \Phi(x_4) \rangle. 
\eea
However, we can at least attempt to identify the interaction terms and write down the Feynman diagrams in the perturbative sense.
We do \emph{not} include the Feynman rules, as it depends on whether we can use the propagators of free fields as the starting point to 
compute the scattering amplitudes. As we mentioned in \Sec{sec:Fracton} 
that the quadrupole (instead of dipole) of interacting matters at four different locations is the better configuration to
think about the low energy physics of the theory --- this observation indeed is consistent with \Eq{eq:4-point-correlation-function}. 
We can be afraid that the free field description of the theory (\Eq{eq:U(1)vector-gauge-matter}) may not be ideal.
In any case, we list the Feynman diagrams for  $|(\Phi \prt_\mu \prt_\nu \Phi - \prt_\mu \Phi \prt_\nu \Phi - \ii g A_{\mu \nu} \Phi^2)|^2$ in 
Figure \ref{fig:Feyman-line-matter}.
\begin{figure}[!h]
	\centering
	(i) \includegraphics[width=3.cm]{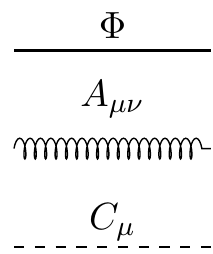}\quad\quad\\[8mm]
	(ii) \includegraphics[width=4.cm]{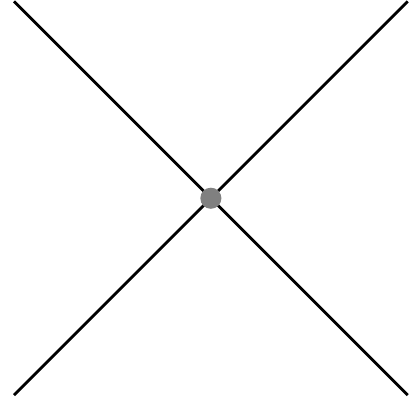}\quad\quad
	(iii) \includegraphics[width=4.cm]{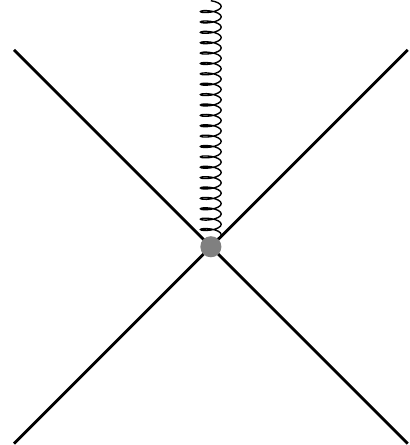}\quad\quad
	(iv) \includegraphics[width=3.5cm]{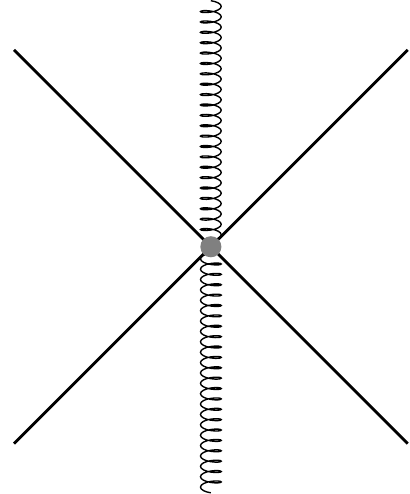}
	\caption[caption]{Feynman diagrams relevant for the tree-level interactions of
	$|(\Phi \prt_\mu \prt_\nu \Phi - \prt_\mu \Phi \prt_\nu \Phi - \ii g A_{\mu \nu} \Phi^2)|^2$.\\ 
	(i) The solid line, the curly spring, and the dashed line are for the propagators for the complex
	scalar $\Phi$, the real symmetric rank-2 tensor $A$, and the 1-form $C$ field respectively. 
	{We however should beware that the free propagator 
	may \emph{not} be an ideal concept for these highly interacting (non-Gaussian integral) field theories.}
	(ii) The $|(\Phi \prt_\mu \prt_\nu \Phi - \prt_\mu \Phi \prt_\nu \Phi )|^2$ in \Eq{eq:U(1)vector-gauge-matter} 
	contains multiple terms involving the 4 points of $\Phi$ field interactions (subfigure (ii)),
	involving their ``analogous momentums.'' (iii)  The $\ii g(\Phi \prt_\mu \prt_\nu \Phi - \prt_\mu \Phi \prt_\nu \Phi ) A_{\mu \nu} \Phi^2$ and its complex conjugation 
	in \Eq{eq:U(1)vector-gauge-matter}
		contains the 5 points of four $\Phi$ field and one $A_{\mu \nu}$ interactions (subfigure (iii)).
		(iv)  The $g^2 A_{\mu \nu}^2 |\Phi|^4$ 
	in \Eq{eq:U(1)vector-gauge-matter}
		contains the 6 points of four $\Phi$ field and two $A_{\mu \nu}$ interactions (subfigure (iv)).
	}
	\label{fig:Feyman-line-matter}
\end{figure}

On the other hand, there are free field descriptions for rank-2 symmetric tensor $A$ in \Eq{eq:ZABC}.
Thus, the propagator of $A$ read from
$|\hat F^c_{\mu \nu \xi}|^2
=\Big((\prt_\mu - \ii g_c C_\mu) A_{\nu \xi} -(\prt_{\nu} - \ii g_c C_\nu ) A_{\mu \xi}\Big)
\Big((\prt^\mu + \ii g_c C^\mu) A^{\nu \xi} -(\prt^{\nu} + \ii g_c C^\nu ) A^{\mu \xi}\Big)$,
and its Feynman diagrams, may be a more tractable problem than that of \Eq{eq:U(1)vector-gauge-matter}.
We list the Feynman diagrams for  $|\hat F^c_{\mu \nu \xi}|^2$ in 
Figure \ref{fig:Feyman-gauge}.

\begin{figure}[!h]
	\centering
	(i) \includegraphics[width=5.cm]{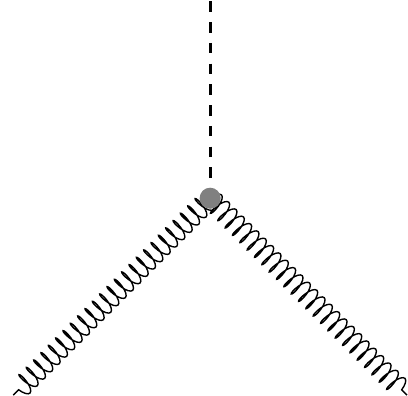}\quad\quad
	(ii) \includegraphics[width=5.cm]{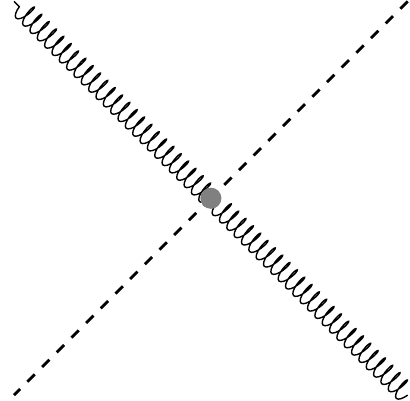}
	\caption[caption]{Feynman diagrams relevant for tree-level interactions of
	$|\hat F^c_{\mu \nu \xi}|^2$. The conventions follow \Fig{fig:Feyman-line-matter}. \\ 
	The propagator of $A$ in \Fig{fig:Feyman-line-matter} is from the $\prt A \prt A$  in \Eq{eq:|F|^2} and \Eq{eq:ZABC}.\\
	(i) The interaction term from $\ii \prt A C A$ actually \emph{vanishes}  in \Eq{eq:|F|^2},
	because $|\hat F^c_{\mu \nu \xi}|^2$ is necessarily real but $\ii \prt A C A$ involves an imaginary term. 
	(ii) The interaction term from $C A C A$ in \Eq{eq:|F|^2}.
	}
	\label{fig:Feyman-gauge}
\end{figure}

In Figure \ref{fig:Feyman-gauge} and in $|\hat F^c_{\mu \nu \xi}|^2$, it is important to notice that there is a kinetic term for $A$ in the Lagrangian $\prt A \prt A$ term.
But there is no kinetic term for $C$. In fact, as we already knew the $B$ and $C$ are dual gauge fields
 of the topological BF theory sector for the TQFT. Thus we expect the physical observables of $C$ fields captured by
the topological links of extend operators listed in Table \ref{table:TQFT} \cite{1602.05951WWY, Putrov2016qdo1612.09298, Wang1901.11537WWY}. 

\item
\emph{Universality of field theory}: We have attempted to develop our theories based on field theories in any dimension. 
We hope that the universality of our field theories can capture the wider universality of a large class of lattice models.
Although in the fracton literature \cite{RahulNandkishore2018sel1803.11196, PretkoReview2020cko2001.01722}, 
many known exactly solvable lattice models are powerful, it is still important for us to re-collect the universal data out of arbitrary lattice models.
 
\item
\emph{Lattice models}: An open question is that whether our field theories can be regularized on the lattice as a Hamiltonian quantum mechanical model?
(This question addresses a different pursuit for quantization: Instead of attempting to define our field theory as a quantum field theory, we may define our theory as a unitary 
many-body quantum mechanic model.) Several previous works can help, for example,
\begin{itemize}
\item By generalizing the Kitaev quantum double model \cite{Kitaev2003}, exactly solvable quantum Hamiltonian models of twisted topological gauge theories and Dijkgraaf-Witten theories are known and constructed in 2+1D \cite{Wan1211.3695} and in 3+1D \cite{Wang1404.7854, Wan2014woaWWH1409.3216}.
\item Exactly solvable quantum Hamiltonian models of twisted fracton models, analogous to twisted topological gauge theories  \cite{Wan1211.3695, Wan2014woaWWH1409.3216},
are constructed in 3+1D \cite{Song2018gbb1805.06899, 2019arXiv190709048SShirleySlagleXieChen}, or in any dimension \cite{LiPengYe2019tje1909.02814}.
\end{itemize}
Since our theory is a hybrid class between 
symmetric higher-rank tensor gauge theory (i.e., higher-spin gauge theory) and anti-symmetric tensor TQFT of Dijkgraaf-Witten/group-cohomology types  \cite{Wan1211.3695, Wan2014woaWWH1409.3216},
combining the above two versions of Hamiltonian models potentially can be related to the quantum Hamiltonians for \Eq{eq:ZABC} and \Eq{eq:ZABC-aniso}.
 
\item
\emph{Time crystal}: We mentioned in \Sec{sec:matter-vector-higher-moment-vector-global-symmetry}
 and in \Eq{eq:time-crystal} about a potential application of our theories to time crystal \cite{Shapere2012nq1202.2537, Wilczek2012jt1202.2539}.
Since \Eq{eq:time-crystal} shows
$$
\Phi \to \Phi e^{\ii  \Lambda_{0} \cdot x_0}=\Phi e^{\ii  \Lambda_{0} \cdot t}, \quad 
$$
while $x_0=t$ is time coordinate.
The field configuration can respect a certain periodic time constraint $t \simeq t + \frac{2\pi}{\Lambda_{0}} \Z$. 
Of course, what we said is only for the possible kinematics from the vector-like global symmetry (say at a UV effective field theory).
In order to know whether there is any time crystal order formed in the ground state (namely in the quantum vacua at low energy), 
we need to know the quantum dynamics (say as the IR fate).
This is again another hard question, related to the perturbative and quantization issue in Remark \ref{eq:Feynman-graph-quantization}
and the further much challenging non-perturbative dynamics \cite{Wilson1974skPhysRevD.10.2445ConfinementofQuarks}.

\item
\emph{Dark matter}: 

There are several possibilities for our theories to play a fundamental physics role for the dark matter:
\begin{enumerate}[label=\textcolor{blue}{(\arabic*).},ref=(\arabic*)]
\item  \label{model1}
\Eq{eq:U(1)vector-gauge-matter} or \eq{eq:U(1)vector-gauge-matter-space}, with matter and with or without gauge fields
\item  \label{model2}
\Eq{eq:ZABC} and \Eq{eq:ZABC-aniso} with gauge fields and TQFT sectors, 
and their higher-moment tensor generalization.
\end{enumerate}
In cosmology, the present research suggests that the total mass-energy of the Universe contains:
\begin{itemize}[leftmargin=2.0mm] 
\item 5\% of ordinary matter (e.g. baryonic and leptonic) and energy, 
\item 27\% dark matter, 
\item 68\% of dark energy as an unknown form of energy.
\end{itemize}
Thus, dark matter constitutes 85\% of the total mass of the Universe, while dark energy plus dark matter constitutes 95\% of the total mass-energy content  of the Universe.

Hypothetically, we can interpret our model \ref{model1} with fractonic matter $\Phi$ of complex scalar bosons as a source of dark matter decoupling from the Standard Model particle physics. We may also interpret our model \ref{model2}'s anyonic particles or anyonic extended objects (including 
anyonic strings and anyonic branes, etc.), that can be embedded or foliated in the spacetime.
In either case, the candidate dark matters from our model have the following properties: 
\begin{enumerate}[leftmargin=2.0mm, label=\arabic*.]

\item Non-baryonic and non-leptonic.

\item Non-supersymmetry (or no need to have supersymmetry).

\item {Weakly interacting massive particles (WIMPs) or extended objects} as hypothetical particles for dark matter. It can be massive if we interpret dark matters as 
anyonic particles. 
But our model also provides non-particles but extended objects as dark matter candidate: 
Anyonic extended objects (anyonic strings/branes, etc.) in \Eq{eq:ZABC} and \Eq{eq:ZABC-aniso}, e.g. in Table \ref{table:TQFT}.
They can be highly energetic gapped excitations that we cannot probe easily at low energy, but they affect the topological degenerate zero energy subspace,
see Table \ref{Table:GSD}.

\item {Weakly interacting (nearly) massless or gapless matters}:
Models of \ref{model1} (\Eq{eq:U(1)vector-gauge-matter} or \eq{eq:U(1)vector-gauge-matter-space}) can have 
 massless or gapless (or nearly depending on our input on the mass) matters, decoupled (or nearly decoupled) from Standard Model sectors.

\item Cold dark matter: Our dark matter candidates are highly quantum and highly entangled states of matter (both for particle-like or extended string/brane-like objects).
They exist and are robust at the absolute zero temperature without thermal effect. Although our candidates can still play a role in the warm and hot environment in the Universe.

\end{enumerate}

\item
\emph{Dark energy}:
Dark energy is an unknown energy form, hypothetically permeating all of space and time. 
It tends to accelerate the expansion of the Universe. 
Dark energy is the accepted hypothesis to explain that the Universe is expanding at an accelerating rate especially after the 1990s observations.
We do \emph{not} yet have a concrete model for producing such energy to expand the space from our model.
However, we should note that there is an enormous amount of static and dynamical energy from the gauge structure of our models
(in both Models \ref{model1} and \ref{model2}). 
For a simpler compact abelian tensor gauge model, the huge amount of static energy is emphasized already in \cite{Pretko2016kxt1604.05329}.
Related issues on how matters interact and their relations through gravity via a Mach's principle is revisited in \cite{Pretko2017fbf1702.07613} for the fracton context.
We expect that more immense energy stored in our non-abelian models with both gapless and gapped TQFTs sectors.
\end{enumerate}
\cred{\emph{Note added}: The sequel of this work as companions include \Ref{Wang2019cbjJWKaiXu1911.01804}
and \Ref{WXY3Wang2019mtt1912.13485}, 
etc. The idea of generating a larger family of theories by coupling to dynamical TQFTs is  pursued and developed since 2016 in 
\cite{Wang2016jdt1602.05569, 1602.05951WWY, Putrov2016qdo1612.09298}.  
The theory presented here is found in July 2019.
After the completion of this work, \Ref{Lozano2019gck1912.11224DarkMatterDarkEnergy} appears on arXiv,
suggesting a fractonic qubit lattice model with emergent matter fields and mimetic gravity. They compare such fracton models to Mimetic Dark Matter
and  Mimetic Tensor-Vector-Scalar Models in the high-energy physics literature.}

\section{Acknowledgements} 


%
JW wishes to thank Edward Witten for a conversation in 2016, 
and thank Stephen L.\;Adler for a conversation in 2019. 
JW thanks Trithep Devakul and Wilbur Shirley 
for a collaboration on the work \cite{Devakul2019dujTrithepShirleyJWang1910.01630} and valuable discussions on sub-dimensional symmetries.
JW thanks Shing-Tung Yau for informing the indispensable impact of Chern's work in 1940s \cite{chern1946characteristic} prior to 
the later formulation of Yang-Mills theory \cite{PhysRev96191YM1954}.
JW thanks the participants of ``Fracton Phases of Matter and Topological Crystalline Order on December 3-5, 2018'' in PCTS Princeton \cite{JWang2018Embeddon}
and ``Simons Collaboration on Ultra-Quantum Matter meeting on September 12-13, 2019'' at Harvard, also at a CMSA seminar \cite{JWangCMSA2019}, for helpful feedbacks.  
JW has been benefited  
by presenting this model informally at these conferences, 
special thanks the valuable feedback of Michael Hermele
 and Michael Pretko. 
JW was supported by
NSF Grant PHY-1606531 and Institute for Advanced Study. 
KX  is supported by Harvard Math Graduate Program and
``The Black Hole Initiative: Towards a Center for Interdisciplinary Research,'' Templeton Foundation. 
This work is also supported by 
NSF Grant DMS-1607871 ``Analysis, Geometry and Mathematical Physics'' 
and Center for Mathematical Sciences and Applications at Harvard University.

\appendix
\section{References}

\bibliographystyle{Yang-Mills.bst}
\bibliography{fracton.bib}

\providecommand{\href}[2]{#2}\begingroup\raggedright\begin{thebibliography}{10}

\bibitem{Maxwell1865zz}
J.~C. Maxwell, \emph{{A dynamical theory of the electromagnetic field}},
  \href{http://dx.doi.org/10.1098/rstl.1865.0008}{\emph{Phil. Trans. Roy. Soc.
  Lond.} {\bf 155} 459--512 (1865)}.

\bibitem{Weyl1929ZPhy}
H.~{Weyl}, \emph{{Elektron und Gravitation. I}},
  \href{http://dx.doi.org/10.1007/BF01339504}{\emph{Zeitschrift fur Physik}
  {\bf 56} 330--352 (1929 May)}.

\bibitem{RevModPhys13.203Pauli}
W.~Pauli, \emph{Relativistic field theories of elementary particles},
  \href{http://dx.doi.org/10.1103/RevModPhys.13.203}{\emph{Rev. Mod. Phys.}
  {\bf 13} 203--232 (1941 Jul)}.

\bibitem{chern1946characteristic}
S.-S. Chern, \emph{Characteristic classes of hermitian manifolds},
  {\emph{Annals of Mathematics} 85--121 (1946)}.

\bibitem{PhysRev96191YM1954}
C.~N. Yang and R.~L. Mills, \emph{{Conservation of Isotopic Spin and Isotopic
  Gauge Invariance}},
  \href{http://dx.doi.org/10.1103/PhysRev.96.191}{\emph{Phys. Rev.} {\bf 96}
  191--195 (1954 Oct)}.

\bibitem{RahulNandkishore2018sel1803.11196}
R.~M. Nandkishore and M.~Hermele, \emph{{Fractons}},
  \href{http://dx.doi.org/10.1146/annurev-conmatphys-031218-013604}{\emph{Ann.
  Rev. Condensed Matter Phys.} {\bf 10} 295--313 (2019)},
  [\href{https://arxiv.org/abs/1803.11196}{{\tt arXiv:1803.11196}}].

\bibitem{PretkoReview2020cko2001.01722}
M.~Pretko, X.~Chen and Y.~You, \emph{{Fracton Phases of Matter}},
  \href{https://arxiv.org/abs/2001.01722}{{\tt arXiv:2001.01722}}.

\bibitem{KalbRamond1974yc}
M.~Kalb and P.~Ramond, \emph{{Classical direct interstring action}},
  \href{http://dx.doi.org/10.1103/PhysRevD.9.2273}{\emph{Phys. Rev.} {\bf D9}
  2273--2284 (1974)}.

\bibitem{Banks2010zn1011.5120}
T.~Banks and N.~Seiberg, \emph{{Symmetries and Strings in Field Theory and
  Gravity}}, \href{http://dx.doi.org/10.1103/PhysRevD.83.084019}{\emph{Phys.
  Rev.} {\bf D83} 084019 (2011)}, [\href{https://arxiv.org/abs/1011.5120}{{\tt
  arXiv:1011.5120}}].

\bibitem{Gaiotto2014kfa1412.5148}
D.~Gaiotto, A.~Kapustin, N.~Seiberg and B.~Willett, \emph{{Generalized Global
  Symmetries}}, \href{http://dx.doi.org/10.1007/JHEP02(2015)172}{\emph{JHEP}
  {\bf 02} 172 (2015)}, [\href{https://arxiv.org/abs/1412.5148}{{\tt
  arXiv:1412.5148}}].

\bibitem{Freed2016rqq1604.06527}
D.~S. Freed and M.~J. Hopkins, \emph{{Reflection positivity and invertible
  topological phases}},  \href{https://arxiv.org/abs/1604.06527}{{\tt
  arXiv:1604.06527}}.

\bibitem{Wan2018bns1812.11967}
Z.~Wan and J.~Wang, \emph{{Higher Anomalies, Higher Symmetries, and Cobordisms
  I: Classification of Higher-Symmetry-Protected Topological States and Their
  Boundary Fermionic/Bosonic Anomalies via a Generalized Cobordism Theory}},
  \href{https://arxiv.org/abs/1812.11967}{{\tt arXiv:1812.11967}}.

\bibitem{Wang1404.7854}
J.~C. {Wang} and X.-G. {Wen}, \emph{{Non-Abelian string and particle braiding
  in topological order: Modular SL (3 ,Z ) representation and (3 +1 )
  -dimensional twisted gauge theory}},
  \href{http://dx.doi.org/10.1103/PhysRevB.91.035134}{\emph{Phys. Rev. B} {\bf
  91} 035134 (2015 Jan.)}, [\href{https://arxiv.org/abs/1404.7854}{{\tt
  arXiv:1404.7854}}].

\bibitem{Wang1405.7689}
J.~C. Wang, Z.-C. Gu and X.-G. Wen, \emph{{Field theory representation of
  gauge-gravity symmetry-protected topological invariants, group cohomology and
  beyond}}, \href{http://dx.doi.org/10.1103/PhysRevLett.114.031601}{\emph{Phys.
  Rev. Lett.} {\bf 114} 031601 (2015)},
  [\href{https://arxiv.org/abs/1405.7689}{{\tt arXiv:1405.7689}}].

\bibitem{Gu2015lfa1503.01768}
Z.-C. Gu, J.~C. Wang and X.-G. Wen, \emph{{Multi-kink topological terms and
  charge-binding domain-wall condensation induced symmetry-protected
  topological states: Beyond Chern-Simons/BF theory}},
  \href{http://dx.doi.org/10.1103/PhysRevB.93.115136}{\emph{Phys. Rev.} {\bf
  B93} 115136 (2016)}, [\href{https://arxiv.org/abs/1503.01768}{{\tt
  arXiv:1503.01768}}].

\bibitem{YeGu2015eba1508.05689}
P.~Ye and Z.-C. Gu, \emph{{Topological quantum field theory of
  three-dimensional bosonic Abelian-symmetry-protected topological phases}},
  \href{http://dx.doi.org/10.1103/PhysRevB.93.205157}{\emph{Phys. Rev.} {\bf
  B93} 205157 (2016)}, [\href{https://arxiv.org/abs/1508.05689}{{\tt
  arXiv:1508.05689}}].

\bibitem{Wang2016jdt1602.05569}
J.~C.-F. Wang, \emph{{Aspects of Symmetry, Topology and Anomalies in Quantum
  Matter}}.
\newblock PhD thesis, MIT, 2015.
\newblock \href{https://arxiv.org/abs/1602.05569}{{\tt arXiv:1602.05569}}.

\bibitem{1602.05951WWY}
J.~Wang, X.-G. Wen and S.-T. Yau, \emph{{Quantum Statistics and Spacetime
  Surgery}},  \href{https://arxiv.org/abs/1602.05951}{{\tt arXiv:1602.05951}}.

\bibitem{Tiwari2016zru1603.08429}
A.~Tiwari, X.~Chen and S.~Ryu, \emph{{Wilson operator algebras and ground
  states of coupled BF theories}},
  \href{http://dx.doi.org/10.1103/PhysRevB.95.245124}{\emph{Phys. Rev.} {\bf
  B95} 245124 (2017)}, [\href{https://arxiv.org/abs/1603.08429}{{\tt
  arXiv:1603.08429}}].

\bibitem{HeYQZheng2016xpi1608.05393}
H.~He, Y.~Zheng and C.~von Keyserlingk, \emph{{Field theories for gauged
  symmetry-protected topological phases: Non-Abelian anyons with Abelian gauge
  group $\mathbb Z_2^{3}$}},
  \href{http://dx.doi.org/10.1103/PhysRevB.95.035131}{\emph{Phys. Rev.} {\bf
  B95} 035131 (2017)}, [\href{https://arxiv.org/abs/1608.05393}{{\tt
  arXiv:1608.05393}}].

\bibitem{Putrov2016qdo1612.09298}
P.~Putrov, J.~Wang and S.-T. Yau, \emph{{Braiding Statistics and Link
  Invariants of Bosonic/Fermionic Topological Quantum Matter in 2+1 and 3+1
  dimensions}}, \href{http://dx.doi.org/10.1016/j.aop.2017.06.019}{\emph{Annals
  Phys.} {\bf 384} 254--287 (2017)},
  [\href{https://arxiv.org/abs/1612.09298}{{\tt arXiv:1612.09298}}].

\bibitem{Wang2018edf1801.05416}
J.~Wang, K.~Ohmori, P.~Putrov, Y.~Zheng, Z.~Wan, M.~Guo et~al.,
  \emph{{Tunneling Topological Vacua via Extended Operators: (Spin-)TQFT
  Spectra and Boundary Deconfinement in Various Dimensions}},
  \href{http://dx.doi.org/10.1093/ptep/pty051}{\emph{PTEP} {\bf 2018} 053A01
  (2018)}, [\href{https://arxiv.org/abs/1801.05416}{{\tt arXiv:1801.05416}}].

\bibitem{Wang1901.11537WWY}
J.~Wang, X.-G. Wen and S.-T. Yau, \emph{{Quantum Statistics and Spacetime
  Topology: Quantum Surgery Formulas}},
  \href{http://dx.doi.org/10.1016/j.aop.2019.06.002}{\emph{Annals Phys.} {\bf
  409} 167904 (2019)}, [\href{https://arxiv.org/abs/1901.11537}{{\tt
  arXiv:1901.11537}}].

\bibitem{DijkgraafWitten1989pz1990}
R.~Dijkgraaf and E.~Witten, \emph{{Topological Gauge Theories and Group
  Cohomology}},
  \href{http://dx.doi.org/10.1007/BF02096988}{\emph{Commun.Math.Phys.} {\bf
  129} 393 (1990)}.

\bibitem{Dirac1936tgProcRoySocLond}
P.~A.~M. Dirac, \emph{{Relativistic wave equations}},
  \href{http://dx.doi.org/10.1098/rspa.1936.0111}{\emph{Proc. Roy. Soc. Lond.}
  {\bf A155} 447--459 (1936)}.

\bibitem{FierzPauli1939ix}
M.~Fierz and W.~Pauli, \emph{{On relativistic wave equations for particles of
  arbitrary spin in an electromagnetic field}},
  \href{http://dx.doi.org/10.1098/rspa.1939.0140}{\emph{Proc. Roy. Soc. Lond.}
  {\bf A173} 211--232 (1939)}.

\bibitem{SinghHagen1974qz}
L.~P.~S. Singh and C.~R. Hagen, \emph{{Lagrangian formulation for arbitrary
  spin. 1. The boson case}},
  \href{http://dx.doi.org/10.1103/PhysRevD.9.898}{\emph{Phys. Rev.} {\bf D9}
  898--909 (1974)}.

\bibitem{Fronsdal1978rbPRD}
C.~Fronsdal, \emph{{Massless Fields with Integer Spin}},
  \href{http://dx.doi.org/10.1103/PhysRevD.18.3624}{\emph{Phys. Rev.} {\bf D18}
  3624 (1978)}.

\bibitem{2016arXiv160108235RRasmussenYouXu}
A.~{Rasmussen}, Y.-Z. {You} and C.~{Xu}, \emph{{Stable Gapless Bose Liquid
  Phases without any Symmetry}}, {\emph{arXiv e-prints} arXiv:1601.08235 (2016
  Jan)}, [\href{https://arxiv.org/abs/1601.08235}{{\tt arXiv:1601.08235}}].

\bibitem{Orland1981kuInstantonsDisorderAntisymmetric1982}
P.~Orland, \emph{{Instantons and Disorder in Antisymmetric Tensor Gauge
  Fields}}, \href{http://dx.doi.org/10.1016/0550-3213(82)90468-0}{\emph{Nucl.
  Phys.} {\bf B205} 107--118 (1982)}.

\bibitem{Pearson1981pkPRDPhaseStructureAntisymmetricTensorGaugeFields}
R.~B. Pearson, \emph{{Phase Structure of Antisymmetric Tensor Gauge Fields}},
  \href{http://dx.doi.org/10.1103/PhysRevD.26.2013}{\emph{Phys. Rev.} {\bf D26}
  2013 (1982)}.

\bibitem{Pretko2016kxt1604.05329}
M.~Pretko, \emph{{Subdimensional Particle Structure of Higher Rank U(1) Spin
  Liquids}}, \href{http://dx.doi.org/10.1103/PhysRevB.95.115139}{\emph{Phys.
  Rev.} {\bf B95} 115139 (2017)}, [\href{https://arxiv.org/abs/1604.05329}{{\tt
  arXiv:1604.05329}}].

\bibitem{Pretko2016lgv1606.08857}
M.~Pretko, \emph{{Generalized Electromagnetism of Subdimensional Particles: A
  Spin Liquid Story}},
  \href{http://dx.doi.org/10.1103/PhysRevB.96.035119}{\emph{Phys. Rev.} {\bf
  B96} 035119 (2017)}, [\href{https://arxiv.org/abs/1606.08857}{{\tt
  arXiv:1606.08857}}].

\bibitem{Pretko2017xar1707.03838}
M.~Pretko, \emph{{Higher-Spin Witten Effect and Two-Dimensional Fracton
  Phases}}, \href{http://dx.doi.org/10.1103/PhysRevB.96.125151}{\emph{Phys.
  Rev.} {\bf B96} 125151 (2017)}, [\href{https://arxiv.org/abs/1707.03838}{{\tt
  arXiv:1707.03838}}].

\bibitem{Slagle2018kqf1807.00827}
K.~Slagle, A.~Prem and M.~Pretko, \emph{{Symmetric Tensor Gauge Theories on
  Curved Spaces}},
  \href{http://dx.doi.org/10.1016/j.aop.2019.167910}{\emph{Annals Phys.} {\bf
  410} 167910 (2019)}, [\href{https://arxiv.org/abs/1807.00827}{{\tt
  arXiv:1807.00827}}].

\bibitem{Pretko2018jbi1807.11479}
M.~Pretko, \emph{{The Fracton Gauge Principle}},
  \href{http://dx.doi.org/10.1103/PhysRevB.98.115134}{\emph{Phys. Rev.} {\bf
  B98} 115134 (2018)}, [\href{https://arxiv.org/abs/1807.11479}{{\tt
  arXiv:1807.11479}}].

\bibitem{Gromov2018nbv1812.05104}
A.~Gromov, \emph{{Towards classification of Fracton phases: the multipole
  algebra}}, \href{http://dx.doi.org/10.1103/PhysRevX.9.031035}{\emph{Phys.
  Rev.} {\bf X9} 031035 (2019)}, [\href{https://arxiv.org/abs/1812.05104}{{\tt
  arXiv:1812.05104}}].

\bibitem{deWildPropitius1995cf9511195}
M.~D.~F. de~Wild~Propitius, \emph{{Topological interactions in broken gauge
  theories}}.
\newblock PhD thesis, Amsterdam U., 1995.
\newblock \href{https://arxiv.org/abs/hep-th/9511195}{{\tt
  arXiv:hep-th/9511195}}.

\bibitem{Shapere2012nq1202.2537}
A.~Shapere and F.~Wilczek, \emph{{Classical Time Crystals}},
  \href{http://dx.doi.org/10.1103/PhysRevLett.109.160402}{\emph{Phys. Rev.
  Lett.} {\bf 109} 160402 (2012)}, [\href{https://arxiv.org/abs/1202.2537}{{\tt
  arXiv:1202.2537}}].

\bibitem{Wilczek2012jt1202.2539}
F.~Wilczek, \emph{{Quantum Time Crystals}},
  \href{http://dx.doi.org/10.1103/PhysRevLett.109.160401}{\emph{Phys. Rev.
  Lett.} {\bf 109} 160401 (2012)}, [\href{https://arxiv.org/abs/1202.2539}{{\tt
  arXiv:1202.2539}}].

\bibitem{Wang2019cbjJWKaiXu1911.01804}
J.~Wang, K.~Xu and S.-T. Yau, \emph{{Higher-Rank Tensor Non-Abelian Field
  Theory: Higher-Moment or Subdimensional Polynomial Global Symmetry, Algebraic
  Variety, Noether's Theorem, and Gauging}},
  \href{https://arxiv.org/abs/1911.01804}{{\tt arXiv:1911.01804}}.

\bibitem{WXY3Wang2019mtt1912.13485}
J.~Wang and S.-T. Yau, \emph{{Non-Abelian Gauged Fractonic Matter Field Theory:
  New Sigma Models, Superfluids and Vortices}},
  \href{https://arxiv.org/abs/1912.13485}{{\tt arXiv:1912.13485}}.

\bibitem{Wang2014wkaLevin1412.1781}
C.~Wang and M.~Levin, \emph{{Topological invariants for gauge theories and
  symmetry-protected topological phases}},
  \href{http://dx.doi.org/10.1103/PhysRevB.91.165119}{\emph{Phys. Rev.} {\bf
  B91} 165119 (2015)}, [\href{https://arxiv.org/abs/1412.1781}{{\tt
  arXiv:1412.1781}}].

\bibitem{Wilson1974skPhysRevD.10.2445ConfinementofQuarks}
K.~G. Wilson, \emph{{Confinement of Quarks}},
  \href{http://dx.doi.org/10.1103/PhysRevD.10.2445}{\emph{Phys. Rev.} {\bf D10}
  2445--2459 (1974)}.

\bibitem{Vijay2015mka1505.02576VijayHaahFu}
S.~Vijay, J.~Haah and L.~Fu, \emph{{A New Kind of Topological Quantum Order: A
  Dimensional Hierarchy of Quasiparticles Built from Stationary Excitations}},
  \href{http://dx.doi.org/10.1103/PhysRevB.92.235136}{\emph{Phys. Rev.} {\bf
  B92} 235136 (2015)}, [\href{https://arxiv.org/abs/1505.02576}{{\tt
  arXiv:1505.02576}}].

\bibitem{Vijay2016phm1603.04442}
S.~Vijay, J.~Haah and L.~Fu, \emph{{Fracton Topological Order, Generalized
  Lattice Gauge Theory and Duality}},
  \href{http://dx.doi.org/10.1103/PhysRevB.94.235157}{\emph{Phys. Rev.} {\bf
  B94} 235157 (2016)}, [\href{https://arxiv.org/abs/1603.04442}{{\tt
  arXiv:1603.04442}}].

\bibitem{2005PRL0404182Chamon}
C.~{Chamon}, \emph{{Quantum Glassiness in Strongly Correlated Clean Systems: An
  Example of Topological Overprotection}},
  \href{http://dx.doi.org/10.1103/PhysRevLett.94.040402}{\emph{"Phys. Rev.
  Lett."} {\bf 94} 040402 (2005 Jan)},
  [\href{https://arxiv.org/abs/cond-mat/0404182}{{\tt
  arXiv:cond-mat/0404182}}].

\bibitem{1108.2051CastelnovoChamon}
C.~{Castelnovo} and C.~{Chamon}, \emph{{Topological quantum glassiness}},
  \href{http://dx.doi.org/10.1080/14786435.2011.609152}{\emph{Philosophical
  Magazine} {\bf 92} 304--323 (2012 Jan)},
  [\href{https://arxiv.org/abs/1108.2051}{{\tt arXiv:1108.2051}}].

\bibitem{2011PhRvAHaah1101.1962}
J.~{Haah}, \emph{{Local stabilizer codes in three dimensions without string
  logical operators}},
  \href{http://dx.doi.org/10.1103/PhysRevA.83.042330}{\emph{"Phys. Rev. A"}
  {\bf 83} 042330 (2011 Apr)}, [\href{https://arxiv.org/abs/1101.1962}{{\tt
  arXiv:1101.1962}}].

\bibitem{JWang2018Embeddon}
J.~C. {Wang}, \emph{{Quantization of Spacetime Emedding: Embeddon (2018)}},
  {\emph{Unpublished notes, drawn inspiration from PCTS Princeton workshop:
  ``David Huse 60th Birthday'' April 2018, and ``Fracton Phases of Matter and
  Topological Crystalline Order'' December 3-5, 2018} (2018)}.

\bibitem{Slagle2017wrc1708.04619}
K.~Slagle and Y.~B. Kim, \emph{{Quantum Field Theory of X-Cube Fracton
  Topological Order and Robust Degeneracy from Geometry}},
  \href{http://dx.doi.org/10.1103/PhysRevB.96.195139}{\emph{Phys. Rev.} {\bf
  B96} 195139 (2017)}, [\href{https://arxiv.org/abs/1708.04619}{{\tt
  arXiv:1708.04619}}].

\bibitem{Shirley2017suz1712.05892}
W.~Shirley, K.~Slagle, Z.~Wang and X.~Chen, \emph{{Fracton Models on General
  Three-Dimensional Manifolds}},
  \href{http://dx.doi.org/10.1103/PhysRevX.8.031051}{\emph{Phys. Rev.} {\bf X8}
  031051 (2018)}, [\href{https://arxiv.org/abs/1712.05892}{{\tt
  arXiv:1712.05892}}].

\bibitem{You2019ciz1904.11530}
Y.~You, T.~Devakul, S.~L. Sondhi and F.~J. Burnell, \emph{{Fractonic
  Chern-Simons and BF theories}},  \href{https://arxiv.org/abs/1904.11530}{{\tt
  arXiv:1904.11530}}.

\bibitem{Fuji2019qea1908.02257}
Y.~Fuji, \emph{{Anisotropic layer construction of anisotropic fracton models}},
   \href{https://arxiv.org/abs/1908.02257}{{\tt arXiv:1908.02257}}.

\bibitem{BulmashMaissamBarkeshli2019taq1905.05771}
D.~Bulmash and M.~Barkeshli, \emph{{Gauging fractons: immobile non-Abelian
  quasiparticles, fractals, and position-dependent degeneracies}},
  \href{https://arxiv.org/abs/1905.05771}{{\tt arXiv:1905.05771}}.

\bibitem{PremWilliamson2019etl1905.06309}
A.~Prem and D.~J. Williamson, \emph{{Gauging permutation symmetries as a route
  to non-Abelian fractons}},  \href{https://arxiv.org/abs/1905.06309}{{\tt
  arXiv:1905.06309}}.

\bibitem{Vijay2017cti1706.07070}
S.~Vijay and L.~Fu, \emph{{A Generalization of Non-Abelian Anyons in Three
  Dimensions}},  \href{https://arxiv.org/abs/1706.07070}{{\tt
  arXiv:1706.07070}}.

\bibitem{Song2018gbb1805.06899}
H.~Song, A.~Prem, S.-J. Huang and M.~A. Martin-Delgado, \emph{{Twisted Fracton
  Models in Three Dimensions}},
  \href{http://dx.doi.org/10.1103/PhysRevB.99.155118}{\emph{Phys. Rev.} {\bf
  B99} 155118 (2019)}, [\href{https://arxiv.org/abs/1805.06899}{{\tt
  arXiv:1805.06899}}].

\bibitem{PremHuangSong2018jsn1806.04687}
A.~Prem, S.-J. Huang, H.~Song and M.~Hermele, \emph{{Cage-Net Fracton Models}},
  \href{http://dx.doi.org/10.1103/PhysRevX.9.021010}{\emph{Phys. Rev.} {\bf X9}
  021010 (2019)}, [\href{https://arxiv.org/abs/1806.04687}{{\tt
  arXiv:1806.04687}}].

\bibitem{2007arXiv0711.0977LJohnLoftin}
J.~{Loftin}, \emph{{Affine Hermitian-Einstein Metrics}}, {\emph{arXiv e-prints}
  arXiv:0711.0977 (2007 Nov)}, [\href{https://arxiv.org/abs/0711.0977}{{\tt
  arXiv:0711.0977}}].

\bibitem{LWXYfracton}
J.~Lofte, J.~Wang, K.~Xu and S.-T. Yau, \emph{In preparation}, {\emph{To
  appear}}.

\bibitem{SeibergF2019}
N.~Seiberg, \emph{{Field Theories With a Vector Global Symmetry}},
  \href{https://arxiv.org/abs/1909.10544}{{\tt arXiv:1909.10544}}.

\bibitem{Kitaev2003}
A.~Y. {Kitaev}, \emph{{Fault-tolerant quantum computation by anyons}},
  \href{http://dx.doi.org/10.1016/S0003-4916(02)00018-0}{\emph{Annals of
  Physics} {\bf 303} 2--30 (2003 Jan.)},
  [\href{https://arxiv.org/abs/quant-ph/9707021}{{\tt
  arXiv:quant-ph/9707021}}].

\bibitem{Wan1211.3695}
Y.~{Hu}, Y.~{Wan} and Y.-S. {Wu}, \emph{{Twisted quantum double model of
  topological phases in two dimensions}},
  \href{http://dx.doi.org/10.1103/PhysRevB.87.125114}{\emph{Phys. Rev. B} {\bf
  87} 125114 (2013 Mar.)}, [\href{https://arxiv.org/abs/1211.3695}{{\tt
  arXiv:1211.3695}}].

\bibitem{Wan2014woaWWH1409.3216}
Y.~Wan, J.~C. Wang and H.~He, \emph{{Twisted Gauge Theory Model of Topological
  Phases in Three Dimensions}},
  \href{http://dx.doi.org/10.1103/PhysRevB.92.045101}{\emph{Phys. Rev.} {\bf
  B92} 045101 (2015)}, [\href{https://arxiv.org/abs/1409.3216}{{\tt
  arXiv:1409.3216}}].

\bibitem{2019arXiv190709048SShirleySlagleXieChen}
W.~{Shirley}, K.~{Slagle} and X.~{Chen}, \emph{{Twisted foliated fracton
  phases}}, {\emph{arXiv e-prints} arXiv:1907.09048 (2019 Jul)},
  [\href{https://arxiv.org/abs/1907.09048}{{\tt arXiv:1907.09048}}].

\bibitem{LiPengYe2019tje1909.02814}
M.-Y. Li and P.~Ye, \emph{{Exactly Solvable Models for Fracton Topological
  Order in All Dimensions}},  \href{https://arxiv.org/abs/1909.02814}{{\tt
  arXiv:1909.02814}}.

\bibitem{Pretko2017fbf1702.07613}
M.~Pretko, \emph{{Emergent gravity of fractons: Mach's principle revisited}},
  \href{http://dx.doi.org/10.1103/PhysRevD.96.024051}{\emph{Phys. Rev.} {\bf
  D96} 024051 (2017)}, [\href{https://arxiv.org/abs/1702.07613}{{\tt
  arXiv:1702.07613}}].

\bibitem{Lozano2019gck1912.11224DarkMatterDarkEnergy}
L.~Lozano and H.~Garcia-Compean, \emph{{Emergent Dark Matter and Dark Energy
  from a Lattice Model}},  \href{https://arxiv.org/abs/1912.11224}{{\tt
  arXiv:1912.11224}}.

\bibitem{Devakul2019dujTrithepShirleyJWang1910.01630}
T.~Devakul, W.~Shirley and J.~Wang, \emph{{Strong planar subsystem
  symmetry-protected topological phases and their dual fracton orders}},
  \href{https://arxiv.org/abs/1910.01630}{{\tt arXiv:1910.01630}}.

\bibitem{JWangCMSA2019}
J.~C. {Wang}, \emph{{Higher-Rank Tensor Field Theory of Non-Abelian Fracton and
  Embeddon}}, {\emph{Quantum Matter/Quantum Field Theory seminar, Quantum
  Matter in Mathematics and Physics at Harvard CMSA, September 25th, 2019
  \href{https://www.youtube.com/watch?v=Ibz3dUq35O8}{https://www.youtube.com/watch?v=Ibz3dUq35O8}}
  (2019)}.

\end{thebibliography}\endgroup

\end{document}